\documentclass[reprint,longbibliography,aps,prb,superscriptaddress]{revtex4-2}
\usepackage[english]{babel}
\usepackage{amsmath,amsthm,amssymb,bbm,amsfonts,color,graphicx}
\usepackage{url}
\usepackage{tikz}
\usepackage{enumerate}
\usepackage{mathrsfs}
\usepackage{relsize}
\usepackage[colorlinks=true, allcolors=blue]{hyperref}
\def\bS{{\bf S}}

\newcommand{\be}{\beta}
\newcommand{\la}{\lambda}

\newcommand{\al}{\alpha}

\newcommand{\HEISal}{{\mrm{HEIS}(\alpha)}}
\newcommand{\HEIS}{{\mrm{HEIS}}}
\newcommand{\RZ}{\mathrm{RZ}}

\newcommand{\fSim}{{\mathrm{fSim}}}
\newcommand{\ii}{\mathrm{i}}
\newcommand{\ee}{\mathrm{e}}
\def\brapsith{{\bra{\theta}}}
\def\psith{{\ket{\theta}}}
\newcommand{\eq}[1]{Eq.~(\ref{eq:#1})}
\newcommand{\fig}[1]{Fig.~\ref{fig:#1}}
\renewcommand{\sec}[1]{Sec.~\ref{sec:#1}}

\newcommand{\app}[1]{Appendix~\ref{sec:#1}}
\newcommand{\SWAP}{\mathrm{SWAP}}
\newcommand{\Eth}{E(\theta)}

\newcommand{\HH}{{\mathcal H}}
\newcommand{\mc}[1]{\mathcal{#1}}
\newcommand{\id}{\mathbbm{1}}
\newcommand{\tr}{\mathrm{tr}}

\newcommand{\nn}{\nonumber}

\newcommand{\ket}[1]{\left\lvert #1 \right\rangle} % for Dirac bras
\newcommand{\bra}[1]{\left\langle #1 \right\rvert} % for Dirac kets
\newcommand{\braket}[2]{\left\langle #1 \vphantom{#2} \right|\hspace{-.26em}
	\left. #2 \vphantom{#1} \right\rangle} % for Dirac brackets
\newcommand{\ketbra}[2]{\ket{#1}\!\!\bra{#2}}

\let\perptmp\perp
\renewcommand{\perp}{{\! \mathsmaller{\perptmp}}}

\newcommand{\mrm}{\mathrm}

\newcommand{\nodagger}{{\phantom{\dagger}}}

\newcommand{\psiinit}{{\ket{\psi_\mrm{init}}}}

\def\bi{{\bf i}}

\newcommand{\poly}{\mathrm{poly}\,}

\newcommand{\KVQE}{\mathrm{KVQE}}

\makeatletter
\def\@bibdataout@aps{%
\immediate\write\@bibdataout{%
@CONTROL{%
apsrev41Control%
\longbibliography@sw{%
    ,author="08",editor="1",pages="1",title="0",year="1"%
    }{%
    ,author="08",editor="1",pages="1",title="",year="1"%
    }%
  }%
}%
\if@filesw \immediate \write \@auxout {\string \citation {apsrev41Control}}\fi
}
\makeatother

\begin{document}
\author{Joris Kattem\"olle}
	\affiliation{Institute for Theoretical Physics, University of Amsterdam,
		Science Park 904, Amsterdam, Netherlands}
	\affiliation{QuSoft, CWI,
	  Science Park 123, Amsterdam, Netherlands}
	\affiliation{Department of Physics, University of Konstanz, D-78457 Konstanz, Germany}
	\author{Jasper van Wezel}
	\title{Variational quantum eigensolver for the \\ Heisenberg antiferromagnet on the kagome lattice}
	\affiliation{Institute for Theoretical Physics, University of Amsterdam,
		Science Park 904, Amsterdam, Netherlands}
	\affiliation{QuSoft, CWI,
	  Science Park 123, Amsterdam, Netherlands}

	\begin{abstract}
	  Establishing the nature of the ground state of the Heisenberg antiferromagnet (HAFM) on the kagome lattice is well known to be a prohibitively difficult problem for classical computers. Here, we give a detailed proposal for a Variational Quantum Eigensolver (VQE) intending to solve this physical problem on a quantum computer. At the same time, this VQE constitutes an explicit experimental proposal for showing a useful quantum advantage on Noisy Intermediate-Scale Quantum (NISQ) devices because of its natural hardware compatibility. We classically emulate noiseless and noisy quantum computers with either 2D-grid or all-to-all connectivity and simulate patches of the kagome HAFM of up to 20 sites. In the noiseless case, the ground-state energy, as found by the VQE, approaches the true ground-state energy exponentially as a function of the circuit depth. Furthermore, VQEs for the HAFM on any graph can inherently perform their quantum computations in a decoherence-free subspace that protects against collective longitudinal and collective transversal noise, adding to the noise-resilience of these algorithms. Nevertheless, the extent of the effects of other noise types suggests the need for error mitigation and performance targets alternative to high-fidelity ground-state preparation, even for essentially hardware-native VQEs.
	\end{abstract}
	\maketitle

	\section{Introduction}
\begin{figure}[b]
		\centering
		$\KVQE_K$: ansatz	\includegraphics[width=1\linewidth]{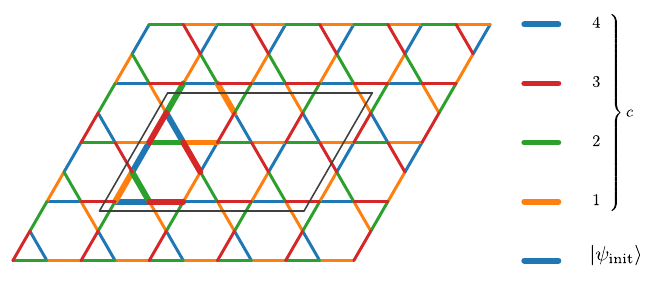}
		\caption{\label{fig:periodic_patch} The kagome lattice, with edges colored according to a  minimal edge coloring with the smallest possible coloring unit cell (thicker lines). The ansatz of $\KVQE_K$, with initial state $\psiinit$ and circuit cycle $c$, is derived from this edge coloring by identifying every subset of edges having the same color with a layer of parametrized gates. The gray parallelogram delineates a periodic patch that is simulated in this paper.}
	\end{figure}

	Despite decades of developments in numerical methods, the ground-state properties of the Heisenberg antiferromagnet (HAFM) on the kagome lattice (\fig{periodic_patch}) remain elusive, owing to its geometrical frustration. The kagome HAFM is the prime candidate for exhibiting a new phase of magnetism, the quantum spin liquid (QSL) \cite{jiang2008density,yan2011spin, anderson1973resonating}, and forms a model for the magnetic properties of minerals like Herbertsmithite \cite{norman2016herbertsmithite}. Approaches towards solving this ground-state problem include
	exact diagonalization of finite-size patches \cite{lauchli2019s} and the density matrix renormalization group (DMRG) method \cite{yan2011spin}. Next to the QSL, a Valence Bond Crystal (VBC) has been proposed as the ground state of the kagome HAFM \cite{ marston1991spin,nikolic2003physics,singh2007ground,singh2008triplet,evenbly2010frustrated}. (See Ref. \cite{lauchli2019s} and references therein for a more complete overview of the techniques and proposals.) All classical methods for finding the ground state of the kagome HAFM are ultimately limited, for example by the inability to treat large patches (exact diagonalization), or the inability to describe highly entangled states (DMRG).

	Quantum computation is a new player in this field that brings with it entirely novel possibilities. One method for finding ground states on a quantum computer is the Variational Quantum Eigensolver (VQE) \cite{mcclean2016theory,peruzzo2014variational}. VQEs are especially suited for Noisy Intermediate-Scale Quantum (NISQ) \cite{preskill2018quantum} devices because of their relatively mild circuit depth requirements and inherent noise resilience \cite{mcclean2016theory,peruzzo2014variational,o2016scalable, reiner2019finding}. A VQE is a variational method. What sets the VQE apart from classical variational methods is that the parametrized state is obtained by applying a parametrized quantum circuit to some easy-to-prepare reference state of the quantum computer's register. The energy of the resulting state is obtained by performing measurements on many copies of that state. (Generally, the classical simulation of state preparation and measurement is intractable.) Parameter variation and optimization are still performed by a classical routine. So, a VQE can be seen as a classical variational method that uses a quantum computer as a subroutine for its function calls to the energy landscape.

	Quantum computers can already outperform classical computers, and have hence obtained what is called quantum supremacy or a quantum advantage \cite{arute2019quantum,zhong2020quantum}. However, the tasks for which quantum computers can currently outperform classical computers have no known application; these tasks were designed purely for showing a quantum advantage. The milestone of a \emph{useful} quantum advantage, where a quantum computer performs a useful task that cannot be performed on any classical computer, is still ahead  \cite{arute2019quantum,zhong2020quantum}.

	In this paper, we design a VQE for the kagome HAFM as an explicit proposal for showing a useful quantum advantage on NISQ devices and for a novel method for finding the ground-state properties of the kagome HAFM. For NISQ VQE algorithms to give an advantage over purely classical methods, it has recently become increasingly clear that the structure of the problem must be close to the quantum hardware the VQE is run on \cite{franca2020limitations}. For this reason, the central point of our paper is careful consideration of hardware compatibility and model choice. We explicitly consider the case where the quantum computer has the limited connectivity of a 2D grid. We refer to the resulting VQE as $\KVQE_{G}$ for short. We refer to the VQE as  $\KVQE_K$ whenever we assume a quantum computer with at least kagome connectivity (this includes architectures with all-to-all connectivity). We tested $\KVQE_{G/K}$ by noiseless and noisy emulation on a classical computer in a host of cases, which are detailed in \sec{classical_implementation}. For both $\KVQE_{G}$ and $\KVQE_K$, the results of one case are reported in this paper. The remainder can be found in the Supplemental Material~\cite{HeisenbergVQE}. $\KVQE_{G/K}$ only gives information about the ground-state properties of finite-size patches. As is standard practice in numerical methods, properties of the infinite system can be derived from a finite-size scaling approach, as in Ref. \cite{lauchli2011ground}. Such an approach is beyond the scope of the current paper, since there are not enough patches for which it is tractable to emulate KVQE classically, but would be fully possible on a quantum computer with on the order of 50 qubits.

\begin{figure}[t]
		\centering
		$\KVQE_{G}$: ansatz

		\vspace{1em}

		\includegraphics[width=\linewidth]{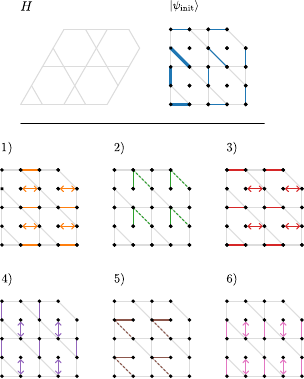}
		\caption{\label{fig:kagome_ansatz}(\textbf{Top left}) The kagome lattice, of which only the 20-site open patch simulated by $\KVQE_{G}$ is shown. For the kagome HAFM, spin-1/2 particles are placed on the vertices and the Heisenberg interaction is defined along the edges [\eq{HAFM}]. (\textbf{Top right}) The initial state $\ket{\psi_\mrm{init}}$. Black dots represent qubits and solid blue lines represent singlets. A sheered kagome lattice is added in the background in gray as a guide to the eye. The bolder solid lines form one unit cell of the dimer covering, which can be used to extend the current ansatz to systems of arbitrary size. For open boundaries, some patching of the regular dimer covering is needed. Here, this consists of the two singlets at the far right. (\textbf{Bottom}) The cycle $c$ that is applied to $\psiinit$ $p$ times to obtain the ansatz state. Solid colored lines represent $\HEIS(\alpha)$ gates. A $\HEIS(\alpha)$ gate evolves two qubits according to the Heisenberg exchange interaction for a time given by $\alpha$ [\eq{HEIS_gate}]. We use one parameter per HEIS gate, leading to $M=30 p$ parameters in total. Two-headed arrows indicate $\SWAP$ gates. Dashed, colored lines indicate along which bond of the kagome lattice the HEIS gates of that layer act effectively. The unit cell of the cycle equals that of the kagome lattice itself, and can hence be straightforwardly extended to larger system sizes.
		}

	\end{figure}

	$\KVQE_{G/K}$ uses the Hamiltonian Variational Ansatz (HVA) \cite{wecker2015progress}. In the HVA, to find the ground state of a Hamiltonian $H$, the ansatz state is obtained by first preparing some known, easy-to-prepare ground state of a Hamiltonian $H_\mrm{init}$. Thereafter, this state is evolved sequentially by terms from $H$ and $H_\mrm{init}$. (Commuting terms may be evolved by simultaneously.) In the resulting circuit, every gate corresponds to time evolution along a term in $H$ or $H_\mrm{init}$, where the parameter of that gate is set by the time duration of the evolution. The HVA itself does not specify $H_\mrm{init}$ nor the sequence of terms the initial state is evolved by. Our choice of $H_\mrm{init}$ and gate sequences for the cases presented in this paper are detailed in Figs. \ref{fig:periodic_patch} and \ref{fig:kagome_ansatz}.

\begin{figure}[t]
		\centering
		CVQE: ansatz
		\vspace{.5em}
		\hrule
		\vspace{1em}
		\includegraphics[width=1\linewidth]{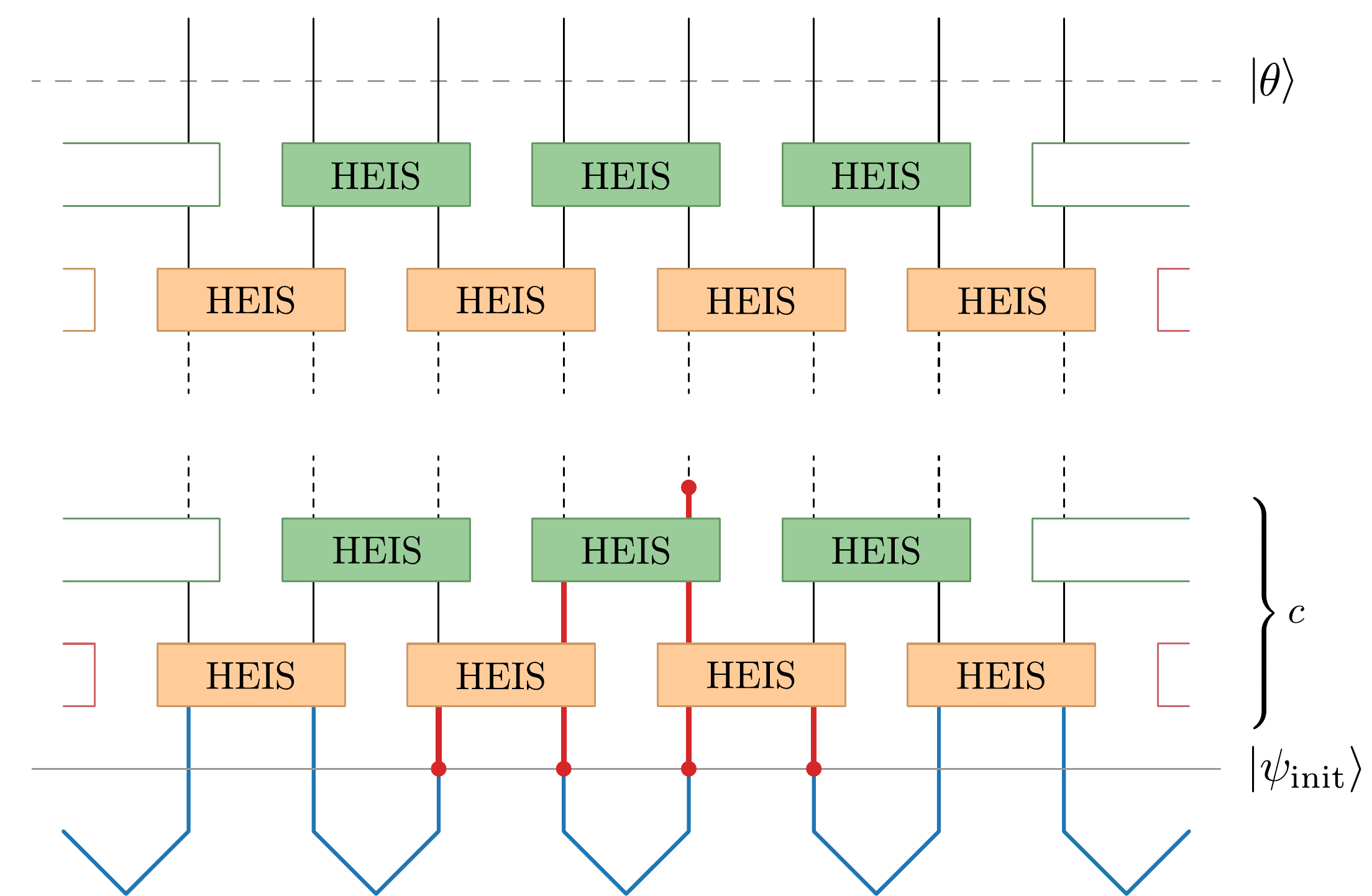}
		\caption{\label{fig:chain_circuit}
			The ansatz for CVQE, with time running from bottom to top. The initial state $\psiinit$ consists of $n/2$ adjacent singlets, displayed in blue. The circuit that is applied to $\psiinit$ consists of $p$ repetitions of a cycle $c$, each time with new parameters, and with one parameter per gate. For one of the qubits, the past light cone (discussed in \sec{chain}) that is due to a single cycle $c$ (thus excluding the singlet generation) is displayed with thicker, red lines.
		}
	\end{figure}

	In addition, we propose to run a similar VQE for the HAFM on the periodic chain as an intermediate goal. We call this VQE CVQE for short. In contrast to $\KVQE_{G/K}$, a system-size resource scaling, as presented in this paper, is possible for CVQE due to the many classically emulatable patches. Furthermore, CVQE is a suitable benchmark problem for quantum hardware because, in contrast to the kagome HAFM, the ground state of the HAFM on the chain can be computed efficiently classically through the Bethe ansatz \cite{bethe1931theorie,franchini2017introduction,caux2009correlation}. This opens the possibility of comparing the optimal energy found by the VQE running on a quantum computer against the exact ground-state energy, even for chains with hundreds of sites. It is only for systems of up to approximately 50 qubits that similar benchmarks can be made for the kagome HAFM  \cite{lauchli2019s}. The explicit gate sequence we use for CVQE is depicted in \fig{chain_circuit}. CVQE is similar to a VQE in Ref.~\cite{ho2019efficient}. Differences with the results in Ref.~\cite{ho2019efficient} are that we simulate a periodic chain instead of an open chain and that we use one parameter per gate instead of one parameter per layer. We go beyond the results in Ref. \cite{ho2019efficient} by the study of larger circuit depths and system sizes, a finite-size resource scaling, and a study of noise effects.

	\subsection{Hardware compatibility}
    $\KVQE_{G}$ is exceptionally compatible with NISQ hardware for three reasons. First, because the Hamiltonian of the kagome HAFM is a spin Hamiltonian, it is directly a Hamiltonian defined on qubits, eliminating any overhead from fermion-to-spin maps. In contrast, many Hamiltonians for which VQEs are proposed are fermionic. Examples include those in quantum chemistry \cite{peruzzo2014variational, o2016scalable, grimsley2019adaptive}, and the Fermi-Hubbard model \cite{wecker2015progress,cade2020strategies,reiner2019finding,arute2020observation}. For a VQE to solve for the ground state of a fermionic Hamiltonian, it first needs to be mapped to a spin Hamiltonian, for example, by the Jordan-Wigner \cite{nielsen2005fermionic}, Bravyi-Kitaev \cite{bravyi2002fermionic}, or ternary-tree \cite{jiang2020optimal} transformations. Fermion to spin maps either increase the nonlocality of terms in the Hamiltonian or introduce additional qubits, in any case leading to an overhead in quantum resources.

	A second reason that $\KVQE_{G}$ is close to NISQ hardware is that its gates are essentially native to multiple NISQ architectures. The HVA requires time evolution generated by terms in the Hamiltonian. For the HAFM, this amounts to turning on an exchange interaction between qubits, which is native to quantum dot architectures \cite{van2021quantum, barthelemy2013quantum,loss1998quantum,hendrickx2021four}. In \sec{hardware_implementation}, we show this interaction can also be realized on the superconducting hardware by Google AI Quantum \cite{foxen2020demonstrating} using a single native two-qubit gate and at most four single-qubit gates. (If a two-qubit gate is equal to a single native gate up to single-qubit rotations, we call the former gate `essentially native'.)

	Next to the HVA, a well-known type of ansatz is the Hardware-Efficient Ansatz (HEA) \cite{kandala2017hardware, peruzzo2014variational}. The HEA is hardware inspired; the circuit generating the ansatz state consists, by definition, of gates native to the hardware, avoiding the need to compile the ansatz into native gates. However, the HEA suffers from the `barren plateau' problem: the gradient of the energy cost function is exponentially small in the number of parameters \cite{mcclean2018barren}. The HVA, on the other hand, is problem inspired, and there is some evidence that it does not suffer from the barren plateau problem \cite{wiersema2020exploring}. However,  for execution on a quantum computer, gates in the HVA generally need to be compiled to gates native to that quantum computer. This increases the circuit depth, which is undesirable for NISQ devices. For $\KVQE_{G}$, such compilation is not required on quantum dot architectures, and only minimal compilation that does not increase the number of two-qubit gates is needed on Google's hardware. So, to summarize, $KVQE_G$ on quantum dot and Google's superconducting hardware has the unique property that the HVA is essentially equal to the HEA.

	Finally, $\KVQE_{G}$ is close to NISQ hardware because it runs on hardware with the connectivity of a 2D grid with minimal overhead. This is the connectivity that is also required for the surface code \cite{fowler2012surface}, and therefore much effort is put into designing platforms with grid connectivity~\cite{hill2015surface,andersen2020repeated,versluis2017scalable,arute2019quantum,hendrickx2021four}.

	Given a quantum computer with kagome connectivity, the hardware compatibility of $\KVQE_{K}$ goes even further because it would require no geometric mapping at all. Quantum architectures with kagome connectivity also form a proposal for fault-tolerant quantum computation \cite{hutter2015parafermions}, and superconducting quantum processors with said connectivity are under theoretical \cite{schmidt2012circuit, kim2017scalable,kim2019quantum} and experimental \cite{underwood2012low,kollar2019hyperbolic} investigation. Unlike superconducting platforms, ion-trap quantum computers feature all-to-all connectivity \cite{cirac1996quantum,molmer1999multiparticle,kielpinski2002architecture,monz201114-qubit}, which naturally accommodates kagome connectivity. CVQE requires line-connectivity, which is a limited connectivity available on most quantum computing platforms.

	All circuits in our paper can directly be used for the dynamic quantum simulation of the HAFM on the kagome lattice, offering an additional route towards a quantum advantage \cite{childs2018towards}. The dynamical simulation of the kagome HAFM has arguably less scientific relevance, and we therefore focus on ground state simulation in the current paper.

	\subsection{Summary of numerical results}\label{sec:sum_num}
	The task for $\KVQE_{G}$ reported here is the simulation of a 20-site open patch of the kagome HAFM. For this case, we assume a 24-qubit quantum computer with grid connectivity, the ability to natively implement the exchange interaction, SWAP, $\sqrt{Z}$, and $X$ gates, and where every exchange interaction gate in the ansatz circuit has its own parameter. We emulate this quantum computer classically, as detailed in \sec{classical_implementation}. The optimal state obtained by KVQE is compared to the exact ground state of the 20-site patch, which is obtained by exact diagonalization. In the noiseless case, we find that the optimal energy obtained by $\KVQE_{G}$ approaches the true ground state energy exponentially as a function of the circuit depth. Also, the fidelity (overlap squared) between the optimal state and the true ground state approaches unity exponentially in the circuit depth. A fidelity of $>$99.9\% is reached at a circuit depth of 99. More details are found in \fig{kagome_results} and \sec{kagome}.

	The task for $\KVQE_K$ reported here is the simulation of an 18-site \emph{periodic} patch. For this case, we assume a noiseless 18-qubit quantum computer with all-to-all connectivity, and the other settings as before \footnote{For the sake of comparison of the gate count, we assume the same native gates as before. To the best of our knowledge, quantum computers that support both the essentially native implementation of the exchange interaction and kagome connectivity do not exist yet, so on current devices with all-to-all connectivity there will be some transpilation overhead.}. Also the optimal energy obtained by $\KVQE_{K}$ approaches the true ground state energy exponentially as a function of the circuit depth. The fidelity initially plateaus, but then continues exponentially towards unity. A fidelity of $>$99.9\% is reached at a circuit of depth 151. We find the increment of the circuit depth required for obtaining a fidelity of  $>$99.9\% (as compared to the previous paragraph) not to be the result of the different ansatz, but rather to be the result of the difference in the systems that are simulated. More details are found in \fig{kagome_on_kagome_results} and \sec{kagome}.

	Under the same hardware assumptions, we let CVQE simulate a 20-site periodic chain. The fidelity and energy of the optimal state found by CVQE improve exponentially as a function of circuit depth, with a sudden improvement of performance after $p_\mrm{crit}=5$ due to the availability of system-wide entanglement after $p_\mrm{crit}$. The optimal state found by CVQE reaches a fidelity of $>$99.9\% at a circuit of depth 19. More details are found in \fig{chain_results} and \sec{chain}.

Considering the effects of noise, we find that without any modification, the HVA for the HAFM on any graph intrinsically performs its quantum computations inside a Decoherence-Free Subspace (DFS) \cite{palma1996quantum,zanardi1997error,duan1998reducing,lidar1998decoherence,kattemolle2019dynamical} that protects against noise that couples to the quantum registers' total spin operator in any direction (\sec{sym}). This type of noise occurs in realistic scenarios, for example, when a single bath of long-wavelength modes of a bosonic bath (like the electromagnetic field) couples to the qubits collectively transversely \cite{dicke1964coherence,kirton2018dicke} and/or collectively longitudinally \cite{palma1996quantum,monz201114-qubit}. A VQE whose states occurring during the ansatz preparation remain in a DFS was treated before in Ref. \cite{kokail2019self}, but differs from our setting in multiple essential ways (\app{symmetry}). The fact that our proposed VQEs naturally perform their computations in a DFS has direct practical consequences, since it adds to the natural noise robustness of these VQEs against realistic noise sources.

We also study the degradation of the fidelity $\mc F$ between the true ground state and the output state of the VQE, as a function of the error probability $p_e$, for local depolarizing and bit-flip noise models (\sec{noisy_results}). We find that to prepare a ground state with fidelity $\mc F$, the error probability must be less than $p_e\approx (1-\mc F)/(nd)$, with $n$ the number of sites and $d$ the depth of the circuit.  Using this relation, we estimate that to obtain the kagome HAFM ground state on a system of a hundred sites with a fidelity of 99.9\%, error rates as low as $p_e=10^{-7}$ may be needed. This is several orders of magnitude below known error thresholds for fault-tolerant quantum computation~\cite{wang2011surface,dalton2022variational}. This makes the high-fidelity preparation of ground states of systems with on the order of a hundred sites unlikely on noisy hardware. Nevertheless, other classically nontrivial performance targets might be achieved by noisy hardware (\sec{discussion}), while error mitigation may alleviate the stringent constraints on the physical error rates~\cite{temme2017error,koczor2021exponential,huggins2021virtual,song2019quantum,bultrini2021quantum,berg2022probabilistic,bravi2022future,bonet2018low-cost,sagastizabal2019experimental}.

	\section{Methods}
	\subsection{VQE}
	In this subsection, we give a more detailed introduction to VQEs. It may be skipped by readers already familiar with VQEs. The problem of finding the ground state energy of general $k$-local Hamiltonians is believed to be intractable even on quantum computers \cite{kempe2006complexity,piddock2017complexity,aharonov2002quantum}. Nevertheless, there may be problem instance classes for which quantum computers could solve for the ground state efficiently. The VQE is a proposed general method for finding ground states on quantum computers.

	Consider a quantum mechanical system with Hilbert space $\HH$ of dimension $N$, Hamiltonian $H$ with ground state energy $E_0$, and a subset of parametrized states $\{\psith\}\subseteq \HH$, with $\theta \in \mathbb R^{m}$. To describe all of $\HH$, it is necessary that $m=O(N)$. The fact that
	\begin{equation*}
		E(\theta)=\bra{\theta}H \ket\theta \geq E_0
	\end{equation*}
	for all $\ket\theta\in\HH$ is called the variational principle. For reasons of scalability, in variational methods, one generally employs a set of states described by $m=\mrm{polylog}\,N$ parameters. Variational methods, like the VQE, seek to minimize $E(\theta)$ to hence establish an upper bound for the ground-state energy.

	As input, a VQE receives a description of a Hamiltonian on $n$ spin-1/2 particles,
	\begin{equation}\label{eq:Hamiltonian1}
		H=\sum_{i=1}^{l} h_i H_i,
	\end{equation}
	with $h_i$ real coefficients and $H_i$ Hermitian operators. For $k$-local Hamiltonians, $l=\poly n$.

	A VQE proceeds by first choosing an initial set of parameters $\theta$. These initial parameters may be chosen at random or may be inspired by a classical approximate solution to the ground state, for example by the Hartree-Fock ground state \cite{szabo2012modern}. Then, a criterion is chosen, for example, that a maximum number of iterations has not been reached or that $E(\theta)$ has not reached a value below a given threshold. The VQE then proceeds as follows.

	\begin{figure}[t]
		\centering
		$\KVQE_{G}$: results
		\vspace{.5 em}
		\hrule
		\vspace{1 em}	\includegraphics[width=1\linewidth]{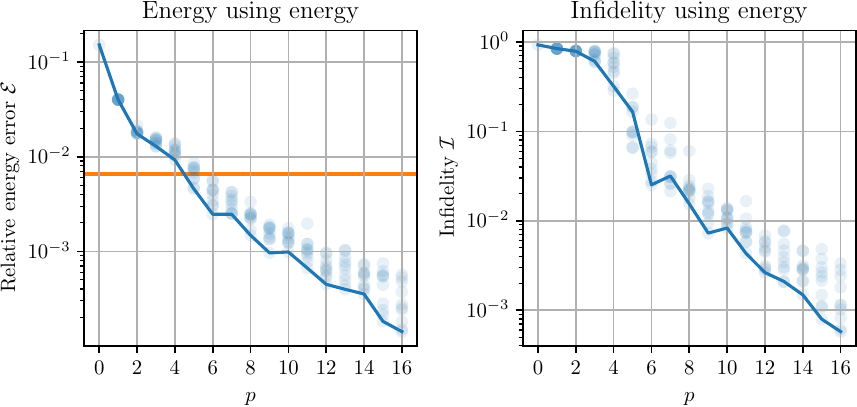}
		\caption{\label{fig:kagome_results} (\textbf{Left}) Semi-log plot of the relative energy error [\eq{rel_en_error}] obtained by $\KVQE_{G}$ for 20 sites (\fig{kagome_ansatz}), as a function of the number of cycles $p$. Translucent points represent the 10 local minima that were found by $\KVQE_G$ per $p$. At $p=0$, only one point is present since no parameters need to be optimized. Every cycle is a circuit of depth 6 and uses  30 parameters. The solid line connects the lowest local minima $E(\theta^*)$. An orange horizontal line is drawn at the value of $\mc E$ corresponding to the energy of the first excited state (not restricting this state to any total spin sector). (\textbf{Right}) Semi-log plot of the infidelities of the states corresponding to the local minima in the left plot. The solid line connects the points that, for a given $p$, are lowest in \emph{energy}. Although it occurs regularly, these points need not have the lowest infidelity.}

	\end{figure}
	\begin{enumerate}
		\item
		While the criterion is true, repeat:
		\begin{enumerate}
			\item \emph{Prepare the ansatz $\ket{\theta}$.}\\ Prepare the initial state $\psiinit$. Apply a
			parametrized circuit $C(\theta)$ to obtain the state $\ket\theta=C(\theta)\psiinit$. The circuit $C(\theta)$ usually consists of gates on fixed positions, where every or some of the gates are parametrized.

			\item \emph{Measure and store $E(\theta)$.} \\By linearity, $E(\theta)=\sum_i h_i \brapsith H_i \psith$.  Each expectation value $\brapsith H_i \psith$ can be estimated by measuring the operator $H_i$ repeatedly (each measurement requires a new preparation of $\psith$) and taking the statistical average. See Ref. \cite{mcclean2016theory} for the expected number of measurements using this method, or Ref. \cite{huang2021efficient}, and references therein, for more efficient methods.

			\item \emph{Update $\theta$.}\\ Based on $\Eth$ and previous outcomes of $\Eth$,  update $\theta$ according to some classical optimization algorithm.

		\end{enumerate}
		\item Return $\theta^*$, which we define as the $\theta$ that achieved the lowest energy.
	\end{enumerate}
	Physically relevant information, such as correlation functions, can now be extracted from $\ket{\theta^*}$ by repeatedly preparing and performing measurements on $\ket{\theta^*}$.	Different VQEs differ in the way circuits are parametrized, how (an estimate for) $E(\theta)$ is obtained, and what specific optimization routine is used. These will be detailed in the subsequent sections.

    	\begin{figure}[t]
		\centering
		$\KVQE_K$: results
		\vspace{.5 em}
		\hrule
		\vspace{1 em}	\includegraphics[width=1\linewidth]{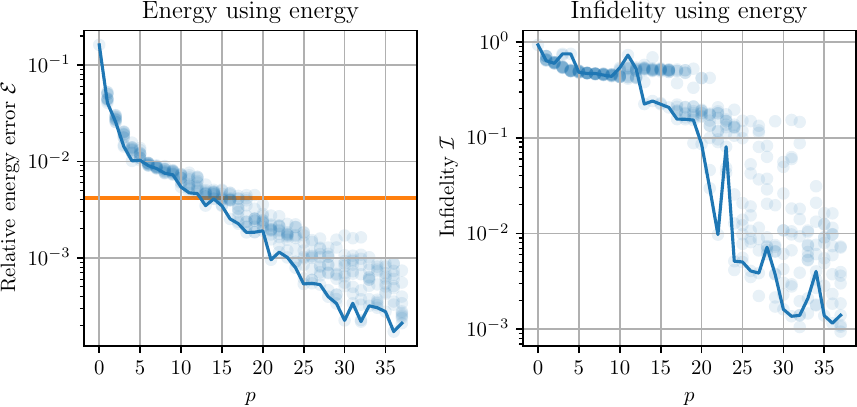}
		\caption{\label{fig:kagome_on_kagome_results} Semi-log plots similar to those in \fig{kagome_results}, but now for $\KVQE_K$, simulating a  periodic patch of 18 sites (\fig{periodic_patch}).  Per $p$, ten local minima are displayed. Every cycle is a circuit of depth 4 and uses 36 parameters.}

	\end{figure}

	\subsection{Ansatz}\label{sec:ansatz}
	\subsubsection{Cyclic HVA}
	In the HVA \cite{wecker2015progress}, the initial state $\psiinit$ is the ground state of a Hamiltonian $H_\mrm{init}$. The Hamiltonian $H_{\mrm{init}}$ is chosen in such a way that its ground state is known and easy to prepare. The ansatz state $\ket\theta$ is obtained by sequentially evolving along terms in $H$, according to some fixed sequence $i$,
	\begin{equation*}
		\ket \theta = C(\theta) \psiinit,
	\end{equation*}
	with
	\begin{equation*}
		C(\theta)=\exp(-\ii \theta_M H_{i_M} )\ldots\exp(-\ii \theta_2 H_{i_2})\exp(-\ii \theta_1 H_{i_1}).
	\end{equation*}
	The $M$ parameters are formed by the time duration of the $M$ evolutions. Often, as for example in, but not limited to, Refs. \cite{ho2019efficient, wecker2015progress, wiersema2020exploring,cade2020strategies}, $C$ consists of $p$ cycles of a smaller circuit $c$, each time defined by the same sequence $i$ of terms in the Hamiltonian. Every cycle gets its own set of  $m$ parameters. It is convenient to write $\theta$ as  $\theta=(\theta_1,\ldots,\theta_p)$, with $\theta_j=(\theta_{j_1},\ldots,\theta_{j_m})$.
	Then, a single cycle reads
	\begin{equation*}
		c(\theta_j)=\exp(-\ii\,\theta_{j_m}H_{i_m} )\ldots\exp(-\ii\, \theta_{j_2} H_{i_2}) \exp(-\ii\, \theta_{j_1} H_{i_1}),
	\end{equation*}
	and so
	\begin{equation}\label{eq:cyclic2}
		\psith = c(\theta_p) \ldots c(\theta_1) \ket{\psi_\mrm{init}}.
	\end{equation}
	We call this type of HVA the \emph{cyclic} HVA. The cyclic HVA shows a close relation between static quantum simulation and dynamic quantum simulation; choosing $i=(1,\ldots,l)$ (or a permutation thereof) and $\theta_j=(t/p,\ldots,t/p)$ for all $j$, the cyclic HVA implements quantum time evolution for a target time $t$ with $p$ Trotter steps. In this way, the cyclic HVA can also mimic (but is more general than) adiabatic time evolution. In the case that no gap closes while adiabatically evolving from $H_\mrm{init}$ to $H$, the HVA thus ensures that the ground state of $H$ can be prepared with the ansatz (but without guaranties on the required circuit depth). This formed the initial motivation for the HVA in VQEs \cite{wecker2015progress}.

	\begin{figure}
		\centering
		CVQE: results
		\vspace{.5em}
		\hrule
		\vspace{1em}
		\includegraphics[width=1 \linewidth]{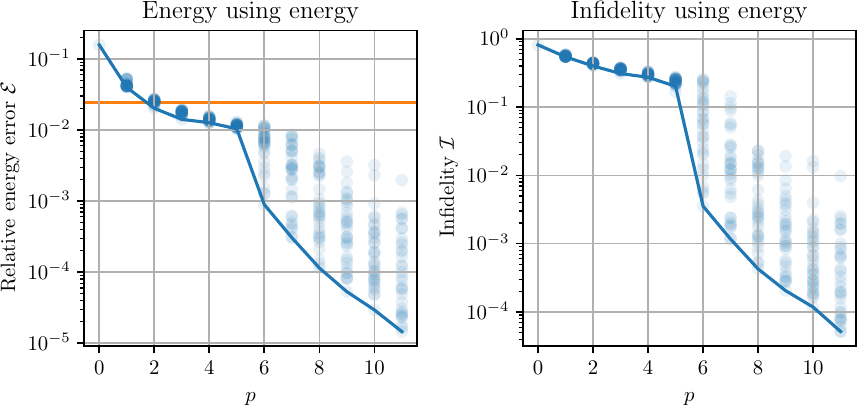}
		\caption{\label{fig:chain_results}
			Semi-log plots similar to those in \fig{kagome_results}, but now for CVQE (\fig{chain_circuit}), simulating the HAFM on a periodic chain of 20 sites. Per $p$, 32 local minima are displayed. Every cycle is a circuit of depth 2 and uses 20 parameters.
		}
	\end{figure}

	\subsubsection{The HVA in this paper}
	We use the cyclic HVA, where the Hamiltonian (in units where $\hbar=1$) is given by the HAFM Hamiltonian [cf. \eq{Hamiltonian1}]
	\begin{equation}\label{eq:HAFM}
		H=\sum_{\langle i,j \rangle} \mathbf{S}^{(i)} \cdot \mathbf{S}^{(j)},
	\end{equation}
	where $\bS^{(i)}=(X_i,Y_i,Z_i)^T/2$, with $X_i,Y_i,Z_i$ the Pauli matrices acting on spin $i$ only, and where the sum runs over the edges $\langle i,j \rangle$ of some graph $G$. In this paper, $G$ is either an open or a periodic patch of the kagome lattice, or a periodic chain, and we take $H_\mrm{init}=\sum_{\langle i,j \rangle'} \mathbf{S}^{(i)} \cdot \mathbf{S}^{(j)}$, with $\langle i,j \rangle'$ edges in a perfect matching of $G$. Since the ground state of a single term in $H$ is the singlet state $\ket s=(\ket{01}-\ket{10})/\sqrt{2}$, this means $\ket{\psi_\mrm{init}}$ is a so-called dimer covering on the relevant physical system, where every dimer is a singlet state.

	We define the gate $\HEIS$ (Heisenberg interaction) as time evolution along a single term in the HAFM Hamiltonian. In the computational basis of two qubits, it reads
	\begin{align}
		\mrm{HEIS}(\al)&\equiv\ee^{-\ii\al/4}\ee^{-\ii \al \bS^{(1)} \cdot \bS^{(2)}}\label{eq:HEIS_gate}\\
		&=\left(
		\begin{array}{cccc}
			\ee^{-\ii \al/2} & 0 & 0 & 0 \\
			0 & \cos(\al/2) & -\ii \sin(\al/2) & 0 \\
			0 &  -\ii \sin(\al/2) &  \cos(\al/2) & 0  \\
			0 & 0 & 0 & \ee^{-\ii \al/2}
		\end{array}\right).\nn
	\end{align}
	In a parametrized circuit [\eq{cyclic2}], every instance of the HEIS gate gets its own parameter $\alpha=\theta_{j_k}$.

	\subsubsection{Symmetry}\label{sec:sym}
	The initial states, Hamiltonian, and the ansatz circuits used in this paper possess an $\mrm{SU}(2)$ rotational symmetry. In this section, we discuss how this leads the following properties of our VQEs: $(i)$ the variational manifold lays within the correct spin sector, $(ii)$ they offer the opportunity of studying the physically relevant spin gap \cite{lecheminant1997order,sindzingre2009low-energy,lauchli2011ground,yan2011spin,lauchli2019s,nakano2011numerical} of the kagome HAFM, and  $(iii)$ they intrinsically perform their computations in a DFS. More details are given in \app{symmetry}.

First, $(i)$ for the two-spin singlet state, the total spin quantum number $S$ and the total magnetization quantum number $S_z$ vanish, and therefore $S=S_z=0$ for our initial states $\ket{\psi_\mrm{init}}$. By the $\mrm{SU}(2)$ symmetry of our ansatz circuits, also $S=S_z=0$ for the ansatz state $\ket{\theta}$ (\app{symmetry}). The ground state of the HAFM on the kagome lattice is believed to have those same quantum numbers (\app{symmetry}). Thus, the additional benefit of using the HVA over the HEA is that using the HVA, the variational manifold $\{\ket \theta\}$ automatically entirely lays within the correct spin sector.

The spin gap ($ii$) is defined as the energy difference between the lowest energy eigenstate in the $S=0$ sector and the lowest energy eigenstate in the $S=1$ sector \cite{lecheminant1997order,sindzingre2009low-energy,lauchli2011ground,yan2011spin,lauchli2019s,nakano2011numerical}. Similar to the ideas in Ref. \cite{seki2020symmetry}, $\KVQE_{G/K}$ and CVQE can be extended to study the spin gap by running them with an $S=1$ initial state, obtaining an optimal energy $E(\theta^*_{S=1})$, in addition to running them with a $S=0$ dimer covering as the initial state, obtaining $E(\theta^*_{S=0})$. An estimate for the spin gap is then given by  $E(\theta^*_{S=1})-E(\theta^*_{S=0})$. $S=1$ initial states can be obtained by changing one of the singlets in our initial state $\ket{\psi_\mrm{init}}$ into one of the three two-spin triplet states. Thus, we obtain a $(S=1,S_z=1)$, $(S=1,S_z=0)$, or $(S=1,S_z=-1)$ dimer covering as the initial state, depending on which triplet state was chosen. In \app{symmetry}, we show that the energy $E(\theta)$ is invariant under the choice of triplet state. Noise may brake the symmetry of the ansatz circuit. $\KVQE_{G/K}$ and CVQE may abuse this noise to put amplitude on $S=0$ states, thus obtaining unjustly low variational energies.  This problem can be handled by symmetry verification \cite{bonet2018low-cost} or by adding a term to the Hamiltonian that penalizes $S\neq 1$ states~\cite{mcclean2016theory,ryabinkin2018constrained,kuroiwa2021penalty} (\app{symmetry}).

For the performance of $\KVQE_{G/K}$ and CVQE using $S=1$ initial states, we expect to obtain results similar to those when using $S=0$ initial states, as already presented in this paper. Correspondingly, the horizontal lines in Figs.~\ref{fig:kagome_results}--\ref{fig:chain_results} represent the first excited state above the ground state, where we put no restriction on the total spin sector of either the ground state or the first excited state. That is, the horizontal line need not represent the ground state in the $S=1$ sector.

Finally ($iii$), similar considerations lead to the conclusion that the VQEs in this paper naturally perform their computations in a DFS.  If HEIS gates are native, or can be compiled in terms of gates that preserve the total spin quantum number $S$ at all times during gate execution, the state $\ket \psi$ occurring at any (nondiscretized) time during the ansatz circuit has total spin quantum number $S=0$. It follows directly that $\mc S \ket \psi=0$ for any $\mc S\in\mc D=\{\bS^2,\bS^{(\mrm{tot})}_x,\bS^{(\mrm{tot})}_y,\bS^{(\mrm{tot})}_z\}$. Here, $\bS^2\equiv\bS^{(\mrm{tot})}\cdot\bS^{(\mrm{tot})}$, with total spin operator $\bS^{(\mrm{tot})}=\sum_{i=1}^n\mathbf{S}^{(i)}$ [cf. \eq{HAFM}]. Therefore, $\ket \psi$ is in a DFS that protects against continuous coupling to (any linear combination of) the operators in $\mc D$~(\app{symmetry}).

	\subsubsection{Parameter multiplicity}
	In this paper, every Heisenberg gate in the ansatz gets its own parameter; we have One parameter Per Gate (OPG). Another possibility would be to have multiple HEIS gates per cycle share the same parameter. We call this One parameter Per Slice (OPS). We say the qubits sharing the same parameter are in the same `slice'.

	A possible advantage of OPS is that, by choosing proper slices, we can ensure that the state produced by the circuit has the lattice symmetries expected to be present in the ground state. This would make the search space smaller by only restricting to states with the desired lattice symmetry.

	Nevertheless, OPG has advantages over OPS. With OPS, we may overlook symmetry-broken ground states. For example, it is unknown whether the ground state of the kagome lattice has a spontaneously broken symmetry \cite{lauchli2019s}. Second, even if the ground state does not break any symmetries, the depth of the OPG circuit for a given state may be lower than the depth of the OPS circuit that produces the same state. For NISQ devices, it is imperative to keep circuit depths as low as possible. Finally, the inherent noise-resilience of VQEs may be compromised by choosing OPS over OPG. As an illustration, say we are given a noiseless quantum computer, a Hamiltonian, and a minimal-depth OPS circuit that produces the ground state of that Hamiltonian. Suppose that now a static but random over-rotation is added to each HEIS gate. Then it is very unlikely that the OPS circuit can still produce the correct ground state, no matter its parameters. When we lift the restriction that the gates in every slice share the same parameter, and hence go to an OPG circuit, the over-rotations can be absorbed into the parameters, and hence the ground state can still be produced without an increase in circuit depth.

	\subsection{Analysis} \label{sec:analysis}
	We assess the effectiveness of $\KVQE_{G/K}$ and CVQE by running them for fixed system sizes but a varying number of cycles $p$. For every $p$, we plot the relative energy error $\mc E$ between the true ground state energy, $E_0$, and the optimal energy found by the VQE, $E(\theta^*)$,
	\begin{equation}\label{eq:rel_en_error}
		\mc E=\left\lvert\frac{E(\theta^*)-E_0}{E_0}\right\rvert.
	\end{equation}

	Additionally, we plot the infidelity $\mc I$ between the true ground state $\ket{E_0}$ and the optimal state obtained by the VQE, $\ket{\theta^*}$,
	\begin{equation}\label{eq:infidel}
		\mc I=1-\mc F\equiv1-\lvert\braket{E_0}{\theta^*}\rvert^2,
	\end{equation}
	with $\mc F$ the fidelity between $\ket{E_0}$ and $\ket{\theta^*}$. Even in plots showing the infidelity, the corresponding VQE optimized the energy, not the infidelity. The infidelity is a useful figure of merit because it upper bounds the relative error in expectation value of any observable~\cite{beach2019making},
	\begin{equation*}
		\frac{\left\lvert\bra{E_0} O \ket{E_0} - \bra{\theta^*}O\ket{\theta^*} \right\rvert}{\lVert O \rVert}\leq4\sqrt{\mc I},
	\end{equation*}\sloppy
	with $\lVert \cdot \rVert$ the operator norm, although this bound may be loose \cite{bosse2021probing}.

	We obtain $E_0$ and $\ket{E_0}$, and thus $\mc E$ and $\mc I$, by exact diagonalization. For large system sizes, such as those needed for obtaining a quantum advantage, this will no longer be possible.

	\subsection{Noise}\label{sec:noise}
	On NISQ devices, noise will lead to errors during the preparation and measurement of the ansatz state. If the entire ansatz state preparation can be performed inside a DFS, these errors have no effect  \cite{palma1996quantum,zanardi1997error,duan1998reducing,lidar1998decoherence} (also see \app{symmetry}). However, perfect DFSs only exists in idealized scenarios, and hence noise will generally degrade the performance of any VQE on a NISQ device~\cite{kattemolle2019dynamical}.

Two common local noise models (against which an encoding into a DFS is generally impossible) are those involving depolarizing and bit-flip noise. Given an $n$--qubit density matrix $\rho$, the depolarizing channel acting on qubit $q$ is given by the map
\begin{equation}\label{eq:depol}
  \rho\mapsto (1-p_{e})\rho + \frac{p_{e}}{3}(X_{q}\rho X_{q}+Y_{q}\rho\, Y_{q}+Z_{q} \rho Z_{q}).
\end{equation}
Another common channel is the bit-flip channel
\begin{equation}\label{eq:bitflip}
  \rho \mapsto (1-p_{e})\rho + p_{e} X_{q}\rho X_{q}.
\end{equation}
These channels have a stochastic interpretation. For example, the latter channel can be interpreted as the following process: with probability $(1-p_e)$, no error happens, and with probability $p_e$, a bit-flip error occurs.

Under stochastic noise of the type $\rho\mapsto \sum_i p_i A_i^\nodagger \rho A_i^\dagger$, with all $A_i$ unitary, such as for the bit-flip and depolarizing channel, the ansatz circuit prepares a mixed state $\rho(\theta)=\sum_\bi p_\bi\ketbra{\psi_\bi(\theta)}{\psi_\bi(\theta)}$, with $p_\bi$ the probability of noise realization $\bi$ and $\ket{\psi_\bi(\theta)}$ the pure state in case of known noise realization $\bi$. Given a $\rho(\theta)$ of the aforementioned form, the energy expectation value becomes $E(\theta)=\tr[H \rho_{}(\theta)]=\sum_\bi p_\bi E^{(\bi)}(\theta)$, with $E^{(\bi)}(\theta)=\bra{\psi_\bi(\theta)}H\ket{\psi_\bi(\theta)}$ the energy in case of noise realization $\bi$. Likewise, the fidelity at given parameters $\theta$ can be written as $\mc F(\theta)=\sum_\bi p_\bi\mc F^{(\bi)}(\theta)$, with $\mc F^{(\bi)}(\theta)=\lvert\braket{E_0}{\psi_\bi(\theta)}\rvert^2$ the fidelity in case of noise realization $\bi$. Because $\sum_\bi p_\bi=1$, this can equivalently be phrased in terms of the infidelity; $\mc I(\theta)=\sum_\bi p_\bi\mc I^{(\bi)}(\theta)$, with $\mc I^{(\bi)}(\theta)=1-\mc F^{(\bi)}(\theta)$ the infidelity in case of noise realization $\bi$.

For the remainder of this paper, we assume that after the preparation of the initial state, and after every layer of HEIS-gates, the depolarizing channel acts separately on all qubits. We separately consider the case where instead the bit-flip channel acts at those spacetime locations. There are $nd$ spacetime locations where an error may occur during a circuit run, with $d=\lvert c \rvert p +1$ the depth of the circuit, where $\lvert c \rvert$ is the number of layers in the cycle $c$. Given $\theta^*$, the fixed optimal parameters output by the VQE, and under the approximation that the occurrence of one or more errors during the execution of the ansatz circuit yields a state that is orthogonal to the true ground state, the noisy fidelity becomes
\begin{equation}\label{eq:fid_ana}
  \mc F(\theta^*)\approx (1-p_e)^{nd}\mc F_0(\theta^*),
\end{equation}
with $\mc F_0\equiv \mc F^{(0^L)}$ the fidelity in the noiseless case. Equivalently, under the same approximation, the infidelity becomes $\mc I(\theta^*)\approx1-(1-p_e)^{nd}[1-\mc I_0(\theta^*)]$, with $\mc I\equiv \mc I^{(0^L)}$.

	\subsection{Classical implementation}\label{sec:classical_implementation}
	We emulate the quantum circuits in this paper using the homegrown, optionally GPU-accelerated, classical quantum emulator \texttt{HeisenbergVQE}. Documentation, source code, and all generated data are available as Supplemental Material \cite{HeisenbergVQE}. \texttt{HeisenbergVQE} is tailored to running VQEs for the Heisenberg model on any graph. It is written in Python \cite{python}, with performance-critical code delegated to C via NumPy \cite{harris2020array} if GPU acceleration is off, and CUDA via CuPy \cite{nishino2017cupy} if GPU acceleration is on.

	We exploit the full access to the wave function, granted by classical emulation, for the computation of $E(\theta)$. We assume a noiseless quantum computer and use a gradient-based optimization method. Gradients are computed using backward-mode automatic differentiation \cite{nielsen2015neural}, as implemented in Chainer~\cite{tokui2015chainer}.

For optimization of the cost function $E(\theta)$, we first choose initial parameters uniformly at random in the interval $[-10^{-3},10^{-3})$. There is some evidence that for the HVA, points close to the origin in parameter space are good starting points for local optimization \cite{wiersema2020exploring}. We then use the BFGS algorithm, as implemented in SciPy \cite{virtanen2020scipy}, to find a local minimum. At every step of the BFGS routine, the energy and gradient of the energy are calculated. Here, we call these two steps together one \emph{function call}. The steps of random parameter generation and local optimization (one `round') are repeated a variable number of times. The parameters that achieve the lowest energy out of all local minimization rounds,  $\theta^*$, are stored together with $E(\theta^*)$. Thus, every data point in Figs. \ref{fig:kagome_results}, \ref{fig:kagome_on_kagome_results} and \ref{fig:chain_results} corresponds to one local optimization, where the initial point of optimization is chosen independently and at random in the interval $[-10^{-3},10^{-3})$. Starting many rounds of local optimization from unrelated starting points has the benefit of straightforward parallelization.  We note that the inner local classical optimization loop may be circumvented by using a Full Quantum Eigensolver (FQE) \cite{wei2020full}, based on linear combinations of unitary operations \cite{long2006general,childs2012hamiltonian} at the cost of increased quantum resources and reduced success probability. We do not use this method here.

	\texttt{HeisenbergVQE} computes exact ground states using SciPy's wrapper of ARPACK,  which implements the Implicitly Restarted Lanczos Method \cite{lehoucq1998arpack}. Operator-vector multiplication is optionally GPU accelerated. The energy of the exact ground state and the exact ground state vector itself are used as a reference for the performance of the VQEs in this paper. Such reference will not be available for system sizes used in experiments showing a quantum advantage.

	For the emulation of noisy quantum computation, we estimate $E(\theta)$ and $\mc I(\theta)$ by sampling $E^{(\bi)}$ and $\mc I^{(\bi)}$ according to the probability distribution $p_i$ (\sec{noise}) and computing the respective sample means. As the number
of shots is increased, the sample means approach the true value of $E(\theta)$ and $\mc I(\theta)$, respectively. We call one stochastic execution of the ansatz circuit one \emph{shot} of that circuit. In our implementation, every shot leads to one sample of $E(\theta)$ and $\mc I(\theta)$. For a finite number of shots, there is a statistical uncertainty in the sample means of $E(\theta)$ and $\mc I(\theta)$, which we characterize by 95\% confidence intervals (CIs). These CIs are obtained by the default implementation of the bootstrapping method \cite{efron1994introduction} in SciPy \cite{virtanen2020scipy}. At a fixed number of shots, reducing $p_e$ or the number $m$ of spacetime locations where an error may occur eventually leads to sets of shots where very few to no errors have occurred. This effect can lead to unjustly low CIs. (If no error occurred during any shot, the CI found by bootstrapping the data has size identically zero.) We exclude this possibility by increasing the number of shots per estimation of $E(\theta)$ and $\mc I(\theta)$ as a function of $p_e$ or $m$ in such a way that the expected number of shots where at least one error occurred equals 1024. Because of the significantly increased cost of noisy simulation, we only consider previously obtained, noiseless, locally optimal parameters for noisy simulation.

The Supplementary Material \cite{HeisenbergVQE} also includes data and plots for systems and ans\"atze not reported in this paper, including simulations of the HAFM on the triangular lattice, other system sizes of the kagome lattice, simulations that use one parameter per slice (see \sec{ansatz}), $\KVQE_{G}$ simulating periodic patches, $\KVQE_K$ simulating open patches, and runs where we use the infidelity as the cost function. Using the infidelity as a cost function is impractical on quantum computers (or even impossible if the ground state is not known), but may be used by classical computers to obtain further data on the expressibility of an ansatz. For all systems, data were stored in a human-readable format, and include the number of calls to the cost function by the BFGS routine, the wall-clock time of the classical emulation, the initial parameters, the parameters, energy, and infidelity of the local minima, as well as the noise type, error rate, and CIs in the case of noisy simulation.

\section{Results and analysis}
\subsection{Noiseless results}
	\subsubsection{Kagome}\label{sec:kagome}
	The explicit initial state and circuit used by $\KVQE_{G}$ for the simulation of the HAFM on a 20-site patch of the kagome lattice is depicted in \fig{kagome_ansatz}. For this patch, $\KVQE_{G}$ uses 20 data qubits to represent the 20 sites of the patch, and an additional 4 qubits as `swapping stations', used to realize kagome connectivity on grid architectures. The restriction of grid connectivity increases the circuit depth per cycle (assuming HEIS gates are native) from 4 to 6, and (in the thermodynamic limit) introduces one auxiliary qubit per three qubits. Because every gate gets its own parameter, the total number of parameters is $M=30p$.

	Noiseless results are displayed in \fig{kagome_results}. The relative energy error $\mc E$ decreases roughly exponentially in the range of all considered $p$. The number of function calls scales polynomially with $p$ (data available at Ref.~\cite{HeisenbergVQE}).
	$\KVQE_{G}$ finds an energy lower than the energy of the first exited state for $p\geq5$. There is no clear critical $p$ after which $\mc E$ and/or $\mc I$ improve drastically. Nevertheless, $\mc I$ transitions to an improved exponential decay rate somewhere between $p=3$ and $p=5$, reaching a fidelity of $>$99.9\% at $p\geq 16$. Under the hardware assumptions of Sec. \ref{sec:sum_num}, $p=16$ amounts to $20/2\times 3=30$ gates for the generation of the singlets, $30\times16=480$ HEIS gates, and $16\times16=256$ SWAP gates, giving a total of $766$ gates and 480 parameters. The total depth equals $3+6\times 16=99$. To obtain the 10 local minima at $p=16$, a total of 82\,466 function calls were made.

	The explicit initial state and circuit used by $\KVQE_K$ for the simulation of an 18-site periodic patch are depicted in \fig{periodic_patch}. For this patch, we assume a quantum computer with 18 qubits and all-to-all connectivity. Noiseless results are displayed in \fig{kagome_on_kagome_results}. Again, $\mc E$ decreases roughly exponentially in the range of all considered $p$, whereas the number of function calls scales polynomially with $p$ \cite{HeisenbergVQE}. $\KVQE_K$ finds an energy lower than the energy of the first excited state for $p\geq13$. After that same $p$, the performance of $\mc I$ improves significantly.

	We note that for $p$ such that the ansatz is not expressive enough to attain energies below the first excited state (in the present case, this is for $p<13$), there need not be a relation between the energy of an ansatz state and its overlap with the ground state; in this regime, the ansatz may, for example, prepare states with large overlap with the first excited state, and zero overlap with the ground state. It is only for $p$ such that the ansatz can reach energies below the first excited state that a decrease in energy must be met with a decrease in infidelity. Data obtained using the infidelity as the cost function do show uniform exponential decay of $\mc I(p)$ \cite{HeisenbergVQE}.

	In the absence of noise, $\KVQE_K$ (using the energy as the cost function) reaches a fidelity of $>$99.9\% at $p\geq 37$. Under the hardware assumptions of Sec. \ref{sec:sum_num}, $p=37$ amounts to $18/2\times 3=27$ gates for the generation of the singlets, and $36\times37=1332$ HEIS gates, giving a total of $1359$ gates and 1332 parameters. The total depth equals $3+4\times 37=151$. To obtain the 10 local minima at $p=37$, a total of 191\,582 function calls were made.

	One striking feature of these results is that the rate of exponential decay of the optimal energy and infidelity as a function of the number of cycles $p$ is significantly smaller (in absolute value) than those same rates obtained by $\KVQE_{G}$. This difference in performance is not because of the difference in the structure of the ansatz, but because of the difference in the systems being simulated; the performance of $\KVQE_{G}$ in simulating the periodic patch of \fig{periodic_patch} (in this case, the ansatz for $\KVQE_{G}$ is obtained by extending the ansatz of \fig{kagome_ansatz} in such a way that its effective HEIS gates cover the periodic patch of \fig{periodic_patch}), does not differ significantly from the performance of $\KVQE_K$ on that same patch \cite{HeisenbergVQE}. Vice versa, the performance of $\KVQE_K$ (in this case, the ansatz for $\KVQE_K$ is obtained by extending the ansatz of \fig{periodic_patch} in such a way that its effective HEIS gates cover the open patch of \fig{kagome_ansatz}) in simulating the 20-site open patch of \fig{kagome_ansatz} (top left) does not differ significantly from the performance of $\KVQE_G$ on that same patch \cite{HeisenbergVQE}.

	At first, the 18-site periodic patch of \fig{periodic_patch} might seem to be easier to simulate than the 20-site open patch of \fig{kagome_ansatz} (top left) because of the difference in the number of sites. However, the graph defining the periodic patch has 36 edges, as opposed to 30 edges for the open patch. Therefore, the optimization problem corresponding to the periodic patch has more constraints, and can therefore naturally be more challenging. Additionally, boundary effects, which are absent in the periodic patch, may lead to a ground state that is easier to prepare. We leave an investigation of the latter issue for further work.

	\subsubsection{Chain}\label{sec:chain}
	The explicit initial state and circuit used by CVQE for the simulation of a 20-site periodic chain are depicted in \fig{chain_circuit}. Any Quantum Processing Unit (QPU) with grid, kagome, or all-to-all connectivity naturally embeds subsets of qubits with (at least) the connectivity of a periodic chain. Because every gate gets its own parameter, the total number of parameters is $M=20p$.

	Noiseless results are displayed in \fig{chain_results}. Both the relative energy error $\mc E$ and the infidelity  $\mc I$ (\sec{analysis}) initially decrease exponentially as a function of $p$, reaching an energy that is below the first excited state for $p\geq2$. Both functions show a sudden improvement after $p_\mrm{crit}=5$. From $p_\mrm{crit}$ to $p_\mrm{crit}+1$, $\mc E$ drops by an order of magnitude, and $\mc I$ drops by two orders of magnitude. For $p>5$, both functions again decrease roughly exponentially with a rate that is greater in magnitude than before. At the same time, the number of function calls (as defined in \sec{classical_implementation}), grows polynomially with $p$ (\fig{scaling}, bottom). A fidelity of $>$99.9\% is reached for $p\geq 8$. Assuming HEIS gates are native, and that singlets can be created with a circuit of depth 3 (\sec{hardware_implementation}), $p=8$ amounts to a circuit with $20/2\times 3=30$ gates for the preparation of singlets and $20\times 8=160$ HEIS gates,  giving a total of $190$ gates and $160$ parameters. The depth of the circuit is $3+8\times 2=19$. The optimization routine for finding the 32 local minima at $p=8$ used 104\,890 function calls. A fidelity of $>$99.99\% is reached at $p=11$ cycles, using a total of 197\,685 function calls.

	A plausible explanation of the sudden improvement of CVQE beyond $p_\mrm{crit}=5$ is in terms of the past light cone. The past light cone of a qubit $q$ after a circuit $C$ consists of all qubits $q'$ for which there exists a past-directed path through $C$ that connects $q$ to $q'$. It is only when $q'$ is in the past light cone of $q$ that $C$ can build up entanglement between $q$ and $q'$. Also see \fig{chain_circuit}. The ground state of the HAFM on the chain is known to possess long-range entanglement \cite{latorre2004ground}. The sudden improvement of performance is a clear sign of a ground state with long-range entanglement. At $p_\mrm{crit}$, there is no qubit whose past light cone at the end of $C$ covers the entire chain. After $p_\mrm{crit}$, the past light cone of \emph{every} qubit at the end of $C$ covers the entire chain.

	For $\KVQE_{G}$, not all qubits' past light cones cover the entire system for the first time at an identical number of cycles \footnote{Let us focus on $\KVQE_{G}$, and lay out a coordinate system over the 24 qubits used in \fig{kagome_ansatz} (top right). We put the origin (0,0) at the bottom left qubit, the qubit directly above at (0,1), and the qubit directly to the right of the origin at (1,0). At $p=2$, there is no qubit whose past light cone in $C$ covers the entire system. At $p=3$, there are qubits, such as the bottom right (4,0), bottom left (0,0), top left (0,4) and middle (2,2) qubits, whose past light cone in $C$ covers the entire system. There are, however, still some qubits for which this is not the case, such as the qubits at (3,4), (4,3) and the top middle (2,0). At $p=4$, the past light cone of the latter qubits covers the entire system, except for the qubit at (3,4). It is only after $p=5$ cycles that its past light cone covers the entire system.}. For $\KVQE_K$, the past light cone of every qubit covers the entire patch for the first time at $p=3$. However, unlike CVQE, the states that can be reached by $\KVQE_K$ at this depth are still very far from the true ground state, and no sudden improvement of performance is observed.

\subsubsection{Sistem-size scaling}
The noiseless simulations carried out in this paper display an exponential decay of the infidelity [\eq{infidel}] as a function of the number of cycles $p$. However, for the scalability of VQEs, it is essential that this rate of decay does not decrease too strongly (e.g., exponentially) with the system size. Alternatively, we may study the number of cycles $\overline p$ that is required to reach a given target infidelity $\mc I$. For $\KVQE_{G/K}$, obtaining a meaningful estimate for $\overline p$ is likely infeasible with classical emulation. This is because there are too few classically tractable and structurally similar patches for a systematic system-size scaling.

\begin{figure}
  		\vspace{1em}
		\centering
		CVQE: system-size scaling
		\vspace{0.5 em}
		\hrule
		\vspace{1em}
		\includegraphics[width=\columnwidth]{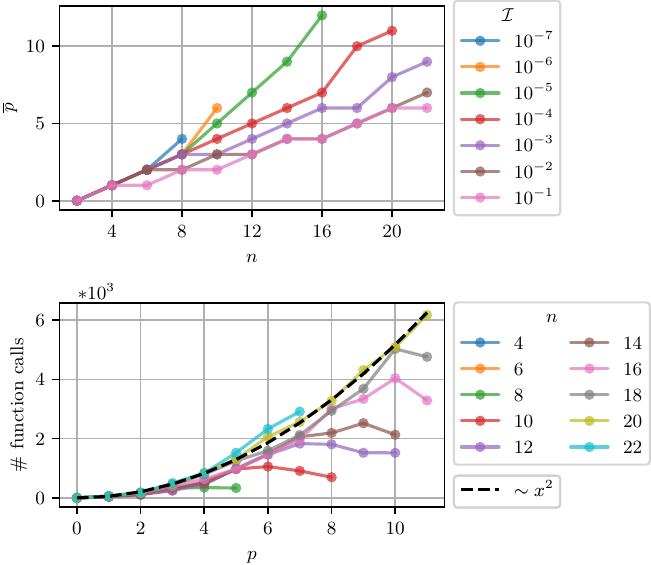}
		\caption{\label{fig:scaling}
		 (\textbf{Top}) The number of cycles $\overline p$ required to prepare the approximate ground state using CVQE, as a function of the number of sites $n$, for various target infidelities $\mc I$.  Results were obtained as in \fig{chain_results}, using the energy as the cost function during classical optimization of the variational parameters. (\textbf{Bottom}) The number of function calls needed to reach a single local optimum, averaged over all 32 local optima that were reached per tuple $(n,p)$. For some $n$, data were not obtained for all $p=0,1,\ldots,11$, either because a negligible infidelity was reached ($n\leq 14$) or because of increasing computational cost ($n=22$). A quadratic fit to the data for $n=20$ (black, dashed) is added as a guide to the eye. We consider one `function call' as one computation of the energy $E(\theta)$ and one computation of the gradient $\nabla E(\theta)$ (\sec{classical_implementation}).
		}
		\end{figure}

For CVQE, however, rings of size $n=2,6,\ldots,22$ give us sufficient classically tractable and structurally similar patches. Results showing $\bar p$, together with the average number of function calls per local optimum, are displayed in \fig{scaling}.

		The data in \fig{scaling} indicate that $\overline p$ is upper bounded by a polynomial of low degree. Additionally, the data indicate that the number of function calls to the energy landscape is upper bounded by a function quadratic in $p$. Therefore, the total run time, excluding any sampling overhead, appears to scale as a polynomial of a polynomial, itself a polynomial.

		\subsection{Noisy results}\label{sec:noisy_results}

\begin{figure}
  \centering
  $\KVQE_{G}$: noisy results
  \vspace{.5 em}
  \hrule
  \vspace{1 em}	\includegraphics[width=1\linewidth]{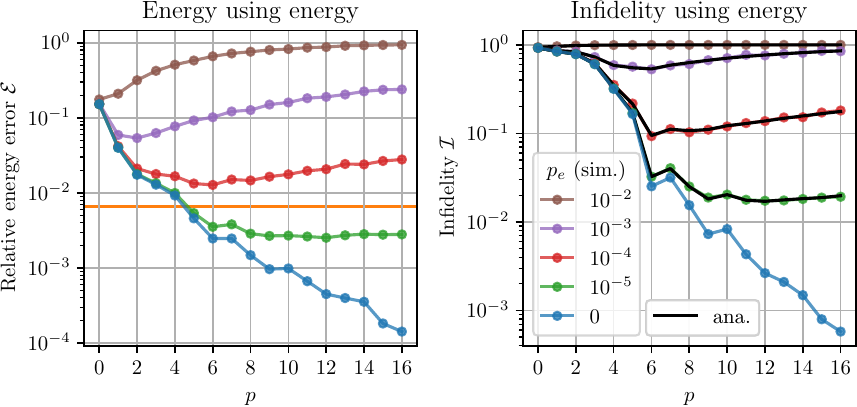}
  \caption{\label{fig:noisy_KVQE_G} (\textbf{Left}) As \fig{kagome_results}, now including the effects of depolarizing noise with error probability $p_{e}$ (legend shown in the right plot). At all $p_e$, data were obtained by computing the energy at $\theta^*(p)$, the noiseless optimal variational parameters as returned by the VQE. The blue line shows the noiseless energy $E(\theta^*)$, and is identical to the blue line in \fig{kagome_results}. Error bars marking the 95\% confidence intervals are smaller than any of the plot points and are therefore omitted. (\textbf{Right}) As the left plot, but now showing the infidelity between the true ground state and the states corresponding to the plot points of the left plot. Additionally, the analytical estimate for the infidelity [\eq{fid_ana}], which is based on the noiseless infidelity $\mc I(\theta^*)$, is shown in black.}
\end{figure}

\begin{figure}
  \centering
  $\KVQE_K$: noisy results
  \vspace{.5 em}
  \hrule
  \vspace{1 em}\includegraphics[width=1\linewidth]{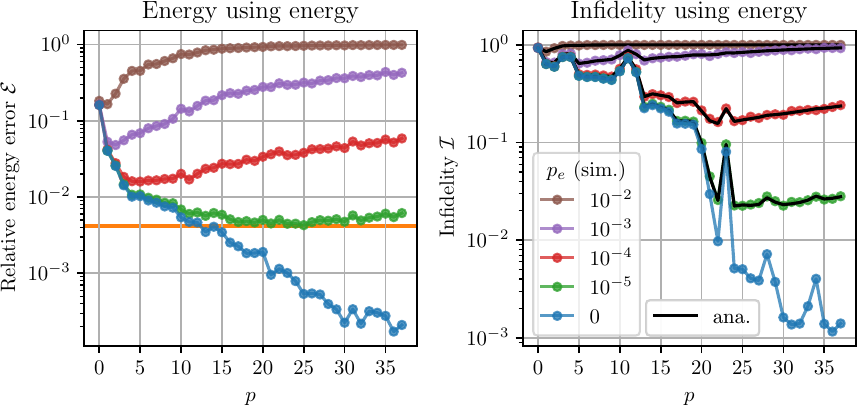}
  \caption{\label{fig:noisy_KVQE_K} As \fig{kagome_on_kagome_results}, now including the effects of depolarizing noise with error probability $p_e$. Definitions for the noisy data are as in \fig{noisy_KVQE_G}.}
\end{figure}

\begin{figure}
  \centering
  CVQE: noisy results
  \vspace{.5 em}
  \hrule
  \vspace{1 em}\includegraphics[width=1\linewidth]{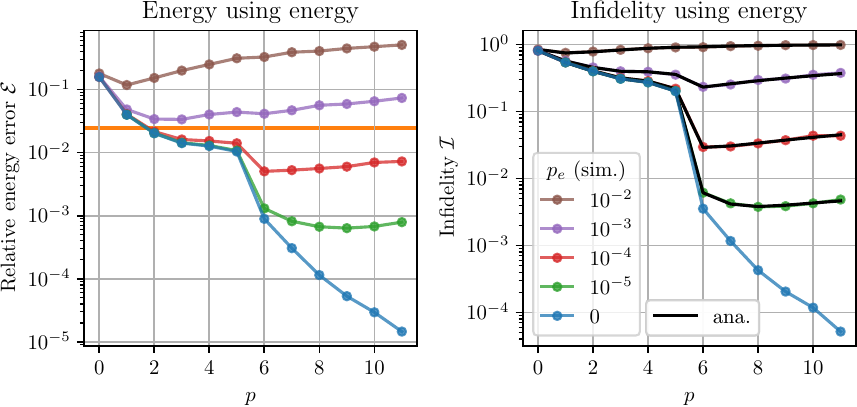}
  \caption{\label{fig:noisy_CVQE} As \fig{chain_results}, now including the effects of depolarizing noise with error probability $p_e$. Definitions for the noisy data are as in \fig{noisy_KVQE_G}.}
\end{figure}

Using the method outlined in \sec{classical_implementation}, we compute the energy and infidelity obtained under the bit-flip and depolarizing noise models (\sec{noise}), at all \emph{locally} optimal noiseless parameters of Figs. \ref{fig:kagome_results}--\ref{fig:chain_results} (corresponding to all translucent plot points in these plots), for $p_{e}\in\{10^{{-2}},10^{{-3}},10^{{-4}},10^{{-5}}\}$. Results for the depolarizing channel, showing the noisy values of the \emph{optimal} noiseless data points, are displayed in Figs. \ref{fig:noisy_KVQE_G}--\ref{fig:noisy_CVQE}.

We observed no significant difference between the effect of the depolarizing and the bit-flip channel. Therefore, only the results regarding the former are shown. The similarity of effect can be understood as follows. The effect of the depolarizing channel on a state on the Bloch sphere is isotropic, while that of the bit-flip channel is not \cite{nielsen2010quantum}. For example, eigenstates of the Pauli-$X$ operator are affected by the depolarizing channel, but are invariant under the bit-flip channel. Nevertheless, given that the input states of the channels have no preferred direction, the detrimental effect of the depolarizing and bit-flip channels are equal on average. (Technically, the twirl of a bit-flip channel with noise parameter $p_e$ over $\mrm{SU}(2)$ yields the depolarizing channel with the same noise parameter, which can be verified straightforwardly using the results of Ref. \cite{emerson2005scalable}.) Therefore, if the states occurring during the ansatz state preparation or our VQEs are sufficiently `isotropic', one expects roughly equal effects of the depolarizing and bit-flip channels.

The locally optimal angles that did not attain the lowest energy in the absence of noise may do so in the presence of noise. We found this effect to make no qualitative difference. More generally, Refs. \cite{kattemolle2022effects,fontana2021evaluating} showed that VQEs can possess \emph{noise adaptivity}; when a noisy ansatz circuit is run, the optimal parameters obtained in the presence of that noise (the noise-aware variational parameters) may attain a lower energy than the optimal parameters obtained in the absence of that noise (the noise-unaware parameters). We have only optimized the energy landscape in the absence of noise, but a constrained form of noise adaptivity my still arise due to the freedom to choose the best noiseless locally optimal parameters in the noisy case. We indeed observed (with numerical significance, as set by a clear separation of the respective CIs) that the locally optimal angles that did not attain the lowest energy in the absence of noise do incidentally obtain the lowest energy in the presence of noise. However, the effect is  insignificant on the scale of the fluctuations of the lines in Figs. \ref{fig:noisy_KVQE_G}--\ref{fig:noisy_CVQE}. The latter variations arise because, per $p$, the VQEs did not always find the global minimum in the noiseless case. In that case, by chance, a lower minimum may be found at $p$ compared to $p+1$, even though the globally minimum energy at $p+1$ is upper bounded by the minimum energy at $p$. Because the overall trend of the energy as a function of $p$ is downward, this must lead to fluctuations in the energy as a function of $p$.

The noisy data for $\KVQE_G$ (periodic $\KVQE_K$) on 20 sites (18 sites) suggest that, already for systems too small to obtain a quantum advantage, error rates between $10^{-4}$ and $10^{-5}$ are necessary to achieve an energy that is lower than the first excited state. This makes it highly unlikely that these energies will be obtained on NISQ hardware without error mitigation. Furthermore, the data show no observable discrepancy between the infidelity obtained by numerical simulation and the estimate of \eq{fid_ana}. Under the reasonable assumption that a similar agreement continues to hold for larger systems, we may use \eq{fid_ana} to obtain estimates for the required error rates for systems too large for classical emulation. Thus, even if a VQE is able to prepare a ground state perfectly, $\mc I \approx 1-(1-p_e)^{nd}$ for depolarizing and bit-flip noise models. Inverting this relation, we have
\begin{equation*}
  p_e\approx \frac{\mc I}{nd}
\end{equation*}
for $p_e\ll 1$. If an infidelity of $10^{-3}$ is demanded for a system of a hundred qubits (as in Refs. \cite{childs2021theory,bravi2022future}), and assuming the ground state can be prepared perfectly with a depth of $d=n=100$, already an error rate of $p_e\approx 10^{-7}$ is required. This is an error rate several orders of magnitude lower than what can currently be achieved, and also several orders of magnitude below known error thresholds for fault-tolerant quantum computation \cite{wang2011surface,dalton2022variational}.

The data for the performance of CVQE on 20 sites under noise (\fig{noisy_CVQE}) indicate that an energy below the first excited state may be reached on quantum computers with linear connectivity and error rates between $10^{-3}$ and $10^{-4}$. Current rates fall between $10^{-2}$ and $10^{-3}$ \cite{ballance2016high,kjaergaard2020superconducting} \footnote{The gate infidelities typically reported in the literature do not correspond directly to the error rates as defined in Eqs. (\ref{eq:depol}) and (\ref{eq:bitflip}), but for the purpose of the order-of-magnitude estimates here these discrepancies are insignificant.}. Thus, if technological improvements are able to bring down the error rates by one order of magnitude, proof-of-principle implementations of CVQE with on the order of 20 qubits can already be carried out. Feasibility is further improved by error mitigation.

\section{Quantum hardware implementation} \label{sec:hardware_implementation}
\begin{figure}
		Singlet preparation
		\vspace{0.5 em}
		\hrule
		\vspace{1em}
		\includegraphics[width=.7 \linewidth]{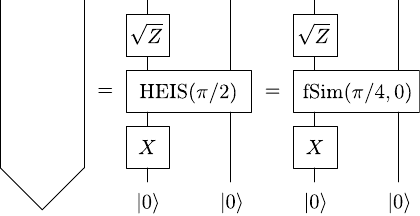}
		\caption{Circuits for preparing the singlet state up to a global phase (left), using gates native to quantum dots (middle) and Google's superconducting hardware (right).
		\label{fig:singlet_circuit}}
	  \centering
	  \vspace{1em}
		$\HEIS$ compilation
		\vspace{0.5 em}
		\hrule
		\vspace{1em}	\includegraphics[width=\linewidth]{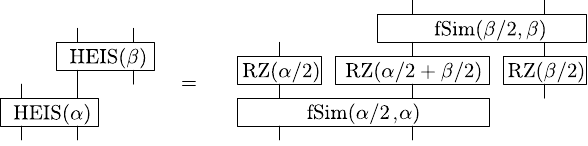}
		\caption{Example of the compilation of $\HEIS$ gates into $\fSim$ gates and single-qubit RZ rotations for $0\leq \alpha/2\leq\pi$.
		\label{fig:HEIS_compilation}}
		\centering
	\end{figure}

The HEIS gate is directly native to quantum dot architectures. This also allows native implementation of the SWAP gate on these devices since $\SWAP=\ii \  \HEIS(\pi)$. Assuming also $X$ and $\sqrt{Z}$ gates are native, singlets can be created with a circuit of depth three (\fig{singlet_circuit}). Thus, on these devices, essentially no compilation is needed for $\KVQE_{G/K}$ and CVQE.

Similarly, on Google Quantum AI's superconducting hardware, compilation of $\KVQE_{G/K}$ and CVQE's gates induces no overhead in the number of two-qubit gates. This is because Google's native parameterized two-qubit gate, the `fermionic simulation' or fSim gate \cite{foxen2020demonstrating}, is related to the HEIS gate by
\begin{equation} \label{eq:HEIS_as_fSim}
		\HEISal=\RZ_0(\al/2)\,\RZ_1(\al/2)\,\fSim(\al/2,\al).
	\end{equation}
Here, $\RZ_0(\theta)=\RZ(\theta)\otimes\id$ and $\RZ_1(\theta)=\id\otimes\RZ(\theta)$, with $\RZ(\theta)=\ee^{-\ii \theta Z/2}$, and $Z$ the Pauli-$Z$ operator. See \app{fSim} for details on the resolution of a technical subtlety regarding \eq{HEIS_as_fSim}. The $\SWAP$ gate is related to the fSim gate by
	\begin{equation*}
		\SWAP=\sqrt{Z_0}\sqrt{Z_1}\fSim(\pi/2,\pi).
	  \end{equation*}
	  Hence, a SWAP gate can be implemented by using one layer of RZ rotations and a single fSim gate. As for the quantum dot architectures, singlets can be created with a circuit of depth three (\fig{singlet_circuit}).

	By adding the angles of subsequent RZ rotations, $\ell$ layers of HEIS and/or SWAP gates can be implemented by at most $\ell+1$ layers of single-qubit RZ rotations and $\ell$ layers of fSim gates. Depending on the specific circuit, further reductions may be possible by using that $\RZ_0(\be) \RZ_1(\be)$ commutes with $\fSim(\theta,\phi)$ and addition of $\RZ$ rotation angles. For an example, see \fig{HEIS_compilation}.

	Current hardware does not yet simultaneously have grid connectivity (for more than 4 qubits) and the ability to essentially natively implement the exchange interaction for all parameter values. quantum dot architectures can natively implement the exchange interaction, but are not yet available with grid connectivity. However, this connectivity may become available in the future, as detailed proposals already exist \cite{barthelemy2013quantum, van2021quantum, hendrickx2021four,li2018crossbar}. Google AI Quantum can implement the exchange interaction essentially natively for all parameter values and can demonstrate grid connectivity \cite{arute2019quantum}, but is not yet able to combine these two features in a single processor. This has, however, been expressed as a future goal. (See Sec. C. of the Supplemental Material of Ref.~\cite{arute2019quantum}.) They are already able to implement any two-qubit gate (so including the HEIS gate for all parameter values) using three native two-qubit gates \cite{harrigan2021quantum}. Current hardware is already capable of efficiently performing CVQE for small problem sizes of open \cite{foxen2020demonstrating} or closed chains \cite{hendrickx2021four}. In such experiments, the observation of a critical circuit depth could form an early goal and would indicate the ability to generate and find ground states with system-wide entanglement.

	\section{Discussion and outlook}\label{sec:discussion}

	In this paper, we have introduced and studied VQEs for the HAFM on the kagome lattice ($\KVQE_{G/K}$) as exceptionally hardware-native algorithms with the potential of showing a useful quantum advantage on NISQ devices. In the noiseless case, the energies found by these VQEs decay exponentially with circuit depth, and the VQEs appear to be scalable, as indicated by the system-size scaling carried out for CVQE. Furthermore, we showed that our VQEs naturally perform their quantum computations in a DFS that protects against collective longitudinal and transversal noise. However, other common noise types put extraordinary demands on the allowed error rates if an energy is to be reached that is below the first excited state, or if the ground state is to be prepared with high fidelity. The following factors alleviate these demands.

  (1) \emph{Performance target.} The requirement of reaching an energy below the first excited state, or an infidelity of $10^{-3}$, may be needlessly demanding. First, arguably a useful quantum advantage can be claimed once a VQE for a system of on the order of a hundred sites finds variational energies lower than those that can be found by state-of-the-art classical variational techniques~\cite{yan2011spin,depenbrock2012nature,he2017signatures}. Second, there are indications that performance targets on physically relevant observables other than the energy, such as spin-spin or dimer-dimer correlation functions, may be much more benign to VQEs \cite{bosse2021probing}. Finally, with minimal adjustments to the classical optimization routine, the VQEs in this paper can be turned into Variational Quantum Thermalizer algorithms \cite{verdon2019quantum, foldager2022noise}. In this setting, a natural performance target for a useful quantum advantage would be to prepare a thermal state with a target temperature inaccessible to classical numerical methods. For systems with on the order of a hundred qubits, the typical energy associated with such a target temperature may be well above that of the first excited state.

  (2) \emph{Noise robustness.} We have computed the energy and infidelity of the output state of a noisy quantum computation at the previously found noiseless locally optimal parameters. This leaves little room for noise robustness. If parameters are optimized in the presence of noise, as on real quantum devices, lower energies may be found due to the noise adaptivity of the parameters \cite{mcclean2017hybrid,fontana2021evaluating,kattemolle2022effects}.

  (3) \emph{Error mitigation.} By combining multiple runs of multiple quantum circuits with classical pre- and postprocessing, the error in estimates of expectation values of observables [such as the error in $E(\theta)$] can be reduced \cite{temme2017error,koczor2021exponential,huggins2021virtual,song2019quantum,bultrini2021quantum}, and can in principle even be made arbitrarily small \cite{berg2022probabilistic}, even in the absence of quantum error correction. Such error-mitigation techniques typically lead to a sampling overhead that increases exponentially with the number of qubits, circuit depth, and error rate. However, the base of the exponent may be brought close to unity and sampling is highly (quantum-) parallelizable \cite{bravi2022future}. Various error-mitigation techniques that exist specifically for VQEs may be used in conjunction \cite{bonet2018low-cost,sagastizabal2019experimental}.

Whether a combination of these factors allows for a useful quantum advantage on pre-error-corrected devices can likely only be answered by the quantum computing community as a whole. Technologically, there is a need for the further reduction of physical error rates. Furthermore, for the implementation of error mitigation and additional techniques such as circuit knitting \cite{bravyi2016trading,peng2020simulating,sun2022perturbative}, the tight and highly parallelized integration of quantum and classical resources, also known as quantum-centric supercomputing \cite{bravi2022future}, is required. Additionally, further theoretical advances in the performance characterization and performance targets of classical quantum algorithms is needed to be able to claim a useful quantum advantage on near term quantum devices.

\begin{figure}[t]
		\centering
		$\KVQE'_{K}$: ansatz
		\vspace{0.5 em}
		\hrule
		\includegraphics[width=\linewidth]{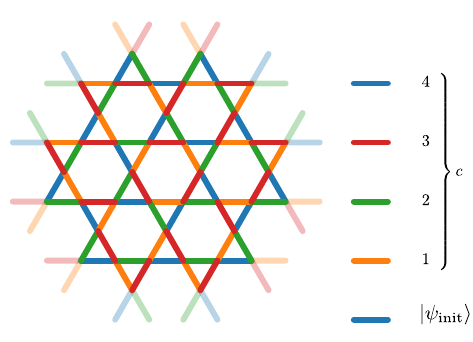}
		\caption{\label{fig:VBC_four_coloring}
			 One unit cell of a four coloring of the kagome lattice (opaque edges), derived from the 36-site VBC (blue lines) of Refs. \cite{marston1991spin,nikolic2003physics,singh2007ground,singh2008triplet,evenbly2010frustrated}. To four-color the entire lattice using this unit cell, a hexagonal tiling of these unit cells needs to be made. This four coloring directly translates to an ansatz for $\KVQE_K$.  On quantum computers with grid connectivity, we propose to use the same initial state, but define the cycle $c$ similarly to the cycle defined for $\KVQE_{G}$ defined in \fig{kagome_ansatz}. }
	\end{figure}

In Refs.~\cite{marston1991spin,nikolic2003physics,singh2007ground,singh2008triplet,evenbly2010frustrated}, the proposed ground state of the kagome HAFM is a 36-site VBC. We propose to use this VBC as the initial VQE state on quantum computers with at least that same number of data qubits (\fig{VBC_four_coloring}). For large patches, this has the potential to answer whether this VBC is the ground state with very low circuit depths.

\emph{Note added.} Shortly after the preprint of this paper was made public, an independent but similar work appeared by Bosse and Montanaro \cite{bosse2021probing}. At a high level, Bosse and Montanaro's work shares the same motivation, methods, and results as the current paper. Differences include the exact mapping from the kagome lattice to a grid, the numerical implementation, and the simulated patches of the kagome lattice. In addition, Bosse and Montanaro report data on the extraction of observables and an investigation of the barren-plateau problem \cite{mcclean2018barren}, where we include an ansatz for quantum computers with kagome connectivity and focus on the experimental realization by offering explicit compilation into native gates and a study of the effects of noise.

\subsection*{Acknowledgements}
	The authors thank C. J. van Diepen, K. L. Groenland, and P. R. Corboz for suggestions and discussions. The numerical simulations in this paper were carried out on the quantum simulation nodes of the Lisa cluster, provided by the Dutch national e-infrastructure with the support of SURF Cooperative, and the Scientific Compute Cluster of the University of Konstanz (SCCKN).

	\bibliography{bib.bib}

\begin{thebibliography}{121}%
\makeatletter
\providecommand \@ifxundefined [1]{%
 \@ifx{#1\undefined}
}%
\providecommand \@ifnum [1]{%
 \ifnum #1\expandafter \@firstoftwo
 \else \expandafter \@secondoftwo
 \fi
}%
\providecommand \@ifx [1]{%
 \ifx #1\expandafter \@firstoftwo
 \else \expandafter \@secondoftwo
 \fi
}%
\providecommand \natexlab [1]{#1}%
\providecommand \enquote  [1]{``#1''}%
\providecommand \bibnamefont  [1]{#1}%
\providecommand \bibfnamefont [1]{#1}%
\providecommand \citenamefont [1]{#1}%
\providecommand \href@noop [0]{\@secondoftwo}%
\providecommand \href [0]{\begingroup \@sanitize@url \@href}%
\providecommand \@href[1]{\@@startlink{#1}\@@href}%
\providecommand \@@href[1]{\endgroup#1\@@endlink}%
\providecommand \@sanitize@url [0]{\catcode `\\12\catcode `\$12\catcode
  `\&12\catcode `\#12\catcode `\^12\catcode `\_12\catcode `\%12\relax}%
\providecommand \@@startlink[1]{}%
\providecommand \@@endlink[0]{}%
\providecommand \url  [0]{\begingroup\@sanitize@url \@url }%
\providecommand \@url [1]{\endgroup\@href {#1}{\urlprefix }}%
\providecommand \urlprefix  [0]{URL }%
\providecommand \Eprint [0]{\href }%
\providecommand \doibase [0]{https://doi.org/}%
\providecommand \selectlanguage [0]{\@gobble}%
\providecommand \bibinfo  [0]{\@secondoftwo}%
\providecommand \bibfield  [0]{\@secondoftwo}%
\providecommand \translation [1]{[#1]}%
\providecommand \BibitemOpen [0]{}%
\providecommand \bibitemStop [0]{}%
\providecommand \bibitemNoStop [0]{.\EOS\space}%
\providecommand \EOS [0]{\spacefactor3000\relax}%
\providecommand \BibitemShut  [1]{\csname bibitem#1\endcsname}%
\let\auto@bib@innerbib\@empty
%</preamble>
\bibitem [{\citenamefont {Jiang}\ \emph {et~al.}(2008)\citenamefont {Jiang},
  \citenamefont {Weng},\ and\ \citenamefont {Sheng}}]{jiang2008density}%
  \BibitemOpen
  \bibfield  {author} {\bibinfo {author} {\bibfnamefont {H.-C.}\ \bibnamefont
  {Jiang}}, \bibinfo {author} {\bibfnamefont {Z.-Y.}\ \bibnamefont {Weng}},\
  and\ \bibinfo {author} {\bibfnamefont {D.~N.}\ \bibnamefont {Sheng}},\
  }\bibfield  {title} {\bibinfo {title} {Density matrix renormalization group
  numerical study of the kagome antiferromagnet},\ }\href
  {https://doi.org/10.1103/PhysRevLett.101.117203} {\bibfield  {journal}
  {\bibinfo  {journal} {Physical Review Letters}\ }\textbf {\bibinfo {volume}
  {101}},\ \bibinfo {pages} {117203} (\bibinfo {year} {2008})}\BibitemShut
  {NoStop}%
\bibitem [{\citenamefont {Yan}\ \emph {et~al.}(2011)\citenamefont {Yan},
  \citenamefont {Huse},\ and\ \citenamefont {White}}]{yan2011spin}%
  \BibitemOpen
  \bibfield  {author} {\bibinfo {author} {\bibfnamefont {S.}~\bibnamefont
  {Yan}}, \bibinfo {author} {\bibfnamefont {D.~A.}\ \bibnamefont {Huse}},\ and\
  \bibinfo {author} {\bibfnamefont {S.~R.}\ \bibnamefont {White}},\ }\bibfield
  {title} {\bibinfo {title} {Spin-liquid ground state of the {$S=1/2$} kagome
  {H}eisenberg antiferromagnet},\ }\href
  {https://doi.org/10.1126/science.1201080} {\bibfield  {journal} {\bibinfo
  {journal} {Science}\ }\textbf {\bibinfo {volume} {332}},\ \bibinfo {pages}
  {1173--1176} (\bibinfo {year} {2011})}\BibitemShut {NoStop}%
\bibitem [{\citenamefont {Anderson}(1973)}]{anderson1973resonating}%
  \BibitemOpen
  \bibfield  {author} {\bibinfo {author} {\bibfnamefont {P.~W.}\ \bibnamefont
  {Anderson}},\ }\bibfield  {title} {\bibinfo {title} {Resonating valence
  bonds: A new kind of insulator?},\ }\href
  {https://doi.org/10.1016/0025-5408(73)90167-0} {\bibfield  {journal}
  {\bibinfo  {journal} {Materials Research Bulletin}\ }\textbf {\bibinfo
  {volume} {8}},\ \bibinfo {pages} {153--160} (\bibinfo {year}
  {1973})}\BibitemShut {NoStop}%
\bibitem [{\citenamefont {Norman}(2016)}]{norman2016herbertsmithite}%
  \BibitemOpen
  \bibfield  {author} {\bibinfo {author} {\bibfnamefont {M.~R.}\ \bibnamefont
  {Norman}},\ }\bibfield  {title} {\bibinfo {title} {Colloquium:
  Herbertsmithite and the search for the quantum spin liquid},\ }\href
  {https://doi.org/10.1103/RevModPhys.88.041002} {\bibfield  {journal}
  {\bibinfo  {journal} {Reviews of Modern Physics}\ }\textbf {\bibinfo {volume}
  {88}},\ \bibinfo {pages} {041002} (\bibinfo {year} {2016})}\BibitemShut
  {NoStop}%
\bibitem [{\citenamefont {L{\"a}uchli}\ \emph {et~al.}(2019)\citenamefont
  {L{\"a}uchli}, \citenamefont {Sudan},\ and\ \citenamefont
  {Moessner}}]{lauchli2019s}%
  \BibitemOpen
  \bibfield  {author} {\bibinfo {author} {\bibfnamefont {A.~M.}\ \bibnamefont
  {L{\"a}uchli}}, \bibinfo {author} {\bibfnamefont {J.}~\bibnamefont {Sudan}},\
  and\ \bibinfo {author} {\bibfnamefont {R.}~\bibnamefont {Moessner}},\
  }\bibfield  {title} {\bibinfo {title} {{$S=1/2$} kagome {H}eisenberg
  antiferromagnet revisited},\ }\href
  {https://doi.org/10.1103/PhysRevB.100.155142} {\bibfield  {journal} {\bibinfo
   {journal} {Physical Review B}\ }\textbf {\bibinfo {volume} {100}},\ \bibinfo
  {pages} {155142} (\bibinfo {year} {2019})}\BibitemShut {NoStop}%
\bibitem [{\citenamefont {Marston}\ and\ \citenamefont
  {Zeng}(1991)}]{marston1991spin}%
  \BibitemOpen
  \bibfield  {author} {\bibinfo {author} {\bibfnamefont {J.}~\bibnamefont
  {Marston}}\ and\ \bibinfo {author} {\bibfnamefont {C.}~\bibnamefont {Zeng}},\
  }\bibfield  {title} {\bibinfo {title} {Spin-{P}eierls and spin-liquid phases
  of kagom{\'e} quantum antiferromagnets},\ }\href
  {https://doi.org/10.1063/1.347830} {\bibfield  {journal} {\bibinfo  {journal}
  {Journal of Applied Physics}\ }\textbf {\bibinfo {volume} {69}},\ \bibinfo
  {pages} {5962--5964} (\bibinfo {year} {1991})}\BibitemShut {NoStop}%
\bibitem [{\citenamefont {Nikolic}\ and\ \citenamefont
  {Senthil}(2003)}]{nikolic2003physics}%
  \BibitemOpen
  \bibfield  {author} {\bibinfo {author} {\bibfnamefont {P.}~\bibnamefont
  {Nikolic}}\ and\ \bibinfo {author} {\bibfnamefont {T.}~\bibnamefont
  {Senthil}},\ }\bibfield  {title} {\bibinfo {title} {Physics of low-energy
  singlet states of the kagome lattice quantum {H}eisenberg antiferromagnet},\
  }\href {https://doi.org/10.1103/physrevb.68.214415} {\bibfield  {journal}
  {\bibinfo  {journal} {Physical Review B}\ }\textbf {\bibinfo {volume} {68}},\
  \bibinfo {pages} {214415} (\bibinfo {year} {2003})}\BibitemShut {NoStop}%
\bibitem [{\citenamefont {Singh}\ and\ \citenamefont
  {Huse}(2007)}]{singh2007ground}%
  \BibitemOpen
  \bibfield  {author} {\bibinfo {author} {\bibfnamefont {R.~R.}\ \bibnamefont
  {Singh}}\ and\ \bibinfo {author} {\bibfnamefont {D.~A.}\ \bibnamefont
  {Huse}},\ }\bibfield  {title} {\bibinfo {title} {Ground state of the spin-1/2
  kagome-lattice {H}eisenberg antiferromagnet},\ }\href
  {https://doi.org/10.1103/PhysRevB.76.180407} {\bibfield  {journal} {\bibinfo
  {journal} {Physical Review B}\ }\textbf {\bibinfo {volume} {76}},\ \bibinfo
  {pages} {180407} (\bibinfo {year} {2007})}\BibitemShut {NoStop}%
\bibitem [{\citenamefont {Singh}\ and\ \citenamefont
  {Huse}(2008)}]{singh2008triplet}%
  \BibitemOpen
  \bibfield  {author} {\bibinfo {author} {\bibfnamefont {R.~R.}\ \bibnamefont
  {Singh}}\ and\ \bibinfo {author} {\bibfnamefont {D.~A.}\ \bibnamefont
  {Huse}},\ }\bibfield  {title} {\bibinfo {title} {Triplet and singlet
  excitations in the valence bond crystal phase of the kagome lattice
  {H}eisenberg model},\ }\href {https://doi.org/10.1103/PhysRevB.77.144415}
  {\bibfield  {journal} {\bibinfo  {journal} {Physical Review B}\ }\textbf
  {\bibinfo {volume} {77}},\ \bibinfo {pages} {144415} (\bibinfo {year}
  {2008})}\BibitemShut {NoStop}%
\bibitem [{\citenamefont {Evenbly}\ and\ \citenamefont
  {Vidal}(2010)}]{evenbly2010frustrated}%
  \BibitemOpen
  \bibfield  {author} {\bibinfo {author} {\bibfnamefont {G.}~\bibnamefont
  {Evenbly}}\ and\ \bibinfo {author} {\bibfnamefont {G.}~\bibnamefont
  {Vidal}},\ }\bibfield  {title} {\bibinfo {title} {Frustrated antiferromagnets
  with entanglement renormalization: Ground state of the spin-1.2 {H}eisenberg
  model on a kagome lattice},\ }\href
  {https://doi.org/10.1103/physrevlett.104.187203} {\bibfield  {journal}
  {\bibinfo  {journal} {Physical Review Letters}\ }\textbf {\bibinfo {volume}
  {104}},\ \bibinfo {pages} {187203} (\bibinfo {year} {2010})}\BibitemShut
  {NoStop}%
\bibitem [{\citenamefont {McClean}\ \emph {et~al.}(2016)\citenamefont
  {McClean}, \citenamefont {Romero}, \citenamefont {Babbush},\ and\
  \citenamefont {Aspuru-Guzik}}]{mcclean2016theory}%
  \BibitemOpen
  \bibfield  {author} {\bibinfo {author} {\bibfnamefont {J.~R.}\ \bibnamefont
  {McClean}}, \bibinfo {author} {\bibfnamefont {J.}~\bibnamefont {Romero}},
  \bibinfo {author} {\bibfnamefont {R.}~\bibnamefont {Babbush}},\ and\ \bibinfo
  {author} {\bibfnamefont {A.}~\bibnamefont {Aspuru-Guzik}},\ }\bibfield
  {title} {\bibinfo {title} {The theory of variational hybrid quantum-classical
  algorithms},\ }\href {https://doi.org/10.1088/1367-2630/18/2/023023}
  {\bibfield  {journal} {\bibinfo  {journal} {New Journal of Physics}\ }\textbf
  {\bibinfo {volume} {18}},\ \bibinfo {pages} {023023} (\bibinfo {year}
  {2016})}\BibitemShut {NoStop}%
\bibitem [{\citenamefont {Peruzzo}\ \emph {et~al.}(2014)\citenamefont
  {Peruzzo}, \citenamefont {McClean}, \citenamefont {Shadbolt}, \citenamefont
  {Yung}, \citenamefont {Zhou}, \citenamefont {Love}, \citenamefont
  {Aspuru-Guzik},\ and\ \citenamefont {O'{B}rien}}]{peruzzo2014variational}%
  \BibitemOpen
  \bibfield  {author} {\bibinfo {author} {\bibfnamefont {A.}~\bibnamefont
  {Peruzzo}}, \bibinfo {author} {\bibfnamefont {J.}~\bibnamefont {McClean}},
  \bibinfo {author} {\bibfnamefont {P.}~\bibnamefont {Shadbolt}}, \bibinfo
  {author} {\bibfnamefont {M.-H.}\ \bibnamefont {Yung}}, \bibinfo {author}
  {\bibfnamefont {X.-Q.}\ \bibnamefont {Zhou}}, \bibinfo {author}
  {\bibfnamefont {P.~J.}\ \bibnamefont {Love}}, \bibinfo {author}
  {\bibfnamefont {A.}~\bibnamefont {Aspuru-Guzik}},\ and\ \bibinfo {author}
  {\bibfnamefont {J.~L.}\ \bibnamefont {O'{B}rien}},\ }\bibfield  {title}
  {\bibinfo {title} {A variational eigenvalue solver on a photonic quantum
  processor},\ }\href {https://doi.org/10.1038/ncomms5213} {\bibfield
  {journal} {\bibinfo  {journal} {Nature Communications}\ }\textbf {\bibinfo
  {volume} {5}},\ \bibinfo {pages} {1--7} (\bibinfo {year} {2014})}\BibitemShut
  {NoStop}%
\bibitem [{\citenamefont {Preskill}(2018)}]{preskill2018quantum}%
  \BibitemOpen
  \bibfield  {author} {\bibinfo {author} {\bibfnamefont {J.}~\bibnamefont
  {Preskill}},\ }\bibfield  {title} {\bibinfo {title} {Quantum {C}omputing in
  the {NISQ} era and beyond},\ }\href
  {https://doi.org/10.22331/q-2018-08-06-79} {\bibfield  {journal} {\bibinfo
  {journal} {{Quantum}}\ }\textbf {\bibinfo {volume} {2}},\ \bibinfo {pages}
  {79} (\bibinfo {year} {2018})}\BibitemShut {NoStop}%
\bibitem [{\citenamefont {O’Malley}\ \emph {et~al.}(2016)\citenamefont
  {O’Malley}, \citenamefont {Babbush}, \citenamefont {Kivlichan},
  \citenamefont {Romero}, \citenamefont {McClean}, \citenamefont {Barends},
  \citenamefont {Kelly}, \citenamefont {Roushan}, \citenamefont {Tranter},
  \citenamefont {Ding} \emph {et~al.}}]{o2016scalable}%
  \BibitemOpen
  \bibfield  {author} {\bibinfo {author} {\bibfnamefont {P.~J.}\ \bibnamefont
  {O’Malley}}, \bibinfo {author} {\bibfnamefont {R.}~\bibnamefont {Babbush}},
  \bibinfo {author} {\bibfnamefont {I.~D.}\ \bibnamefont {Kivlichan}}, \bibinfo
  {author} {\bibfnamefont {J.}~\bibnamefont {Romero}}, \bibinfo {author}
  {\bibfnamefont {J.~R.}\ \bibnamefont {McClean}}, \bibinfo {author}
  {\bibfnamefont {R.}~\bibnamefont {Barends}}, \bibinfo {author} {\bibfnamefont
  {J.}~\bibnamefont {Kelly}}, \bibinfo {author} {\bibfnamefont
  {P.}~\bibnamefont {Roushan}}, \bibinfo {author} {\bibfnamefont
  {A.}~\bibnamefont {Tranter}}, \bibinfo {author} {\bibfnamefont
  {N.}~\bibnamefont {Ding}}, \emph {et~al.},\ }\bibfield  {title} {\bibinfo
  {title} {Scalable quantum simulation of molecular energies},\ }\href
  {https://doi.org/10.1103/PhysRevX.6.031007} {\bibfield  {journal} {\bibinfo
  {journal} {Physical Review X}\ }\textbf {\bibinfo {volume} {6}},\ \bibinfo
  {pages} {031007} (\bibinfo {year} {2016})}\BibitemShut {NoStop}%
\bibitem [{\citenamefont {Reiner}\ \emph {et~al.}(2019)\citenamefont {Reiner},
  \citenamefont {Wilhelm-Mauch}, \citenamefont {Sch{\"o}n},\ and\ \citenamefont
  {Marthaler}}]{reiner2019finding}%
  \BibitemOpen
  \bibfield  {author} {\bibinfo {author} {\bibfnamefont {J.-M.}\ \bibnamefont
  {Reiner}}, \bibinfo {author} {\bibfnamefont {F.}~\bibnamefont
  {Wilhelm-Mauch}}, \bibinfo {author} {\bibfnamefont {G.}~\bibnamefont
  {Sch{\"o}n}},\ and\ \bibinfo {author} {\bibfnamefont {M.}~\bibnamefont
  {Marthaler}},\ }\bibfield  {title} {\bibinfo {title} {Finding the ground
  state of the {H}ubbard model by variational methods on a quantum computer
  with gate errors},\ }\href {https://doi.org/10.1088/2058-9565/ab1e85}
  {\bibfield  {journal} {\bibinfo  {journal} {Quantum Science and Technology}\
  }\textbf {\bibinfo {volume} {4}},\ \bibinfo {pages} {035005} (\bibinfo {year}
  {2019})}\BibitemShut {NoStop}%
\bibitem [{\citenamefont {Arute}\ \emph {et~al.}(2019)\citenamefont {Arute},
  \citenamefont {Arya}, \citenamefont {Babbush}, \citenamefont {Bacon},
  \citenamefont {Bardin}, \citenamefont {Barends}, \citenamefont {Biswas},
  \citenamefont {Boixo}, \citenamefont {Brandao}, \citenamefont {Buell} \emph
  {et~al.}}]{arute2019quantum}%
  \BibitemOpen
  \bibfield  {author} {\bibinfo {author} {\bibfnamefont {F.}~\bibnamefont
  {Arute}}, \bibinfo {author} {\bibfnamefont {K.}~\bibnamefont {Arya}},
  \bibinfo {author} {\bibfnamefont {R.}~\bibnamefont {Babbush}}, \bibinfo
  {author} {\bibfnamefont {D.}~\bibnamefont {Bacon}}, \bibinfo {author}
  {\bibfnamefont {J.~C.}\ \bibnamefont {Bardin}}, \bibinfo {author}
  {\bibfnamefont {R.}~\bibnamefont {Barends}}, \bibinfo {author} {\bibfnamefont
  {R.}~\bibnamefont {Biswas}}, \bibinfo {author} {\bibfnamefont
  {S.}~\bibnamefont {Boixo}}, \bibinfo {author} {\bibfnamefont {F.~G. S.~L.}\
  \bibnamefont {Brandao}}, \bibinfo {author} {\bibfnamefont {D.~A.}\
  \bibnamefont {Buell}}, \emph {et~al.},\ }\bibfield  {title} {\bibinfo {title}
  {Quantum supremacy using a programmable superconducting processor},\ }\href
  {https://doi.org/10.1038/s41586-019-1666-5} {\bibfield  {journal} {\bibinfo
  {journal} {Nature}\ }\textbf {\bibinfo {volume} {574}},\ \bibinfo {pages}
  {505--510} (\bibinfo {year} {2019})}\BibitemShut {NoStop}%
\bibitem [{\citenamefont {Zhong}\ \emph {et~al.}(2020)\citenamefont {Zhong},
  \citenamefont {Wang}, \citenamefont {Deng}, \citenamefont {Chen},
  \citenamefont {Peng}, \citenamefont {Luo}, \citenamefont {Qin}, \citenamefont
  {Wu}, \citenamefont {Ding}, \citenamefont {Hu} \emph
  {et~al.}}]{zhong2020quantum}%
  \BibitemOpen
  \bibfield  {author} {\bibinfo {author} {\bibfnamefont {H.-S.}\ \bibnamefont
  {Zhong}}, \bibinfo {author} {\bibfnamefont {H.}~\bibnamefont {Wang}},
  \bibinfo {author} {\bibfnamefont {Y.-H.}\ \bibnamefont {Deng}}, \bibinfo
  {author} {\bibfnamefont {M.-C.}\ \bibnamefont {Chen}}, \bibinfo {author}
  {\bibfnamefont {L.-C.}\ \bibnamefont {Peng}}, \bibinfo {author}
  {\bibfnamefont {Y.-H.}\ \bibnamefont {Luo}}, \bibinfo {author} {\bibfnamefont
  {J.}~\bibnamefont {Qin}}, \bibinfo {author} {\bibfnamefont {D.}~\bibnamefont
  {Wu}}, \bibinfo {author} {\bibfnamefont {X.}~\bibnamefont {Ding}}, \bibinfo
  {author} {\bibfnamefont {Y.}~\bibnamefont {Hu}}, \emph {et~al.},\ }\bibfield
  {title} {\bibinfo {title} {Quantum computational advantage using photons},\
  }\href {https://doi.org/10.1126/science.abe8770} {\bibfield  {journal}
  {\bibinfo  {journal} {Science}\ }\textbf {\bibinfo {volume} {370}},\ \bibinfo
  {pages} {1460--1463} (\bibinfo {year} {2020})}\BibitemShut {NoStop}%
\bibitem [{\citenamefont {Fran{\c{c}}a}\ and\ \citenamefont
  {Garc{\'{\i}}a-Patr{\'{o}}n}(2021)}]{franca2020limitations}%
  \BibitemOpen
  \bibfield  {author} {\bibinfo {author} {\bibfnamefont {D.~S.}\ \bibnamefont
  {Fran{\c{c}}a}}\ and\ \bibinfo {author} {\bibfnamefont {R.}~\bibnamefont
  {Garc{\'{\i}}a-Patr{\'{o}}n}},\ }\bibfield  {title} {\bibinfo {title}
  {Limitations of optimization algorithms on noisy quantum devices},\ }\href
  {https://doi.org/10.1038/s41567-021-01356-3} {\bibfield  {journal} {\bibinfo
  {journal} {Nature Physics}\ }\textbf {\bibinfo {volume} {17}},\ \bibinfo
  {pages} {1221--1227} (\bibinfo {year} {2021})}\BibitemShut {NoStop}%
\bibitem [{\citenamefont {Kattemölle}()}]{HeisenbergVQE}%
  \BibitemOpen
  \bibfield  {author} {\bibinfo {author} {\bibfnamefont {J.}~\bibnamefont
  {Kattemölle}},\ }\href {https://github.com/barbireau/HVQE} {\bibinfo {title}
  {{H}eisenberg {VQE}}},\ \bibinfo {howpublished}
  {https://github.com/barbireau/HVQE}\BibitemShut {NoStop}%
\bibitem [{\citenamefont {L\"auchli}\ \emph {et~al.}(2011)\citenamefont
  {L\"auchli}, \citenamefont {Sudan},\ and\ \citenamefont
  {S\o{}rensen}}]{lauchli2011ground}%
  \BibitemOpen
  \bibfield  {author} {\bibinfo {author} {\bibfnamefont {A.~M.}\ \bibnamefont
  {L\"auchli}}, \bibinfo {author} {\bibfnamefont {J.}~\bibnamefont {Sudan}},\
  and\ \bibinfo {author} {\bibfnamefont {E.~S.}\ \bibnamefont {S\o{}rensen}},\
  }\bibfield  {title} {\bibinfo {title} {Ground-state energy and spin gap of
  spin-$\frac{1}{2}$ kagom\'e-{H}eisenberg antiferromagnetic clusters:
  Large-scale exact diagonalization results},\ }\href
  {https://doi.org/10.1103/PhysRevB.83.212401} {\bibfield  {journal} {\bibinfo
  {journal} {Physical Review B}\ }\textbf {\bibinfo {volume} {83}},\ \bibinfo
  {pages} {212401} (\bibinfo {year} {2011})}\BibitemShut {NoStop}%
\bibitem [{\citenamefont {Wecker}\ \emph {et~al.}(2015)\citenamefont {Wecker},
  \citenamefont {Hastings},\ and\ \citenamefont {Troyer}}]{wecker2015progress}%
  \BibitemOpen
  \bibfield  {author} {\bibinfo {author} {\bibfnamefont {D.}~\bibnamefont
  {Wecker}}, \bibinfo {author} {\bibfnamefont {M.~B.}\ \bibnamefont
  {Hastings}},\ and\ \bibinfo {author} {\bibfnamefont {M.}~\bibnamefont
  {Troyer}},\ }\bibfield  {title} {\bibinfo {title} {Progress towards practical
  quantum variational algorithms},\ }\href
  {https://doi.org/10.1103/physreva.92.042303} {\bibfield  {journal} {\bibinfo
  {journal} {Physical Review A}\ }\textbf {\bibinfo {volume} {92}},\ \bibinfo
  {pages} {042303} (\bibinfo {year} {2015})}\BibitemShut {NoStop}%
\bibitem [{\citenamefont {Bethe}(1931)}]{bethe1931theorie}%
  \BibitemOpen
  \bibfield  {author} {\bibinfo {author} {\bibfnamefont {H.}~\bibnamefont
  {Bethe}},\ }\bibfield  {title} {\bibinfo {title} {Zur theorie der metalle},\
  }\href {https://doi.org/10.1007/bf01341708} {\bibfield  {journal} {\bibinfo
  {journal} {Zeitschrift f{\"u}r Physik}\ }\textbf {\bibinfo {volume} {71}},\
  \bibinfo {pages} {205--226} (\bibinfo {year} {1931})}\BibitemShut {NoStop}%
\bibitem [{\citenamefont {Franchini}(2017)}]{franchini2017introduction}%
  \BibitemOpen
  \bibfield  {author} {\bibinfo {author} {\bibfnamefont {F.}~\bibnamefont
  {Franchini}},\ }\href {https://doi.org/10.1007/978-3-319-48487-7} {\emph
  {\bibinfo {title} {An Introduction to Integrable Techniques for
  One-Dimensional Quantum Systems}}}\ (\bibinfo  {publisher} {Springer
  International Publishing},\ \bibinfo {year} {2017})\BibitemShut {NoStop}%
\bibitem [{\citenamefont {Caux}(2009)}]{caux2009correlation}%
  \BibitemOpen
  \bibfield  {author} {\bibinfo {author} {\bibfnamefont {J.-S.}\ \bibnamefont
  {Caux}},\ }\bibfield  {title} {\bibinfo {title} {Correlation functions of
  integrable models: A description of the {ABACUS} algorithm},\ }\href
  {https://doi.org/10.1063/1.3216474} {\bibfield  {journal} {\bibinfo
  {journal} {Journal of Mathematical Physics}\ }\textbf {\bibinfo {volume}
  {50}},\ \bibinfo {pages} {095214} (\bibinfo {year} {2009})}\BibitemShut
  {NoStop}%
\bibitem [{\citenamefont {Ho}\ and\ \citenamefont
  {Hsieh}(2019)}]{ho2019efficient}%
  \BibitemOpen
  \bibfield  {author} {\bibinfo {author} {\bibfnamefont {W.~W.}\ \bibnamefont
  {Ho}}\ and\ \bibinfo {author} {\bibfnamefont {T.~H.}\ \bibnamefont {Hsieh}},\
  }\bibfield  {title} {\bibinfo {title} {{Efficient variational simulation of
  non-trivial quantum states}},\ }\href
  {https://doi.org/10.21468/SciPostPhys.6.3.029} {\bibfield  {journal}
  {\bibinfo  {journal} {SciPost Physics}\ }\textbf {\bibinfo {volume} {6}},\
  \bibinfo {pages} {29} (\bibinfo {year} {2019})}\BibitemShut {NoStop}%
\bibitem [{\citenamefont {Grimsley}\ \emph {et~al.}(2019)\citenamefont
  {Grimsley}, \citenamefont {Economou}, \citenamefont {Barnes},\ and\
  \citenamefont {Mayhall}}]{grimsley2019adaptive}%
  \BibitemOpen
  \bibfield  {author} {\bibinfo {author} {\bibfnamefont {H.~R.}\ \bibnamefont
  {Grimsley}}, \bibinfo {author} {\bibfnamefont {S.~E.}\ \bibnamefont
  {Economou}}, \bibinfo {author} {\bibfnamefont {E.}~\bibnamefont {Barnes}},\
  and\ \bibinfo {author} {\bibfnamefont {N.~J.}\ \bibnamefont {Mayhall}},\
  }\bibfield  {title} {\bibinfo {title} {An adaptive variational algorithm for
  exact molecular simulations on a quantum computer},\ }\href
  {https://doi.org/10.1038/s41467-019-10988-2} {\bibfield  {journal} {\bibinfo
  {journal} {Nature Communications}\ }\textbf {\bibinfo {volume} {10}},\
  \bibinfo {pages} {1--9} (\bibinfo {year} {2019})}\BibitemShut {NoStop}%
\bibitem [{\citenamefont {Cade}\ \emph {et~al.}(2020)\citenamefont {Cade},
  \citenamefont {Mineh}, \citenamefont {Montanaro},\ and\ \citenamefont
  {Stanisic}}]{cade2020strategies}%
  \BibitemOpen
  \bibfield  {author} {\bibinfo {author} {\bibfnamefont {C.}~\bibnamefont
  {Cade}}, \bibinfo {author} {\bibfnamefont {L.}~\bibnamefont {Mineh}},
  \bibinfo {author} {\bibfnamefont {A.}~\bibnamefont {Montanaro}},\ and\
  \bibinfo {author} {\bibfnamefont {S.}~\bibnamefont {Stanisic}},\ }\bibfield
  {title} {\bibinfo {title} {Strategies for solving the {F}ermi-{H}ubbard model
  on near-term quantum computers},\ }\href
  {https://doi.org/10.1103/physrevb.102.235122} {\bibfield  {journal} {\bibinfo
   {journal} {Physical Review B}\ }\textbf {\bibinfo {volume} {102}},\ \bibinfo
  {pages} {235122} (\bibinfo {year} {2020})}\BibitemShut {NoStop}%
\bibitem [{\citenamefont {Arute}\ \emph {et~al.}(2020)\citenamefont {Arute},
  \citenamefont {Arya}, \citenamefont {Babbush}, \citenamefont {Bacon},
  \citenamefont {Bardin}, \citenamefont {Barends}, \citenamefont {Bengtsson},
  \citenamefont {Boixo}, \citenamefont {Broughton}, \citenamefont {Buckley}
  \emph {et~al.}}]{arute2020observation}%
  \BibitemOpen
  \bibfield  {author} {\bibinfo {author} {\bibfnamefont {F.}~\bibnamefont
  {Arute}}, \bibinfo {author} {\bibfnamefont {K.}~\bibnamefont {Arya}},
  \bibinfo {author} {\bibfnamefont {R.}~\bibnamefont {Babbush}}, \bibinfo
  {author} {\bibfnamefont {D.}~\bibnamefont {Bacon}}, \bibinfo {author}
  {\bibfnamefont {J.~C.}\ \bibnamefont {Bardin}}, \bibinfo {author}
  {\bibfnamefont {R.}~\bibnamefont {Barends}}, \bibinfo {author} {\bibfnamefont
  {A.}~\bibnamefont {Bengtsson}}, \bibinfo {author} {\bibfnamefont
  {S.}~\bibnamefont {Boixo}}, \bibinfo {author} {\bibfnamefont
  {M.}~\bibnamefont {Broughton}}, \bibinfo {author} {\bibfnamefont {B.~B.}\
  \bibnamefont {Buckley}}, \emph {et~al.},\ }\href@noop {} {\bibinfo {title}
  {Observation of separated dynamics of charge and spin in the
  {F}ermi-{H}ubbard model}} (\bibinfo {year} {2020}),\ \Eprint
  {https://arxiv.org/abs/2010.07965} {arXiv:2010.07965 [quant-ph]} \BibitemShut
  {NoStop}%
\bibitem [{\citenamefont {Nielsen}(2005)}]{nielsen2005fermionic}%
  \BibitemOpen
  \bibfield  {author} {\bibinfo {author} {\bibfnamefont {M.}~\bibnamefont
  {Nielsen}},\ }\bibfield  {title} {\bibinfo {title} {The fermionic canonical
  commutation relations and the {J}ordan-{W}igner transform},\ }\href@noop {}
  {\bibfield  {journal} {\bibinfo  {journal} {School of Physical Sciences, The
  University of Queensland}\ }\textbf {\bibinfo {volume} {59}} (\bibinfo {year}
  {2005})}\BibitemShut {NoStop}%
\bibitem [{\citenamefont {Bravyi}\ and\ \citenamefont
  {Kitaev}(2002)}]{bravyi2002fermionic}%
  \BibitemOpen
  \bibfield  {author} {\bibinfo {author} {\bibfnamefont {S.~B.}\ \bibnamefont
  {Bravyi}}\ and\ \bibinfo {author} {\bibfnamefont {A.~Y.}\ \bibnamefont
  {Kitaev}},\ }\bibfield  {title} {\bibinfo {title} {Fermionic quantum
  computation},\ }\href {https://doi.org/10.1006/aphy.2002.6254} {\bibfield
  {journal} {\bibinfo  {journal} {Annals of Physics}\ }\textbf {\bibinfo
  {volume} {298}},\ \bibinfo {pages} {210--226} (\bibinfo {year}
  {2002})}\BibitemShut {NoStop}%
\bibitem [{\citenamefont {Jiang}\ \emph {et~al.}(2020)\citenamefont {Jiang},
  \citenamefont {Kalev}, \citenamefont {Mruczkiewicz},\ and\ \citenamefont
  {Neven}}]{jiang2020optimal}%
  \BibitemOpen
  \bibfield  {author} {\bibinfo {author} {\bibfnamefont {Z.}~\bibnamefont
  {Jiang}}, \bibinfo {author} {\bibfnamefont {A.}~\bibnamefont {Kalev}},
  \bibinfo {author} {\bibfnamefont {W.}~\bibnamefont {Mruczkiewicz}},\ and\
  \bibinfo {author} {\bibfnamefont {H.}~\bibnamefont {Neven}},\ }\bibfield
  {title} {\bibinfo {title} {Optimal fermion-to-qubit mapping via ternary trees
  with applications to reduced quantum states learning},\ }\href
  {https://doi.org/10.22331/q-2020-06-04-276} {\bibfield  {journal} {\bibinfo
  {journal} {Quantum}\ }\textbf {\bibinfo {volume} {4}},\ \bibinfo {pages}
  {276} (\bibinfo {year} {2020})}\BibitemShut {NoStop}%
\bibitem [{\citenamefont {van Diepen}\ \emph {et~al.}(2021)\citenamefont {van
  Diepen}, \citenamefont {Hsiao}, \citenamefont {Mukhopadhyay}, \citenamefont
  {Reichl}, \citenamefont {Wegscheider},\ and\ \citenamefont
  {Vandersypen}}]{van2021quantum}%
  \BibitemOpen
  \bibfield  {author} {\bibinfo {author} {\bibfnamefont {C.~J.}\ \bibnamefont
  {van Diepen}}, \bibinfo {author} {\bibfnamefont {T.-K.}\ \bibnamefont
  {Hsiao}}, \bibinfo {author} {\bibfnamefont {U.}~\bibnamefont {Mukhopadhyay}},
  \bibinfo {author} {\bibfnamefont {C.}~\bibnamefont {Reichl}}, \bibinfo
  {author} {\bibfnamefont {W.}~\bibnamefont {Wegscheider}},\ and\ \bibinfo
  {author} {\bibfnamefont {L.~M.~K.}\ \bibnamefont {Vandersypen}},\ }\bibfield
  {title} {\bibinfo {title} {Quantum simulation of antiferromagnetic
  {H}eisenberg chain with gate-defined quantum dots},\ }\href
  {https://doi.org/10.1103/PhysRevX.11.041025} {\bibfield  {journal} {\bibinfo
  {journal} {Physical Review X}\ }\textbf {\bibinfo {volume} {11}},\ \bibinfo
  {pages} {041025} (\bibinfo {year} {2021})}\BibitemShut {NoStop}%
\bibitem [{\citenamefont {Barthelemy}\ and\ \citenamefont
  {Vandersypen}(2013)}]{barthelemy2013quantum}%
  \BibitemOpen
  \bibfield  {author} {\bibinfo {author} {\bibfnamefont {P.}~\bibnamefont
  {Barthelemy}}\ and\ \bibinfo {author} {\bibfnamefont {L.~M.~K.}\ \bibnamefont
  {Vandersypen}},\ }\bibfield  {title} {\bibinfo {title} {Quantum dot systems:
  a versatile platform for quantum simulations},\ }\href
  {https://doi.org/10.1002/andp.201300124} {\bibfield  {journal} {\bibinfo
  {journal} {Annalen der Physik}\ }\textbf {\bibinfo {volume} {525}},\ \bibinfo
  {pages} {808--826} (\bibinfo {year} {2013})}\BibitemShut {NoStop}%
\bibitem [{\citenamefont {Loss}\ and\ \citenamefont
  {DiVincenzo}(1998)}]{loss1998quantum}%
  \BibitemOpen
  \bibfield  {author} {\bibinfo {author} {\bibfnamefont {D.}~\bibnamefont
  {Loss}}\ and\ \bibinfo {author} {\bibfnamefont {D.~P.}\ \bibnamefont
  {DiVincenzo}},\ }\bibfield  {title} {\bibinfo {title} {Quantum computation
  with quantum dots},\ }\href {https://doi.org/10.1103/physreva.57.120}
  {\bibfield  {journal} {\bibinfo  {journal} {Physical Review A}\ }\textbf
  {\bibinfo {volume} {57}},\ \bibinfo {pages} {120} (\bibinfo {year}
  {1998})}\BibitemShut {NoStop}%
\bibitem [{\citenamefont {Hendrickx}\ \emph {et~al.}(2021)\citenamefont
  {Hendrickx}, \citenamefont {Lawrie}, \citenamefont {Russ}, \citenamefont {van
  Riggelen}, \citenamefont {de~Snoo}, \citenamefont {Schouten}, \citenamefont
  {Sammak}, \citenamefont {Scappucci},\ and\ \citenamefont
  {Veldhorst}}]{hendrickx2021four}%
  \BibitemOpen
  \bibfield  {author} {\bibinfo {author} {\bibfnamefont {N.~W.}\ \bibnamefont
  {Hendrickx}}, \bibinfo {author} {\bibfnamefont {W.~I.}\ \bibnamefont
  {Lawrie}}, \bibinfo {author} {\bibfnamefont {M.}~\bibnamefont {Russ}},
  \bibinfo {author} {\bibfnamefont {F.}~\bibnamefont {van Riggelen}}, \bibinfo
  {author} {\bibfnamefont {S.~L.}\ \bibnamefont {de~Snoo}}, \bibinfo {author}
  {\bibfnamefont {R.~N.}\ \bibnamefont {Schouten}}, \bibinfo {author}
  {\bibfnamefont {A.}~\bibnamefont {Sammak}}, \bibinfo {author} {\bibfnamefont
  {G.}~\bibnamefont {Scappucci}},\ and\ \bibinfo {author} {\bibfnamefont
  {M.}~\bibnamefont {Veldhorst}},\ }\bibfield  {title} {\bibinfo {title} {A
  four-qubit germanium quantum processor},\ }\href
  {https://doi.org/10.1038/s41586-021-03332-6} {\bibfield  {journal} {\bibinfo
  {journal} {Nature}\ }\textbf {\bibinfo {volume} {591}},\ \bibinfo {pages}
  {580--585} (\bibinfo {year} {2021})}\BibitemShut {NoStop}%
\bibitem [{\citenamefont {Foxen}\ \emph {et~al.}(2020)\citenamefont {Foxen},
  \citenamefont {Neill}, \citenamefont {Dunsworth}, \citenamefont {Roushan},
  \citenamefont {Chiaro}, \citenamefont {Megrant}, \citenamefont {Kelly},
  \citenamefont {Chen}, \citenamefont {Satzinger}, \citenamefont {Barends}
  \emph {et~al.}}]{foxen2020demonstrating}%
  \BibitemOpen
  \bibfield  {author} {\bibinfo {author} {\bibfnamefont {B.}~\bibnamefont
  {Foxen}}, \bibinfo {author} {\bibfnamefont {C.}~\bibnamefont {Neill}},
  \bibinfo {author} {\bibfnamefont {A.}~\bibnamefont {Dunsworth}}, \bibinfo
  {author} {\bibfnamefont {P.}~\bibnamefont {Roushan}}, \bibinfo {author}
  {\bibfnamefont {B.}~\bibnamefont {Chiaro}}, \bibinfo {author} {\bibfnamefont
  {A.}~\bibnamefont {Megrant}}, \bibinfo {author} {\bibfnamefont
  {J.}~\bibnamefont {Kelly}}, \bibinfo {author} {\bibfnamefont
  {Z.}~\bibnamefont {Chen}}, \bibinfo {author} {\bibfnamefont {K.}~\bibnamefont
  {Satzinger}}, \bibinfo {author} {\bibfnamefont {R.}~\bibnamefont {Barends}},
  \emph {et~al.},\ }\bibfield  {title} {\bibinfo {title} {Demonstrating a
  continuous set of two-qubit gates for near-term quantum algorithms},\ }\href
  {https://doi.org/10.1103/PhysRevLett.125.120504} {\bibfield  {journal}
  {\bibinfo  {journal} {Physical Review Letters}\ }\textbf {\bibinfo {volume}
  {125}},\ \bibinfo {pages} {120504} (\bibinfo {year} {2020})}\BibitemShut
  {NoStop}%
\bibitem [{\citenamefont {Kandala}\ \emph {et~al.}(2017)\citenamefont
  {Kandala}, \citenamefont {Mezzacapo}, \citenamefont {Temme}, \citenamefont
  {Takita}, \citenamefont {Brink}, \citenamefont {Chow},\ and\ \citenamefont
  {Gambetta}}]{kandala2017hardware}%
  \BibitemOpen
  \bibfield  {author} {\bibinfo {author} {\bibfnamefont {A.}~\bibnamefont
  {Kandala}}, \bibinfo {author} {\bibfnamefont {A.}~\bibnamefont {Mezzacapo}},
  \bibinfo {author} {\bibfnamefont {K.}~\bibnamefont {Temme}}, \bibinfo
  {author} {\bibfnamefont {M.}~\bibnamefont {Takita}}, \bibinfo {author}
  {\bibfnamefont {M.}~\bibnamefont {Brink}}, \bibinfo {author} {\bibfnamefont
  {J.~M.}\ \bibnamefont {Chow}},\ and\ \bibinfo {author} {\bibfnamefont
  {J.~M.}\ \bibnamefont {Gambetta}},\ }\bibfield  {title} {\bibinfo {title}
  {Hardware-efficient variational quantum eigensolver for small molecules and
  quantum magnets},\ }\href {https://doi.org/10.1038/nature23879} {\bibfield
  {journal} {\bibinfo  {journal} {Nature}\ }\textbf {\bibinfo {volume} {549}},\
  \bibinfo {pages} {242--246} (\bibinfo {year} {2017})}\BibitemShut {NoStop}%
\bibitem [{\citenamefont {McClean}\ \emph {et~al.}(2018)\citenamefont
  {McClean}, \citenamefont {Boixo}, \citenamefont {Smelyanskiy}, \citenamefont
  {Babbush},\ and\ \citenamefont {Neven}}]{mcclean2018barren}%
  \BibitemOpen
  \bibfield  {author} {\bibinfo {author} {\bibfnamefont {J.~R.}\ \bibnamefont
  {McClean}}, \bibinfo {author} {\bibfnamefont {S.}~\bibnamefont {Boixo}},
  \bibinfo {author} {\bibfnamefont {V.~N.}\ \bibnamefont {Smelyanskiy}},
  \bibinfo {author} {\bibfnamefont {R.}~\bibnamefont {Babbush}},\ and\ \bibinfo
  {author} {\bibfnamefont {H.}~\bibnamefont {Neven}},\ }\bibfield  {title}
  {\bibinfo {title} {Barren plateaus in quantum neural network training
  landscapes},\ }\href {https://doi.org/10.1038/s41467-018-07090-4} {\bibfield
  {journal} {\bibinfo  {journal} {Nature Communications}\ }\textbf {\bibinfo
  {volume} {9}},\ \bibinfo {pages} {1--6} (\bibinfo {year} {2018})}\BibitemShut
  {NoStop}%
\bibitem [{\citenamefont {Wiersema}\ \emph {et~al.}(2020)\citenamefont
  {Wiersema}, \citenamefont {Zhou}, \citenamefont {de~Sereville}, \citenamefont
  {Carrasquilla}, \citenamefont {Kim},\ and\ \citenamefont
  {Yuen}}]{wiersema2020exploring}%
  \BibitemOpen
  \bibfield  {author} {\bibinfo {author} {\bibfnamefont {R.}~\bibnamefont
  {Wiersema}}, \bibinfo {author} {\bibfnamefont {C.}~\bibnamefont {Zhou}},
  \bibinfo {author} {\bibfnamefont {Y.}~\bibnamefont {de~Sereville}}, \bibinfo
  {author} {\bibfnamefont {J.~F.}\ \bibnamefont {Carrasquilla}}, \bibinfo
  {author} {\bibfnamefont {Y.~B.}\ \bibnamefont {Kim}},\ and\ \bibinfo {author}
  {\bibfnamefont {H.}~\bibnamefont {Yuen}},\ }\bibfield  {title} {\bibinfo
  {title} {Exploring entanglement and optimization within the hamiltonian
  variational ansatz},\ }\href {https://doi.org/10.1103/prxquantum.1.020319}
  {\bibfield  {journal} {\bibinfo  {journal} {PRX Quantum}\ }\textbf {\bibinfo
  {volume} {1}},\ \bibinfo {pages} {020319} (\bibinfo {year}
  {2020})}\BibitemShut {NoStop}%
\bibitem [{\citenamefont {Fowler}\ \emph {et~al.}(2012)\citenamefont {Fowler},
  \citenamefont {Mariantoni}, \citenamefont {Martinis},\ and\ \citenamefont
  {Cleland}}]{fowler2012surface}%
  \BibitemOpen
  \bibfield  {author} {\bibinfo {author} {\bibfnamefont {A.~G.}\ \bibnamefont
  {Fowler}}, \bibinfo {author} {\bibfnamefont {M.}~\bibnamefont {Mariantoni}},
  \bibinfo {author} {\bibfnamefont {J.~M.}\ \bibnamefont {Martinis}},\ and\
  \bibinfo {author} {\bibfnamefont {A.~N.}\ \bibnamefont {Cleland}},\
  }\bibfield  {title} {\bibinfo {title} {Surface codes: Towards practical
  large-scale quantum computation},\ }\href
  {https://doi.org/10.1103/PhysRevA.86.032324} {\bibfield  {journal} {\bibinfo
  {journal} {Physical Review A}\ }\textbf {\bibinfo {volume} {86}},\ \bibinfo
  {pages} {032324} (\bibinfo {year} {2012})}\BibitemShut {NoStop}%
\bibitem [{\citenamefont {Hill}\ \emph {et~al.}(2015)\citenamefont {Hill},
  \citenamefont {Peretz}, \citenamefont {Hile}, \citenamefont {House},
  \citenamefont {Fuechsle}, \citenamefont {Rogge}, \citenamefont {Simmons},\
  and\ \citenamefont {Hollenberg}}]{hill2015surface}%
  \BibitemOpen
  \bibfield  {author} {\bibinfo {author} {\bibfnamefont {C.~D.}\ \bibnamefont
  {Hill}}, \bibinfo {author} {\bibfnamefont {E.}~\bibnamefont {Peretz}},
  \bibinfo {author} {\bibfnamefont {S.~J.}\ \bibnamefont {Hile}}, \bibinfo
  {author} {\bibfnamefont {M.~G.}\ \bibnamefont {House}}, \bibinfo {author}
  {\bibfnamefont {M.}~\bibnamefont {Fuechsle}}, \bibinfo {author}
  {\bibfnamefont {S.}~\bibnamefont {Rogge}}, \bibinfo {author} {\bibfnamefont
  {M.~Y.}\ \bibnamefont {Simmons}},\ and\ \bibinfo {author} {\bibfnamefont
  {L.~C.}\ \bibnamefont {Hollenberg}},\ }\bibfield  {title} {\bibinfo {title}
  {A surface code quantum computer in silicon},\ }\href
  {https://doi.org/10.1126/sciadv.1500707} {\bibfield  {journal} {\bibinfo
  {journal} {Science advances}\ }\textbf {\bibinfo {volume} {1}},\ \bibinfo
  {pages} {e1500707} (\bibinfo {year} {2015})}\BibitemShut {NoStop}%
\bibitem [{\citenamefont {Andersen}\ \emph {et~al.}(2020)\citenamefont
  {Andersen}, \citenamefont {Remm}, \citenamefont {Lazar}, \citenamefont
  {Krinner}, \citenamefont {Lacroix}, \citenamefont {Norris}, \citenamefont
  {Gabureac}, \citenamefont {Eichler},\ and\ \citenamefont
  {Wallraff}}]{andersen2020repeated}%
  \BibitemOpen
  \bibfield  {author} {\bibinfo {author} {\bibfnamefont {C.~K.}\ \bibnamefont
  {Andersen}}, \bibinfo {author} {\bibfnamefont {A.}~\bibnamefont {Remm}},
  \bibinfo {author} {\bibfnamefont {S.}~\bibnamefont {Lazar}}, \bibinfo
  {author} {\bibfnamefont {S.}~\bibnamefont {Krinner}}, \bibinfo {author}
  {\bibfnamefont {N.}~\bibnamefont {Lacroix}}, \bibinfo {author} {\bibfnamefont
  {G.~J.}\ \bibnamefont {Norris}}, \bibinfo {author} {\bibfnamefont
  {M.}~\bibnamefont {Gabureac}}, \bibinfo {author} {\bibfnamefont
  {C.}~\bibnamefont {Eichler}},\ and\ \bibinfo {author} {\bibfnamefont
  {A.}~\bibnamefont {Wallraff}},\ }\bibfield  {title} {\bibinfo {title}
  {Repeated quantum error detection in a surface code},\ }\href
  {https://doi.org/10.1038/s41567-020-0920-y} {\bibfield  {journal} {\bibinfo
  {journal} {Nature Physics}\ }\textbf {\bibinfo {volume} {16}},\ \bibinfo
  {pages} {875--880} (\bibinfo {year} {2020})}\BibitemShut {NoStop}%
\bibitem [{\citenamefont {Versluis}\ \emph {et~al.}(2017)\citenamefont
  {Versluis}, \citenamefont {Poletto}, \citenamefont {Khammassi}, \citenamefont
  {Tarasinski}, \citenamefont {Haider}, \citenamefont {Michalak}, \citenamefont
  {Bruno}, \citenamefont {Bertels},\ and\ \citenamefont
  {DiCarlo}}]{versluis2017scalable}%
  \BibitemOpen
  \bibfield  {author} {\bibinfo {author} {\bibfnamefont {R.}~\bibnamefont
  {Versluis}}, \bibinfo {author} {\bibfnamefont {S.}~\bibnamefont {Poletto}},
  \bibinfo {author} {\bibfnamefont {N.}~\bibnamefont {Khammassi}}, \bibinfo
  {author} {\bibfnamefont {B.}~\bibnamefont {Tarasinski}}, \bibinfo {author}
  {\bibfnamefont {N.}~\bibnamefont {Haider}}, \bibinfo {author} {\bibfnamefont
  {D.~J.}\ \bibnamefont {Michalak}}, \bibinfo {author} {\bibfnamefont
  {A.}~\bibnamefont {Bruno}}, \bibinfo {author} {\bibfnamefont
  {K.}~\bibnamefont {Bertels}},\ and\ \bibinfo {author} {\bibfnamefont
  {L.}~\bibnamefont {DiCarlo}},\ }\bibfield  {title} {\bibinfo {title}
  {Scalable quantum circuit and control for a superconducting surface code},\
  }\href {https://doi.org/10.1103/PhysRevApplied.8.034021} {\bibfield
  {journal} {\bibinfo  {journal} {Physical Review Applied}\ }\textbf {\bibinfo
  {volume} {8}},\ \bibinfo {pages} {034021} (\bibinfo {year}
  {2017})}\BibitemShut {NoStop}%
\bibitem [{\citenamefont {Hutter}\ \emph {et~al.}(2015)\citenamefont {Hutter},
  \citenamefont {Wootton},\ and\ \citenamefont
  {Loss}}]{hutter2015parafermions}%
  \BibitemOpen
  \bibfield  {author} {\bibinfo {author} {\bibfnamefont {A.}~\bibnamefont
  {Hutter}}, \bibinfo {author} {\bibfnamefont {J.~R.}\ \bibnamefont
  {Wootton}},\ and\ \bibinfo {author} {\bibfnamefont {D.}~\bibnamefont
  {Loss}},\ }\bibfield  {title} {\bibinfo {title} {Parafermions in a kagome
  lattice of qubits for topological quantum computation},\ }\href
  {https://doi.org/10.1103/PhysRevX.5.041040} {\bibfield  {journal} {\bibinfo
  {journal} {Physical Review X}\ }\textbf {\bibinfo {volume} {5}},\ \bibinfo
  {pages} {041040} (\bibinfo {year} {2015})}\BibitemShut {NoStop}%
\bibitem [{\citenamefont {Schmidt}\ and\ \citenamefont
  {Koch}(2013)}]{schmidt2012circuit}%
  \BibitemOpen
  \bibfield  {author} {\bibinfo {author} {\bibfnamefont {S.}~\bibnamefont
  {Schmidt}}\ and\ \bibinfo {author} {\bibfnamefont {J.}~\bibnamefont {Koch}},\
  }\bibfield  {title} {\bibinfo {title} {{Circuit QED lattices: towards quantum
  simulation with superconducting circuits}},\ }\href
  {https://doi.org/10.1002/andp.201200261} {\bibfield  {journal} {\bibinfo
  {journal} {Annalen der Physik}\ }\textbf {\bibinfo {volume} {525}},\ \bibinfo
  {pages} {395--412} (\bibinfo {year} {2013})}\BibitemShut {NoStop}%
\bibitem [{\citenamefont {{Kim}}\ and\ \citenamefont
  {{Kim}}(2017)}]{kim2017scalable}%
  \BibitemOpen
  \bibfield  {author} {\bibinfo {author} {\bibfnamefont {M.~D.}\ \bibnamefont
  {{Kim}}}\ and\ \bibinfo {author} {\bibfnamefont {J.}~\bibnamefont {{Kim}}},\
  }\bibfield  {title} {\bibinfo {title} {{Scalable quantum computing model in
  the circuit-QED lattice with circulator function}},\ }\href
  {https://doi.org/10.1007/s11128-017-1644-5} {\bibfield  {journal} {\bibinfo
  {journal} {Quantum Information Processing}\ }\textbf {\bibinfo {volume}
  {16}},\ \bibinfo {eid} {192} (\bibinfo {year} {2017})}\BibitemShut {NoStop}%
\bibitem [{\citenamefont {Kim}(2019)}]{kim2019quantum}%
  \BibitemOpen
  \bibfield  {author} {\bibinfo {author} {\bibfnamefont {M.~D.}\ \bibnamefont
  {Kim}},\ }\bibfield  {title} {\bibinfo {title} {Quantum simulation scheme of
  two-dimensional xy-model hamiltonian with controllable coupling},\ }\href
  {https://doi.org/10.1007/s11128-019-2173-1} {\bibfield  {journal} {\bibinfo
  {journal} {Quantum Information Processing}\ }\textbf {\bibinfo {volume}
  {18}},\ \bibinfo {pages} {1--9} (\bibinfo {year} {2019})}\BibitemShut
  {NoStop}%
\bibitem [{\citenamefont {Underwood}\ \emph {et~al.}(2012)\citenamefont
  {Underwood}, \citenamefont {Shanks}, \citenamefont {Koch},\ and\
  \citenamefont {Houck}}]{underwood2012low}%
  \BibitemOpen
  \bibfield  {author} {\bibinfo {author} {\bibfnamefont {D.~L.}\ \bibnamefont
  {Underwood}}, \bibinfo {author} {\bibfnamefont {W.~E.}\ \bibnamefont
  {Shanks}}, \bibinfo {author} {\bibfnamefont {J.}~\bibnamefont {Koch}},\ and\
  \bibinfo {author} {\bibfnamefont {A.~A.}\ \bibnamefont {Houck}},\ }\bibfield
  {title} {\bibinfo {title} {Low-disorder microwave cavity lattices for quantum
  simulation with photons},\ }\href
  {https://doi.org/10.1103/PhysRevA.86.023837} {\bibfield  {journal} {\bibinfo
  {journal} {Physical Review A}\ }\textbf {\bibinfo {volume} {86}},\ \bibinfo
  {pages} {023837} (\bibinfo {year} {2012})}\BibitemShut {NoStop}%
\bibitem [{\citenamefont {Koll{\'a}r}\ \emph {et~al.}(2019)\citenamefont
  {Koll{\'a}r}, \citenamefont {Fitzpatrick},\ and\ \citenamefont
  {Houck}}]{kollar2019hyperbolic}%
  \BibitemOpen
  \bibfield  {author} {\bibinfo {author} {\bibfnamefont {A.~J.}\ \bibnamefont
  {Koll{\'a}r}}, \bibinfo {author} {\bibfnamefont {M.}~\bibnamefont
  {Fitzpatrick}},\ and\ \bibinfo {author} {\bibfnamefont {A.~A.}\ \bibnamefont
  {Houck}},\ }\bibfield  {title} {\bibinfo {title} {Hyperbolic lattices in
  circuit quantum electrodynamics},\ }\href
  {https://doi.org/10.1038/s41586-019-1348-3} {\bibfield  {journal} {\bibinfo
  {journal} {Nature}\ }\textbf {\bibinfo {volume} {571}},\ \bibinfo {pages}
  {45--50} (\bibinfo {year} {2019})}\BibitemShut {NoStop}%
\bibitem [{\citenamefont {Cirac}\ and\ \citenamefont
  {Zoller}(1995)}]{cirac1996quantum}%
  \BibitemOpen
  \bibfield  {author} {\bibinfo {author} {\bibfnamefont {J.~I.}\ \bibnamefont
  {Cirac}}\ and\ \bibinfo {author} {\bibfnamefont {P.}~\bibnamefont {Zoller}},\
  }\bibfield  {title} {\bibinfo {title} {Quantum computations with cold trapped
  ions},\ }\href {https://doi.org/10.1103/PhysRevLett.74.4091} {\bibfield
  {journal} {\bibinfo  {journal} {Physical Review Letters}\ }\textbf {\bibinfo
  {volume} {74}},\ \bibinfo {pages} {4091--4094} (\bibinfo {year}
  {1995})}\BibitemShut {NoStop}%
\bibitem [{\citenamefont {M\o{}lmer}\ and\ \citenamefont
  {S\o{}rensen}(1999)}]{molmer1999multiparticle}%
  \BibitemOpen
  \bibfield  {author} {\bibinfo {author} {\bibfnamefont {K.}~\bibnamefont
  {M\o{}lmer}}\ and\ \bibinfo {author} {\bibfnamefont {A.}~\bibnamefont
  {S\o{}rensen}},\ }\bibfield  {title} {\bibinfo {title} {Multiparticle
  entanglement of hot trapped ions},\ }\href
  {https://doi.org/10.1103/PhysRevLett.82.1835} {\bibfield  {journal} {\bibinfo
   {journal} {Physical Review Letters}\ }\textbf {\bibinfo {volume} {82}},\
  \bibinfo {pages} {1835--1838} (\bibinfo {year} {1999})}\BibitemShut {NoStop}%
\bibitem [{\citenamefont {Kielpinski}\ \emph {et~al.}(2002)\citenamefont
  {Kielpinski}, \citenamefont {Monroe},\ and\ \citenamefont
  {Wineland}}]{kielpinski2002architecture}%
  \BibitemOpen
  \bibfield  {author} {\bibinfo {author} {\bibfnamefont {D.}~\bibnamefont
  {Kielpinski}}, \bibinfo {author} {\bibfnamefont {C.}~\bibnamefont {Monroe}},\
  and\ \bibinfo {author} {\bibfnamefont {D.~J.}\ \bibnamefont {Wineland}},\
  }\bibfield  {title} {\bibinfo {title} {Architecture for a large-scale
  ion-trap quantum computer},\ }\href {https://doi.org/10.1038/nature00784}
  {\bibfield  {journal} {\bibinfo  {journal} {Nature}\ }\textbf {\bibinfo
  {volume} {417}},\ \bibinfo {pages} {709} (\bibinfo {year}
  {2002})}\BibitemShut {NoStop}%
\bibitem [{\citenamefont {Monz}\ \emph {et~al.}(2011)\citenamefont {Monz},
  \citenamefont {Schindler}, \citenamefont {Barreiro}, \citenamefont {Chwalla},
  \citenamefont {Nigg}, \citenamefont {Coish}, \citenamefont {Harlander},
  \citenamefont {H\"ansel}, \citenamefont {Hennrich},\ and\ \citenamefont
  {Blatt}}]{monz201114-qubit}%
  \BibitemOpen
  \bibfield  {author} {\bibinfo {author} {\bibfnamefont {T.}~\bibnamefont
  {Monz}}, \bibinfo {author} {\bibfnamefont {P.}~\bibnamefont {Schindler}},
  \bibinfo {author} {\bibfnamefont {J.~T.}\ \bibnamefont {Barreiro}}, \bibinfo
  {author} {\bibfnamefont {M.}~\bibnamefont {Chwalla}}, \bibinfo {author}
  {\bibfnamefont {D.}~\bibnamefont {Nigg}}, \bibinfo {author} {\bibfnamefont
  {W.~A.}\ \bibnamefont {Coish}}, \bibinfo {author} {\bibfnamefont
  {M.}~\bibnamefont {Harlander}}, \bibinfo {author} {\bibfnamefont
  {W.}~\bibnamefont {H\"ansel}}, \bibinfo {author} {\bibfnamefont
  {M.}~\bibnamefont {Hennrich}},\ and\ \bibinfo {author} {\bibfnamefont
  {R.}~\bibnamefont {Blatt}},\ }\bibfield  {title} {\bibinfo {title} {14-qubit
  entanglement: Creation and coherence},\ }\href
  {https://doi.org/10.1103/PhysRevLett.106.130506} {\bibfield  {journal}
  {\bibinfo  {journal} {Physical Review Letters}\ }\textbf {\bibinfo {volume}
  {106}},\ \bibinfo {pages} {130506} (\bibinfo {year} {2011})}\BibitemShut
  {NoStop}%
\bibitem [{\citenamefont {Childs}\ \emph {et~al.}(2018)\citenamefont {Childs},
  \citenamefont {Maslov}, \citenamefont {Nam}, \citenamefont {Ross},\ and\
  \citenamefont {Su}}]{childs2018towards}%
  \BibitemOpen
  \bibfield  {author} {\bibinfo {author} {\bibfnamefont {A.~M.}\ \bibnamefont
  {Childs}}, \bibinfo {author} {\bibfnamefont {D.}~\bibnamefont {Maslov}},
  \bibinfo {author} {\bibfnamefont {Y.}~\bibnamefont {Nam}}, \bibinfo {author}
  {\bibfnamefont {N.~J.}\ \bibnamefont {Ross}},\ and\ \bibinfo {author}
  {\bibfnamefont {Y.}~\bibnamefont {Su}},\ }\bibfield  {title} {\bibinfo
  {title} {Toward the first quantum simulation with quantum speedup},\ }\href
  {https://doi.org/10.1073/pnas.1801723115} {\bibfield  {journal} {\bibinfo
  {journal} {Proceedings of the National Academy of Sciences}\ }\textbf
  {\bibinfo {volume} {115}},\ \bibinfo {pages} {9456--9461} (\bibinfo {year}
  {2018})}\BibitemShut {NoStop}%
\bibitem [{Note1()}]{Note1}%
  \BibitemOpen
  \bibinfo {note} {For the sake of comparison of the gate count, we assume the
  same native gates as before. To the best of our knowledge, quantum computers
  that support both the essentially native implementation of the exchange
  interaction and kagome connectivity do not exist yet, so on current devices
  with all-to-all connectivity there will be some transpilation
  overhead.}\BibitemShut {Stop}%
\bibitem [{\citenamefont {Palma}\ \emph {et~al.}(1996)\citenamefont {Palma},
  \citenamefont {Suominen},\ and\ \citenamefont {Ekert}}]{palma1996quantum}%
  \BibitemOpen
  \bibfield  {author} {\bibinfo {author} {\bibfnamefont {G.~M.}\ \bibnamefont
  {Palma}}, \bibinfo {author} {\bibfnamefont {K.-A.}\ \bibnamefont
  {Suominen}},\ and\ \bibinfo {author} {\bibfnamefont {A.~K.}\ \bibnamefont
  {Ekert}},\ }\bibfield  {title} {\bibinfo {title} {Quantum computers and
  dissipation},\ }\href {https://doi.org/10.1098/rspa.1996.0029} {\bibfield
  {journal} {\bibinfo  {journal} {Proceedings of the Royal Society of London
  A}\ }\textbf {\bibinfo {volume} {452}},\ \bibinfo {pages} {567--584}
  (\bibinfo {year} {1996})}\BibitemShut {NoStop}%
\bibitem [{\citenamefont {Zanardi}\ and\ \citenamefont
  {Rasetti}(1997)}]{zanardi1997error}%
  \BibitemOpen
  \bibfield  {author} {\bibinfo {author} {\bibfnamefont {P.}~\bibnamefont
  {Zanardi}}\ and\ \bibinfo {author} {\bibfnamefont {M.}~\bibnamefont
  {Rasetti}},\ }\bibfield  {title} {\bibinfo {title} {Error avoiding quantum
  codes},\ }\href {https://doi.org/10.1142/S0217984997001304} {\bibfield
  {journal} {\bibinfo  {journal} {Modern Physics Letters B}\ }\textbf {\bibinfo
  {volume} {11}},\ \bibinfo {pages} {1085--1093} (\bibinfo {year}
  {1997})}\BibitemShut {NoStop}%
\bibitem [{\citenamefont {Duan}\ and\ \citenamefont
  {Guo}(1998)}]{duan1998reducing}%
  \BibitemOpen
  \bibfield  {author} {\bibinfo {author} {\bibfnamefont {L.-M.}\ \bibnamefont
  {Duan}}\ and\ \bibinfo {author} {\bibfnamefont {G.-C.}\ \bibnamefont {Guo}},\
  }\bibfield  {title} {\bibinfo {title} {Reducing decoherence in
  quantum-computer memory with all quantum bits coupling to the same
  environment},\ }\href {https://doi.org/10.1103/PhysRevA.57.737} {\bibfield
  {journal} {\bibinfo  {journal} {Physical Review A}\ }\textbf {\bibinfo
  {volume} {57}},\ \bibinfo {pages} {737--741} (\bibinfo {year}
  {1998})}\BibitemShut {NoStop}%
\bibitem [{\citenamefont {Lidar}\ \emph {et~al.}(1998)\citenamefont {Lidar},
  \citenamefont {Chuang},\ and\ \citenamefont {Whaley}}]{lidar1998decoherence}%
  \BibitemOpen
  \bibfield  {author} {\bibinfo {author} {\bibfnamefont {D.~A.}\ \bibnamefont
  {Lidar}}, \bibinfo {author} {\bibfnamefont {I.~L.}\ \bibnamefont {Chuang}},\
  and\ \bibinfo {author} {\bibfnamefont {K.~B.}\ \bibnamefont {Whaley}},\
  }\bibfield  {title} {\bibinfo {title} {Decoherence-free subspaces for quantum
  computation},\ }\href {https://doi.org/10.1103/PhysRevLett.81.2594}
  {\bibfield  {journal} {\bibinfo  {journal} {Physical Review Letters}\
  }\textbf {\bibinfo {volume} {81}},\ \bibinfo {pages} {2594--2597} (\bibinfo
  {year} {1998})}\BibitemShut {NoStop}%
\bibitem [{\citenamefont {Kattem\"olle}\ and\ \citenamefont {van
  Wezel}(2019)}]{kattemolle2019dynamical}%
  \BibitemOpen
  \bibfield  {author} {\bibinfo {author} {\bibfnamefont {J.}~\bibnamefont
  {Kattem\"olle}}\ and\ \bibinfo {author} {\bibfnamefont {J.}~\bibnamefont {van
  Wezel}},\ }\bibfield  {title} {\bibinfo {title} {Dynamical fidelity
  susceptibility of decoherence-free subspaces},\ }\href
  {https://doi.org/10.1103/PhysRevA.99.062340} {\bibfield  {journal} {\bibinfo
  {journal} {Physical Review A}\ }\textbf {\bibinfo {volume} {99}},\ \bibinfo
  {pages} {062340} (\bibinfo {year} {2019})}\BibitemShut {NoStop}%
\bibitem [{\citenamefont {Dicke}(1954)}]{dicke1964coherence}%
  \BibitemOpen
  \bibfield  {author} {\bibinfo {author} {\bibfnamefont {R.~H.}\ \bibnamefont
  {Dicke}},\ }\bibfield  {title} {\bibinfo {title} {Coherence in spontaneous
  radiation processes},\ }\href {https://doi.org/10.1103/PhysRev.93.99}
  {\bibfield  {journal} {\bibinfo  {journal} {Physical Review}\ }\textbf
  {\bibinfo {volume} {93}},\ \bibinfo {pages} {99--110} (\bibinfo {year}
  {1954})}\BibitemShut {NoStop}%
\bibitem [{\citenamefont {Kirton}\ \emph {et~al.}(2018)\citenamefont {Kirton},
  \citenamefont {Roses}, \citenamefont {Keeling},\ and\ \citenamefont
  {Torre}}]{kirton2018dicke}%
  \BibitemOpen
  \bibfield  {author} {\bibinfo {author} {\bibfnamefont {P.}~\bibnamefont
  {Kirton}}, \bibinfo {author} {\bibfnamefont {M.~M.}\ \bibnamefont {Roses}},
  \bibinfo {author} {\bibfnamefont {J.}~\bibnamefont {Keeling}},\ and\ \bibinfo
  {author} {\bibfnamefont {E.~G.~D.}\ \bibnamefont {Torre}},\ }\bibfield
  {title} {\bibinfo {title} {Introduction to the dicke model: From equilibrium
  to nonequilibrium, and vice versa},\ }\href
  {https://doi.org/10.1002/qute.201800043} {\bibfield  {journal} {\bibinfo
  {journal} {Advanced Quantum Technologies}\ }\textbf {\bibinfo {volume} {2}},\
  \bibinfo {pages} {1800043} (\bibinfo {year} {2018})}\BibitemShut {NoStop}%
\bibitem [{\citenamefont {Kokail}\ \emph {et~al.}(2019)\citenamefont {Kokail},
  \citenamefont {Maier}, \citenamefont {van Bijnen}, \citenamefont {Brydges},
  \citenamefont {Joshi}, \citenamefont {Jurcevic}, \citenamefont {Muschik},
  \citenamefont {Silvi}, \citenamefont {Blatt}, \citenamefont {Roos},\ and\
  \citenamefont {Zoller}}]{kokail2019self}%
  \BibitemOpen
  \bibfield  {author} {\bibinfo {author} {\bibfnamefont {C.}~\bibnamefont
  {Kokail}}, \bibinfo {author} {\bibfnamefont {C.}~\bibnamefont {Maier}},
  \bibinfo {author} {\bibfnamefont {R.}~\bibnamefont {van Bijnen}}, \bibinfo
  {author} {\bibfnamefont {T.}~\bibnamefont {Brydges}}, \bibinfo {author}
  {\bibfnamefont {M.~K.}\ \bibnamefont {Joshi}}, \bibinfo {author}
  {\bibfnamefont {P.}~\bibnamefont {Jurcevic}}, \bibinfo {author}
  {\bibfnamefont {C.~A.}\ \bibnamefont {Muschik}}, \bibinfo {author}
  {\bibfnamefont {P.}~\bibnamefont {Silvi}}, \bibinfo {author} {\bibfnamefont
  {R.}~\bibnamefont {Blatt}}, \bibinfo {author} {\bibfnamefont {C.~F.}\
  \bibnamefont {Roos}},\ and\ \bibinfo {author} {\bibfnamefont
  {P.}~\bibnamefont {Zoller}},\ }\bibfield  {title} {\bibinfo {title}
  {Self-verifying variational quantum simulation of lattice models},\ }\href
  {https://doi.org/10.1038/s41586-019-1177-4} {\bibfield  {journal} {\bibinfo
  {journal} {Nature}\ }\textbf {\bibinfo {volume} {569}},\ \bibinfo {pages}
  {355--360} (\bibinfo {year} {2019})}\BibitemShut {NoStop}%
\bibitem [{\citenamefont {Wang}\ \emph {et~al.}(2011)\citenamefont {Wang},
  \citenamefont {Fowler},\ and\ \citenamefont {Hollenberg}}]{wang2011surface}%
  \BibitemOpen
  \bibfield  {author} {\bibinfo {author} {\bibfnamefont {D.~S.}\ \bibnamefont
  {Wang}}, \bibinfo {author} {\bibfnamefont {A.~G.}\ \bibnamefont {Fowler}},\
  and\ \bibinfo {author} {\bibfnamefont {L.~C.~L.}\ \bibnamefont
  {Hollenberg}},\ }\bibfield  {title} {\bibinfo {title} {Surface code quantum
  computing with error rates over 1\%},\ }\href
  {https://doi.org/10.1103/PhysRevA.83.020302} {\bibfield  {journal} {\bibinfo
  {journal} {Physical Review A}\ }\textbf {\bibinfo {volume} {83}},\ \bibinfo
  {pages} {020302} (\bibinfo {year} {2011})}\BibitemShut {NoStop}%
\bibitem [{\citenamefont {{Dalton}}\ \emph {et~al.}(2022)\citenamefont
  {{Dalton}}, \citenamefont {{Long}}, \citenamefont {{Yordanov}}, \citenamefont
  {{Smith}}, \citenamefont {{Barnes}}, \citenamefont {{Mertig}},\ and\
  \citenamefont {{Arvidsson-Shukur}}}]{dalton2022variational}%
  \BibitemOpen
  \bibfield  {author} {\bibinfo {author} {\bibfnamefont {K.}~\bibnamefont
  {{Dalton}}}, \bibinfo {author} {\bibfnamefont {C.~K.}\ \bibnamefont
  {{Long}}}, \bibinfo {author} {\bibfnamefont {Y.~S.}\ \bibnamefont
  {{Yordanov}}}, \bibinfo {author} {\bibfnamefont {C.~G.}\ \bibnamefont
  {{Smith}}}, \bibinfo {author} {\bibfnamefont {C.~H.~W.}\ \bibnamefont
  {{Barnes}}}, \bibinfo {author} {\bibfnamefont {N.}~\bibnamefont {{Mertig}}},\
  and\ \bibinfo {author} {\bibfnamefont {D.~R.~M.}\ \bibnamefont
  {{Arvidsson-Shukur}}},\ }\href@noop {} {\bibinfo {title} {{Variational
  quantum chemistry requires gate-error probabilities below the fault-tolerance
  threshold}}} (\bibinfo {year} {2022}),\ \Eprint
  {https://arxiv.org/abs/2211.04505} {arXiv:2211.04505 [quant-ph]} \BibitemShut
  {NoStop}%
\bibitem [{\citenamefont {Temme}\ \emph {et~al.}(2017)\citenamefont {Temme},
  \citenamefont {Bravyi},\ and\ \citenamefont {Gambetta}}]{temme2017error}%
  \BibitemOpen
  \bibfield  {author} {\bibinfo {author} {\bibfnamefont {K.}~\bibnamefont
  {Temme}}, \bibinfo {author} {\bibfnamefont {S.}~\bibnamefont {Bravyi}},\ and\
  \bibinfo {author} {\bibfnamefont {J.~M.}\ \bibnamefont {Gambetta}},\
  }\bibfield  {title} {\bibinfo {title} {Error mitigation for short-depth
  quantum circuits},\ }\href {https://doi.org/10.1103/PhysRevLett.119.180509}
  {\bibfield  {journal} {\bibinfo  {journal} {Physical Review Letters}\
  }\textbf {\bibinfo {volume} {119}},\ \bibinfo {pages} {180509} (\bibinfo
  {year} {2017})}\BibitemShut {NoStop}%
\bibitem [{\citenamefont {Koczor}(2021)}]{koczor2021exponential}%
  \BibitemOpen
  \bibfield  {author} {\bibinfo {author} {\bibfnamefont {B.}~\bibnamefont
  {Koczor}},\ }\bibfield  {title} {\bibinfo {title} {Exponential error
  suppression for near-term quantum devices},\ }\href
  {https://doi.org/10.1103/PhysRevX.11.031057} {\bibfield  {journal} {\bibinfo
  {journal} {Physical Review X}\ }\textbf {\bibinfo {volume} {11}},\ \bibinfo
  {pages} {031057} (\bibinfo {year} {2021})}\BibitemShut {NoStop}%
\bibitem [{\citenamefont {Huggins}\ \emph {et~al.}(2021)\citenamefont
  {Huggins}, \citenamefont {McArdle}, \citenamefont {O'{B}rien}, \citenamefont
  {Lee}, \citenamefont {Rubin}, \citenamefont {Boixo}, \citenamefont {Whaley},
  \citenamefont {Babbush},\ and\ \citenamefont {McClean}}]{huggins2021virtual}%
  \BibitemOpen
  \bibfield  {author} {\bibinfo {author} {\bibfnamefont {W.~J.}\ \bibnamefont
  {Huggins}}, \bibinfo {author} {\bibfnamefont {S.}~\bibnamefont {McArdle}},
  \bibinfo {author} {\bibfnamefont {T.~E.}\ \bibnamefont {O'{B}rien}}, \bibinfo
  {author} {\bibfnamefont {J.}~\bibnamefont {Lee}}, \bibinfo {author}
  {\bibfnamefont {N.~C.}\ \bibnamefont {Rubin}}, \bibinfo {author}
  {\bibfnamefont {S.}~\bibnamefont {Boixo}}, \bibinfo {author} {\bibfnamefont
  {K.~B.}\ \bibnamefont {Whaley}}, \bibinfo {author} {\bibfnamefont
  {R.}~\bibnamefont {Babbush}},\ and\ \bibinfo {author} {\bibfnamefont {J.~R.}\
  \bibnamefont {McClean}},\ }\bibfield  {title} {\bibinfo {title} {Virtual
  distillation for quantum error mitigation},\ }\href
  {https://doi.org/10.1103/PhysRevX.11.041036} {\bibfield  {journal} {\bibinfo
  {journal} {Physical Review X}\ }\textbf {\bibinfo {volume} {11}},\ \bibinfo
  {pages} {041036} (\bibinfo {year} {2021})}\BibitemShut {NoStop}%
\bibitem [{\citenamefont {Song}\ \emph {et~al.}(2019)\citenamefont {Song},
  \citenamefont {Cui}, \citenamefont {Wang}, \citenamefont {Hao}, \citenamefont
  {Feng},\ and\ \citenamefont {Li}}]{song2019quantum}%
  \BibitemOpen
  \bibfield  {author} {\bibinfo {author} {\bibfnamefont {C.}~\bibnamefont
  {Song}}, \bibinfo {author} {\bibfnamefont {J.}~\bibnamefont {Cui}}, \bibinfo
  {author} {\bibfnamefont {H.}~\bibnamefont {Wang}}, \bibinfo {author}
  {\bibfnamefont {J.}~\bibnamefont {Hao}}, \bibinfo {author} {\bibfnamefont
  {H.}~\bibnamefont {Feng}},\ and\ \bibinfo {author} {\bibfnamefont
  {Y.}~\bibnamefont {Li}},\ }\bibfield  {title} {\bibinfo {title} {Quantum
  computation with universal error mitigation on a superconducting quantum
  processor},\ }\href {https://doi.org/10.1126/sciadv.aaw5686} {\bibfield
  {journal} {\bibinfo  {journal} {Science Advances}\ }\textbf {\bibinfo
  {volume} {5}},\ \bibinfo {pages} {eaaw5686} (\bibinfo {year}
  {2019})}\BibitemShut {NoStop}%
\bibitem [{\citenamefont {{Bultrini}}\ \emph {et~al.}(2021)\citenamefont
  {{Bultrini}}, \citenamefont {{Gordon}}, \citenamefont {{Czarnik}},
  \citenamefont {{Arrasmith}}, \citenamefont {{Coles}},\ and\ \citenamefont
  {{Cincio}}}]{bultrini2021quantum}%
  \BibitemOpen
  \bibfield  {author} {\bibinfo {author} {\bibfnamefont {D.}~\bibnamefont
  {{Bultrini}}}, \bibinfo {author} {\bibfnamefont {M.~H.}\ \bibnamefont
  {{Gordon}}}, \bibinfo {author} {\bibfnamefont {P.}~\bibnamefont {{Czarnik}}},
  \bibinfo {author} {\bibfnamefont {A.}~\bibnamefont {{Arrasmith}}}, \bibinfo
  {author} {\bibfnamefont {P.~J.}\ \bibnamefont {{Coles}}},\ and\ \bibinfo
  {author} {\bibfnamefont {L.}~\bibnamefont {{Cincio}}},\ }\href@noop {}
  {\bibinfo {title} {{Unifying and benchmarking state-of-the-art quantum error
  mitigation techniques}}} (\bibinfo {year} {2021}),\ \Eprint
  {https://arxiv.org/abs/2107.13470} {arXiv:2107.13470 [quant-ph]} \BibitemShut
  {NoStop}%
\bibitem [{\citenamefont {{van den Berg}}\ \emph {et~al.}(2022)\citenamefont
  {{van den Berg}}, \citenamefont {{Minev}}, \citenamefont {{Kandala}},\ and\
  \citenamefont {{Temme}}}]{berg2022probabilistic}%
  \BibitemOpen
  \bibfield  {author} {\bibinfo {author} {\bibfnamefont {E.}~\bibnamefont {{van
  den Berg}}}, \bibinfo {author} {\bibfnamefont {Z.~K.}\ \bibnamefont
  {{Minev}}}, \bibinfo {author} {\bibfnamefont {A.}~\bibnamefont {{Kandala}}},\
  and\ \bibinfo {author} {\bibfnamefont {K.}~\bibnamefont {{Temme}}},\
  }\href@noop {} {\bibinfo {title} {{Probabilistic error cancellation with
  sparse Pauli-Lindblad models on noisy quantum processors}}} (\bibinfo {year}
  {2022}),\ \Eprint {https://arxiv.org/abs/2201.09866} {arXiv:2201.09866
  [quant-ph]} \BibitemShut {NoStop}%
\bibitem [{\citenamefont {{Bravyi}}\ \emph {et~al.}(2022)\citenamefont
  {{Bravyi}}, \citenamefont {{Dial}}, \citenamefont {{Gambetta}}, \citenamefont
  {{Gil}},\ and\ \citenamefont {{Nazario}}}]{bravi2022future}%
  \BibitemOpen
  \bibfield  {author} {\bibinfo {author} {\bibfnamefont {S.}~\bibnamefont
  {{Bravyi}}}, \bibinfo {author} {\bibfnamefont {O.}~\bibnamefont {{Dial}}},
  \bibinfo {author} {\bibfnamefont {J.~M.}\ \bibnamefont {{Gambetta}}},
  \bibinfo {author} {\bibfnamefont {D.}~\bibnamefont {{Gil}}},\ and\ \bibinfo
  {author} {\bibfnamefont {Z.}~\bibnamefont {{Nazario}}},\ }\bibfield  {title}
  {\bibinfo {title} {{The future of quantum computing with superconducting
  qubits}},\ }\href {https://doi.org/10.1063/5.0082975} {\bibfield  {journal}
  {\bibinfo  {journal} {Journal of Applied Physics}\ }\textbf {\bibinfo
  {volume} {132}},\ \bibinfo {eid} {160902} (\bibinfo {year}
  {2022})}\BibitemShut {NoStop}%
\bibitem [{\citenamefont {Bonet-Monroig}\ \emph {et~al.}(2018)\citenamefont
  {Bonet-Monroig}, \citenamefont {Sagastizabal}, \citenamefont {Singh},\ and\
  \citenamefont {O'{B}rien}}]{bonet2018low-cost}%
  \BibitemOpen
  \bibfield  {author} {\bibinfo {author} {\bibfnamefont {X.}~\bibnamefont
  {Bonet-Monroig}}, \bibinfo {author} {\bibfnamefont {R.}~\bibnamefont
  {Sagastizabal}}, \bibinfo {author} {\bibfnamefont {M.}~\bibnamefont
  {Singh}},\ and\ \bibinfo {author} {\bibfnamefont {T.~E.}\ \bibnamefont
  {O'{B}rien}},\ }\bibfield  {title} {\bibinfo {title} {Low-cost error
  mitigation by symmetry verification},\ }\href
  {https://doi.org/10.1103/PhysRevA.98.062339} {\bibfield  {journal} {\bibinfo
  {journal} {Physical Review A}\ }\textbf {\bibinfo {volume} {98}},\ \bibinfo
  {pages} {062339} (\bibinfo {year} {2018})}\BibitemShut {NoStop}%
\bibitem [{\citenamefont {Sagastizabal}\ \emph {et~al.}(2019)\citenamefont
  {Sagastizabal}, \citenamefont {Bonet-Monroig}, \citenamefont {Singh},
  \citenamefont {Rol}, \citenamefont {Bultink}, \citenamefont {Fu},
  \citenamefont {Price}, \citenamefont {Ostroukh}, \citenamefont
  {Muthusubramanian}, \citenamefont {Bruno}, \citenamefont {Beekman},
  \citenamefont {Haider}, \citenamefont {O'{B}rien},\ and\ \citenamefont
  {DiCarlo}}]{sagastizabal2019experimental}%
  \BibitemOpen
  \bibfield  {author} {\bibinfo {author} {\bibfnamefont {R.}~\bibnamefont
  {Sagastizabal}}, \bibinfo {author} {\bibfnamefont {X.}~\bibnamefont
  {Bonet-Monroig}}, \bibinfo {author} {\bibfnamefont {M.}~\bibnamefont
  {Singh}}, \bibinfo {author} {\bibfnamefont {M.~A.}\ \bibnamefont {Rol}},
  \bibinfo {author} {\bibfnamefont {C.~C.}\ \bibnamefont {Bultink}}, \bibinfo
  {author} {\bibfnamefont {X.}~\bibnamefont {Fu}}, \bibinfo {author}
  {\bibfnamefont {C.~H.}\ \bibnamefont {Price}}, \bibinfo {author}
  {\bibfnamefont {V.~P.}\ \bibnamefont {Ostroukh}}, \bibinfo {author}
  {\bibfnamefont {N.}~\bibnamefont {Muthusubramanian}}, \bibinfo {author}
  {\bibfnamefont {A.}~\bibnamefont {Bruno}}, \bibinfo {author} {\bibfnamefont
  {M.}~\bibnamefont {Beekman}}, \bibinfo {author} {\bibfnamefont
  {N.}~\bibnamefont {Haider}}, \bibinfo {author} {\bibfnamefont {T.~E.}\
  \bibnamefont {O'{B}rien}},\ and\ \bibinfo {author} {\bibfnamefont
  {L.}~\bibnamefont {DiCarlo}},\ }\bibfield  {title} {\bibinfo {title}
  {Experimental error mitigation via symmetry verification in a variational
  quantum eigensolver},\ }\href {https://doi.org/10.1103/PhysRevA.100.010302}
  {\bibfield  {journal} {\bibinfo  {journal} {Physical Review A}\ }\textbf
  {\bibinfo {volume} {100}},\ \bibinfo {pages} {010302} (\bibinfo {year}
  {2019})}\BibitemShut {NoStop}%
\bibitem [{\citenamefont {Kempe}\ \emph {et~al.}(2006)\citenamefont {Kempe},
  \citenamefont {Kitaev},\ and\ \citenamefont {Regev}}]{kempe2006complexity}%
  \BibitemOpen
  \bibfield  {author} {\bibinfo {author} {\bibfnamefont {J.}~\bibnamefont
  {Kempe}}, \bibinfo {author} {\bibfnamefont {A.}~\bibnamefont {Kitaev}},\ and\
  \bibinfo {author} {\bibfnamefont {O.}~\bibnamefont {Regev}},\ }\bibfield
  {title} {\bibinfo {title} {The complexity of the local hamiltonian problem},\
  }\href {https://doi.org/10.1137/s0097539704445226} {\bibfield  {journal}
  {\bibinfo  {journal} {SIAM Journal on Computing}\ }\textbf {\bibinfo {volume}
  {35}},\ \bibinfo {pages} {1070--1097} (\bibinfo {year} {2006})}\BibitemShut
  {NoStop}%
\bibitem [{\citenamefont {Piddock}\ and\ \citenamefont
  {Montanaro}(2017)}]{piddock2017complexity}%
  \BibitemOpen
  \bibfield  {author} {\bibinfo {author} {\bibfnamefont {S.}~\bibnamefont
  {Piddock}}\ and\ \bibinfo {author} {\bibfnamefont {A.}~\bibnamefont
  {Montanaro}},\ }\bibfield  {title} {\bibinfo {title} {The complexity of
  antiferromagnetic interactions and 2d lattices},\ }\href
  {https://doi.org/10.26421/qic17.7-8-6} {\bibfield  {journal} {\bibinfo
  {journal} {Quantum Information and Computation}\ }\textbf {\bibinfo {volume}
  {17}},\ \bibinfo {pages} {636--672} (\bibinfo {year} {2017})}\BibitemShut
  {NoStop}%
\bibitem [{\citenamefont {{Aharonov}}\ and\ \citenamefont
  {{Naveh}}(2002)}]{aharonov2002quantum}%
  \BibitemOpen
  \bibfield  {author} {\bibinfo {author} {\bibfnamefont {D.}~\bibnamefont
  {{Aharonov}}}\ and\ \bibinfo {author} {\bibfnamefont {T.}~\bibnamefont
  {{Naveh}}},\ }\href@noop {} {\bibinfo {title} {{Quantum NP - A Survey}}}
  (\bibinfo {year} {2002}),\ \Eprint {https://arxiv.org/abs/quant-ph/0210077}
  {arXiv:quant-ph/0210077 [quant-ph]} \BibitemShut {NoStop}%
\bibitem [{\citenamefont {Szabo}\ and\ \citenamefont
  {Ostlund}(2012)}]{szabo2012modern}%
  \BibitemOpen
  \bibfield  {author} {\bibinfo {author} {\bibfnamefont {A.}~\bibnamefont
  {Szabo}}\ and\ \bibinfo {author} {\bibfnamefont {N.~S.}\ \bibnamefont
  {Ostlund}},\ }\href@noop {} {\emph {\bibinfo {title} {Modern quantum
  chemistry: introduction to advanced electronic structure theory}}}\ (\bibinfo
   {publisher} {Courier Corporation},\ \bibinfo {year} {2012})\BibitemShut
  {NoStop}%
\bibitem [{\citenamefont {Huang}\ \emph {et~al.}(2021)\citenamefont {Huang},
  \citenamefont {Kueng},\ and\ \citenamefont {Preskill}}]{huang2021efficient}%
  \BibitemOpen
  \bibfield  {author} {\bibinfo {author} {\bibfnamefont {H.-Y.}\ \bibnamefont
  {Huang}}, \bibinfo {author} {\bibfnamefont {R.}~\bibnamefont {Kueng}},\ and\
  \bibinfo {author} {\bibfnamefont {J.}~\bibnamefont {Preskill}},\ }\bibfield
  {title} {\bibinfo {title} {Efficient estimation of pauli observables by
  derandomization},\ }\href {https://doi.org/10.1103/PhysRevLett.127.030503}
  {\bibfield  {journal} {\bibinfo  {journal} {Physical Review Letters}\
  }\textbf {\bibinfo {volume} {127}},\ \bibinfo {pages} {030503} (\bibinfo
  {year} {2021})}\BibitemShut {NoStop}%
\bibitem [{\citenamefont {Lecheminant}\ \emph {et~al.}(1997)\citenamefont
  {Lecheminant}, \citenamefont {Bernu}, \citenamefont {Lhuillier},
  \citenamefont {Pierre},\ and\ \citenamefont
  {Sindzingre}}]{lecheminant1997order}%
  \BibitemOpen
  \bibfield  {author} {\bibinfo {author} {\bibfnamefont {P.}~\bibnamefont
  {Lecheminant}}, \bibinfo {author} {\bibfnamefont {B.}~\bibnamefont {Bernu}},
  \bibinfo {author} {\bibfnamefont {C.}~\bibnamefont {Lhuillier}}, \bibinfo
  {author} {\bibfnamefont {L.}~\bibnamefont {Pierre}},\ and\ \bibinfo {author}
  {\bibfnamefont {P.}~\bibnamefont {Sindzingre}},\ }\bibfield  {title}
  {\bibinfo {title} {Order versus disorder in the quantum {H}eisenberg
  antiferromagnet on the kagom{\'{e}} lattice using exact spectra analysis},\
  }\href {https://doi.org/10.1103/physrevb.56.2521} {\bibfield  {journal}
  {\bibinfo  {journal} {Physical Review B}\ }\textbf {\bibinfo {volume} {56}},\
  \bibinfo {pages} {2521--2529} (\bibinfo {year} {1997})}\BibitemShut {NoStop}%
\bibitem [{\citenamefont {Sindzingre}\ and\ \citenamefont
  {Lhuillier}(2009)}]{sindzingre2009low-energy}%
  \BibitemOpen
  \bibfield  {author} {\bibinfo {author} {\bibfnamefont {P.}~\bibnamefont
  {Sindzingre}}\ and\ \bibinfo {author} {\bibfnamefont {C.}~\bibnamefont
  {Lhuillier}},\ }\bibfield  {title} {\bibinfo {title} {Low-energy excitations
  of the kagom{\'{e}} antiferromagnet and the spin-gap issue},\ }\href
  {https://doi.org/10.1209/0295-5075/88/27009} {\bibfield  {journal} {\bibinfo
  {journal} {Europhysics Letters}\ }\textbf {\bibinfo {volume} {88}},\ \bibinfo
  {pages} {27009} (\bibinfo {year} {2009})}\BibitemShut {NoStop}%
\bibitem [{\citenamefont {Nakano}\ and\ \citenamefont
  {Sakai}(2011)}]{nakano2011numerical}%
  \BibitemOpen
  \bibfield  {author} {\bibinfo {author} {\bibfnamefont {H.}~\bibnamefont
  {Nakano}}\ and\ \bibinfo {author} {\bibfnamefont {T.}~\bibnamefont {Sakai}},\
  }\bibfield  {title} {\bibinfo {title} {Numerical-diagonalization study of
  spin gap issue of the kagome lattice {H}eisenberg antiferromagnet},\ }\href
  {https://doi.org/10.1143/jpsj.80.053704} {\bibfield  {journal} {\bibinfo
  {journal} {Journal of the Physical Society of Japan}\ }\textbf {\bibinfo
  {volume} {80}},\ \bibinfo {pages} {053704} (\bibinfo {year}
  {2011})}\BibitemShut {NoStop}%
\bibitem [{\citenamefont {Seki}\ \emph {et~al.}(2020)\citenamefont {Seki},
  \citenamefont {Shirakawa},\ and\ \citenamefont {Yunoki}}]{seki2020symmetry}%
  \BibitemOpen
  \bibfield  {author} {\bibinfo {author} {\bibfnamefont {K.}~\bibnamefont
  {Seki}}, \bibinfo {author} {\bibfnamefont {T.}~\bibnamefont {Shirakawa}},\
  and\ \bibinfo {author} {\bibfnamefont {S.}~\bibnamefont {Yunoki}},\
  }\bibfield  {title} {\bibinfo {title} {Symmetry-adapted variational quantum
  eigensolver},\ }\href {https://doi.org/10.1103/PhysRevA.101.052340}
  {\bibfield  {journal} {\bibinfo  {journal} {Physical Review A}\ }\textbf
  {\bibinfo {volume} {101}},\ \bibinfo {pages} {052340} (\bibinfo {year}
  {2020})}\BibitemShut {NoStop}%
\bibitem [{\citenamefont {Ryabinkin}\ \emph {et~al.}(2018)\citenamefont
  {Ryabinkin}, \citenamefont {Genin},\ and\ \citenamefont
  {Izmaylov}}]{ryabinkin2018constrained}%
  \BibitemOpen
  \bibfield  {author} {\bibinfo {author} {\bibfnamefont {I.~G.}\ \bibnamefont
  {Ryabinkin}}, \bibinfo {author} {\bibfnamefont {S.~N.}\ \bibnamefont
  {Genin}},\ and\ \bibinfo {author} {\bibfnamefont {A.~F.}\ \bibnamefont
  {Izmaylov}},\ }\bibfield  {title} {\bibinfo {title} {Constrained variational
  quantum eigensolver: Quantum computer search engine in the {F}ock space},\
  }\href {https://doi.org/10.1021/acs.jctc.8b00943} {\bibfield  {journal}
  {\bibinfo  {journal} {Journal of Chemical Theory and Computation}\ }\textbf
  {\bibinfo {volume} {15}},\ \bibinfo {pages} {249--255} (\bibinfo {year}
  {2018})}\BibitemShut {NoStop}%
\bibitem [{\citenamefont {Kuroiwa}\ and\ \citenamefont
  {Nakagawa}(2021)}]{kuroiwa2021penalty}%
  \BibitemOpen
  \bibfield  {author} {\bibinfo {author} {\bibfnamefont {K.}~\bibnamefont
  {Kuroiwa}}\ and\ \bibinfo {author} {\bibfnamefont {Y.~O.}\ \bibnamefont
  {Nakagawa}},\ }\bibfield  {title} {\bibinfo {title} {Penalty methods for a
  variational quantum eigensolver},\ }\href
  {https://doi.org/10.1103/PhysRevResearch.3.013197} {\bibfield  {journal}
  {\bibinfo  {journal} {Physical Review Research}\ }\textbf {\bibinfo {volume}
  {3}},\ \bibinfo {pages} {013197} (\bibinfo {year} {2021})}\BibitemShut
  {NoStop}%
\bibitem [{\citenamefont {Beach}\ \emph {et~al.}(2019)\citenamefont {Beach},
  \citenamefont {Melko}, \citenamefont {Grover},\ and\ \citenamefont
  {Hsieh}}]{beach2019making}%
  \BibitemOpen
  \bibfield  {author} {\bibinfo {author} {\bibfnamefont {M.~J.~S.}\
  \bibnamefont {Beach}}, \bibinfo {author} {\bibfnamefont {R.~G.}\ \bibnamefont
  {Melko}}, \bibinfo {author} {\bibfnamefont {T.}~\bibnamefont {Grover}},\ and\
  \bibinfo {author} {\bibfnamefont {T.~H.}\ \bibnamefont {Hsieh}},\ }\bibfield
  {title} {\bibinfo {title} {Making trotters sprint: A variational imaginary
  time ansatz for quantum many-body systems},\ }\href
  {https://doi.org/10.1103/PhysRevB.100.094434} {\bibfield  {journal} {\bibinfo
   {journal} {Physical Review B}\ }\textbf {\bibinfo {volume} {100}},\ \bibinfo
  {pages} {094434} (\bibinfo {year} {2019})}\BibitemShut {NoStop}%
\bibitem [{\citenamefont {Bosse}\ and\ \citenamefont
  {Montanaro}(2021)}]{bosse2021probing}%
  \BibitemOpen
  \bibfield  {author} {\bibinfo {author} {\bibfnamefont {J.~L.}\ \bibnamefont
  {Bosse}}\ and\ \bibinfo {author} {\bibfnamefont {A.}~\bibnamefont
  {Montanaro}},\ }\href@noop {} {\bibinfo {title} {Probing ground state
  properties of the kagome antiferromagnetic {H}eisenberg model using the
  variational quantum eigensolver}} (\bibinfo {year} {2021}),\ \Eprint
  {https://arxiv.org/abs/2108.08086} {arXiv:2108.08086 [quant-ph]} \BibitemShut
  {NoStop}%
\bibitem [{\citenamefont {{Python Software Foundation}}(2019)}]{python}%
  \BibitemOpen
  \bibfield  {author} {\bibinfo {author} {\bibnamefont {{Python Software
  Foundation}}},\ }\href {https://docs.python.org/3.8/} {\bibinfo {title}
  {Python language reference, version 2.8}},\ \bibinfo {howpublished}
  {\url{https://docs.python.org/3.8/}} (\bibinfo {year} {2019})\BibitemShut
  {NoStop}%
\bibitem [{\citenamefont {Harris}\ \emph {et~al.}(2020)\citenamefont {Harris},
  \citenamefont {Millman}, \citenamefont {van~der Walt}, \citenamefont
  {Gommers}, \citenamefont {Virtanen}, \citenamefont {Cournapeau},
  \citenamefont {Wieser}, \citenamefont {Taylor}, \citenamefont {Berg},
  \citenamefont {Smith} \emph {et~al.}}]{harris2020array}%
  \BibitemOpen
  \bibfield  {author} {\bibinfo {author} {\bibfnamefont {C.~R.}\ \bibnamefont
  {Harris}}, \bibinfo {author} {\bibfnamefont {K.~J.}\ \bibnamefont {Millman}},
  \bibinfo {author} {\bibfnamefont {S.~J.}\ \bibnamefont {van~der Walt}},
  \bibinfo {author} {\bibfnamefont {R.}~\bibnamefont {Gommers}}, \bibinfo
  {author} {\bibfnamefont {P.}~\bibnamefont {Virtanen}}, \bibinfo {author}
  {\bibfnamefont {D.}~\bibnamefont {Cournapeau}}, \bibinfo {author}
  {\bibfnamefont {E.}~\bibnamefont {Wieser}}, \bibinfo {author} {\bibfnamefont
  {J.}~\bibnamefont {Taylor}}, \bibinfo {author} {\bibfnamefont
  {S.}~\bibnamefont {Berg}}, \bibinfo {author} {\bibfnamefont {N.~J.}\
  \bibnamefont {Smith}}, \emph {et~al.},\ }\bibfield  {title} {\bibinfo {title}
  {Array programming with {NumPy}},\ }\href
  {https://doi.org/10.1038/s41586-020-2649-2} {\bibfield  {journal} {\bibinfo
  {journal} {Nature}\ }\textbf {\bibinfo {volume} {585}},\ \bibinfo {pages}
  {357--362} (\bibinfo {year} {2020})}\BibitemShut {NoStop}%
\bibitem [{\citenamefont {Okuta}\ \emph {et~al.}(2017)\citenamefont {Okuta},
  \citenamefont {Unno}, \citenamefont {Nishino}, \citenamefont {Hido},\ and\
  \citenamefont {Loomis}}]{nishino2017cupy}%
  \BibitemOpen
  \bibfield  {author} {\bibinfo {author} {\bibfnamefont {R.}~\bibnamefont
  {Okuta}}, \bibinfo {author} {\bibfnamefont {Y.}~\bibnamefont {Unno}},
  \bibinfo {author} {\bibfnamefont {D.}~\bibnamefont {Nishino}}, \bibinfo
  {author} {\bibfnamefont {S.}~\bibnamefont {Hido}},\ and\ \bibinfo {author}
  {\bibfnamefont {C.}~\bibnamefont {Loomis}},\ }\bibfield  {title} {\bibinfo
  {title} {{CuPy: A NumPy-Compatible Library for NVIDIA GPU Calculations}},\
  }in\ \href {http://learningsys.org/nips17/assets/papers/paper_16.pdf} {\emph
  {\bibinfo {booktitle} {Proceedings of Workshop on Machine Learning Systems
  (LearningSys) in The Thirty-first Annual Conference on Neural Information
  Processing Systems (NIPS)}}}\ (\bibinfo {year} {2017})\BibitemShut {NoStop}%
\bibitem [{\citenamefont {Nielsen}(2015)}]{nielsen2015neural}%
  \BibitemOpen
  \bibfield  {author} {\bibinfo {author} {\bibfnamefont {M.~A.}\ \bibnamefont
  {Nielsen}},\ }\href {http://neuralnetworksanddeeplearning.com/} {\emph
  {\bibinfo {title} {Neural networks and deep learning}}}\ (\bibinfo
  {publisher} {Determination press},\ \bibinfo {year} {2015})\BibitemShut
  {NoStop}%
\bibitem [{\citenamefont {Tokui}\ \emph {et~al.}(2015)\citenamefont {Tokui},
  \citenamefont {Oono}, \citenamefont {Hido},\ and\ \citenamefont
  {Clayton}}]{tokui2015chainer}%
  \BibitemOpen
  \bibfield  {author} {\bibinfo {author} {\bibfnamefont {S.}~\bibnamefont
  {Tokui}}, \bibinfo {author} {\bibfnamefont {K.}~\bibnamefont {Oono}},
  \bibinfo {author} {\bibfnamefont {S.}~\bibnamefont {Hido}},\ and\ \bibinfo
  {author} {\bibfnamefont {J.}~\bibnamefont {Clayton}},\ }\bibfield  {title}
  {\bibinfo {title} {{Chainer: a Next-Generation Open Source Framework for Deep
  Learning}},\ }in\ \href
  {http://learningsys.org/papers/LearningSys_2015_paper_33.pdf} {\emph
  {\bibinfo {booktitle} {Proceedings of Workshop on Machine Learning Systems
  (LearningSys) in The Twenty-ninth Annual Conference on Neural Information
  Processing Systems (NIPS)}}}\ (\bibinfo {year} {2015})\BibitemShut {NoStop}%
\bibitem [{\citenamefont {Virtanen}\ \emph {et~al.}(2020)\citenamefont
  {Virtanen}, \citenamefont {Gommers}, \citenamefont {Oliphant}, \citenamefont
  {Haberland}, \citenamefont {Reddy}, \citenamefont {Cournapeau}, \citenamefont
  {Burovski}, \citenamefont {Peterson}, \citenamefont {Weckesser},
  \citenamefont {Bright} \emph {et~al.}}]{virtanen2020scipy}%
  \BibitemOpen
  \bibfield  {author} {\bibinfo {author} {\bibfnamefont {P.}~\bibnamefont
  {Virtanen}}, \bibinfo {author} {\bibfnamefont {R.}~\bibnamefont {Gommers}},
  \bibinfo {author} {\bibfnamefont {T.~E.}\ \bibnamefont {Oliphant}}, \bibinfo
  {author} {\bibfnamefont {M.}~\bibnamefont {Haberland}}, \bibinfo {author}
  {\bibfnamefont {T.}~\bibnamefont {Reddy}}, \bibinfo {author} {\bibfnamefont
  {D.}~\bibnamefont {Cournapeau}}, \bibinfo {author} {\bibfnamefont
  {E.}~\bibnamefont {Burovski}}, \bibinfo {author} {\bibfnamefont
  {P.}~\bibnamefont {Peterson}}, \bibinfo {author} {\bibfnamefont
  {W.}~\bibnamefont {Weckesser}}, \bibinfo {author} {\bibfnamefont
  {J.}~\bibnamefont {Bright}}, \emph {et~al.},\ }\bibfield  {title} {\bibinfo
  {title} {{{SciPy} 1.0: Fundamental Algorithms for Scientific Computing in
  Python}},\ }\href {https://doi.org/10.1038/s41592-019-0686-2} {\bibfield
  {journal} {\bibinfo  {journal} {Nature Methods}\ }\textbf {\bibinfo {volume}
  {17}},\ \bibinfo {pages} {261--272} (\bibinfo {year} {2020})}\BibitemShut
  {NoStop}%
\bibitem [{\citenamefont {Wei}\ \emph {et~al.}(2020)\citenamefont {Wei},
  \citenamefont {Li},\ and\ \citenamefont {Long}}]{wei2020full}%
  \BibitemOpen
  \bibfield  {author} {\bibinfo {author} {\bibfnamefont {S.}~\bibnamefont
  {Wei}}, \bibinfo {author} {\bibfnamefont {H.}~\bibnamefont {Li}},\ and\
  \bibinfo {author} {\bibfnamefont {G.-L.}\ \bibnamefont {Long}},\ }\bibfield
  {title} {\bibinfo {title} {A full quantum eigensolver for quantum chemistry
  simulations},\ }\href {https://doi.org/10.34133/2020/1486935} {\bibfield
  {journal} {\bibinfo  {journal} {Research}\ }\textbf {\bibinfo {volume}
  {2020}},\ \bibinfo {pages} {1--11} (\bibinfo {year} {2020})}\BibitemShut
  {NoStop}%
\bibitem [{\citenamefont {Long}(2006)}]{long2006general}%
  \BibitemOpen
  \bibfield  {author} {\bibinfo {author} {\bibfnamefont {G.-L.}\ \bibnamefont
  {Long}},\ }\bibfield  {title} {\bibinfo {title} {General quantum interference
  principle and duality computer},\ }\href
  {https://doi.org/10.1088/0253-6102/45/5/013} {\bibfield  {journal} {\bibinfo
  {journal} {Communications in Theoretical Physics}\ }\textbf {\bibinfo
  {volume} {45}},\ \bibinfo {pages} {825} (\bibinfo {year} {2006})}\BibitemShut
  {NoStop}%
\bibitem [{\citenamefont {Childs}\ and\ \citenamefont
  {Wiebe}(2012)}]{childs2012hamiltonian}%
  \BibitemOpen
  \bibfield  {author} {\bibinfo {author} {\bibfnamefont {A.~M.}\ \bibnamefont
  {Childs}}\ and\ \bibinfo {author} {\bibfnamefont {N.}~\bibnamefont {Wiebe}},\
  }\bibfield  {title} {\bibinfo {title} {Hamiltonian simulation using linear
  combinations of unitary operations},\ }\href
  {https://doi.org/10.26421/qic12.11-12-1} {\bibfield  {journal} {\bibinfo
  {journal} {Quantum Information and Computation}\ }\textbf {\bibinfo {volume}
  {12}},\ \bibinfo {pages} {901--924} (\bibinfo {year} {2012})}\BibitemShut
  {NoStop}%
\bibitem [{\citenamefont {Lehoucq}\ \emph {et~al.}(1998)\citenamefont
  {Lehoucq}, \citenamefont {Sorensen},\ and\ \citenamefont
  {Yang}}]{lehoucq1998arpack}%
  \BibitemOpen
  \bibfield  {author} {\bibinfo {author} {\bibfnamefont {R.~B.}\ \bibnamefont
  {Lehoucq}}, \bibinfo {author} {\bibfnamefont {D.~C.}\ \bibnamefont
  {Sorensen}},\ and\ \bibinfo {author} {\bibfnamefont {C.}~\bibnamefont
  {Yang}},\ }\href {https://doi.org/10.1137/1.9780898719628} {\emph {\bibinfo
  {title} {ARPACK Users' Guide}}}\ (\bibinfo  {publisher} {Society for
  Industrial and Applied Mathematics},\ \bibinfo {year} {1998})\BibitemShut
  {NoStop}%
\bibitem [{\citenamefont {Efron}\ and\ \citenamefont
  {Tibshirani}(1994)}]{efron1994introduction}%
  \BibitemOpen
  \bibfield  {author} {\bibinfo {author} {\bibfnamefont {B.}~\bibnamefont
  {Efron}}\ and\ \bibinfo {author} {\bibfnamefont {R.}~\bibnamefont
  {Tibshirani}},\ }\href {https://doi.org/10.1201/9780429246593} {\emph
  {\bibinfo {title} {An Introduction to the Bootstrap}}}\ (\bibinfo
  {publisher} {Chapman and Hall/{CRC}},\ \bibinfo {year} {1994})\BibitemShut
  {NoStop}%
\bibitem [{\citenamefont {Latorre}\ \emph {et~al.}(2004)\citenamefont
  {Latorre}, \citenamefont {Rico},\ and\ \citenamefont
  {Vidal}}]{latorre2004ground}%
  \BibitemOpen
  \bibfield  {author} {\bibinfo {author} {\bibfnamefont {J.}~\bibnamefont
  {Latorre}}, \bibinfo {author} {\bibfnamefont {E.}~\bibnamefont {Rico}},\ and\
  \bibinfo {author} {\bibfnamefont {G.}~\bibnamefont {Vidal}},\ }\bibfield
  {title} {\bibinfo {title} {Ground state entanglement in quantum spin
  chains},\ }\href {https://doi.org/10.26421/qic4.1-4} {\bibfield  {journal}
  {\bibinfo  {journal} {Quantum Information and Computation}\ }\textbf
  {\bibinfo {volume} {4}},\ \bibinfo {pages} {48--92} (\bibinfo {year}
  {2004})}\BibitemShut {NoStop}%
\bibitem [{Note2()}]{Note2}%
  \BibitemOpen
  \bibinfo {note} {Let us focus on $\protect \mathrm {KVQE}_{G}$, and lay out a
  coordinate system over the 24 qubits used in Fig.~\ref {fig:kagome_ansatz}
  (top right). We put the origin (0,0) at the bottom left qubit, the qubit
  directly above at (0,1), and the qubit directly to the right of the origin at
  (1,0). At $p=2$, there is no qubit whose past light cone in $C$ covers the
  entire system. At $p=3$, there are qubits, such as the bottom right (4,0),
  bottom left (0,0), top left (0,4) and middle (2,2) qubits, whose past light
  cone in $C$ covers the entire system. There are, however, still some qubits
  for which this is not the case, such as the qubits at (3,4), (4,3) and the
  top middle (2,0). At $p=4$, the past light cone of the latter qubits covers
  the entire system, except for the qubit at (3,4). It is only after $p=5$
  cycles that its past light cone covers the entire system.}\BibitemShut
  {Stop}%
\bibitem [{\citenamefont {Nielsen}\ and\ \citenamefont
  {Chuang}(2010)}]{nielsen2010quantum}%
  \BibitemOpen
  \bibfield  {author} {\bibinfo {author} {\bibfnamefont {M.~A.}\ \bibnamefont
  {Nielsen}}\ and\ \bibinfo {author} {\bibfnamefont {I.~L.}\ \bibnamefont
  {Chuang}},\ }\href {https://doi.org/10.1017/CBO9780511976667} {\emph
  {\bibinfo {title} {Quantum Computation and Quantum Information: 10th
  Anniversary Edition}}}\ (\bibinfo  {publisher} {Cambridge University Press},\
  \bibinfo {year} {2010})\BibitemShut {NoStop}%
\bibitem [{\citenamefont {Emerson}\ \emph {et~al.}(2005)\citenamefont
  {Emerson}, \citenamefont {Alicki},\ and\ \citenamefont
  {{\.{Z}}yczkowski}}]{emerson2005scalable}%
  \BibitemOpen
  \bibfield  {author} {\bibinfo {author} {\bibfnamefont {J.}~\bibnamefont
  {Emerson}}, \bibinfo {author} {\bibfnamefont {R.}~\bibnamefont {Alicki}},\
  and\ \bibinfo {author} {\bibfnamefont {K.}~\bibnamefont {{\.{Z}}yczkowski}},\
  }\bibfield  {title} {\bibinfo {title} {Scalable noise estimation with random
  unitary operators},\ }\href {https://doi.org/10.1088/1464-4266/7/10/021}
  {\bibfield  {journal} {\bibinfo  {journal} {Journal of Optics B: Quantum and
  Semiclassical Optics}\ }\textbf {\bibinfo {volume} {7}},\ \bibinfo {pages}
  {S347--S352} (\bibinfo {year} {2005})}\BibitemShut {NoStop}%
\bibitem [{\citenamefont {{Kattem{\"o}lle}}\ and\ \citenamefont
  {{Burkard}}(2022)}]{kattemolle2022effects}%
  \BibitemOpen
  \bibfield  {author} {\bibinfo {author} {\bibfnamefont {J.}~\bibnamefont
  {{Kattem{\"o}lle}}}\ and\ \bibinfo {author} {\bibfnamefont {G.}~\bibnamefont
  {{Burkard}}},\ }\href@noop {} {\bibinfo {title} {{Effects of correlated
  errors on the Quantum Approximate Optimization Algorithm}}} (\bibinfo {year}
  {2022}),\ \Eprint {https://arxiv.org/abs/2207.10622} {arXiv:2207.10622
  [quant-ph]} \BibitemShut {NoStop}%
\bibitem [{\citenamefont {Fontana}\ \emph {et~al.}(2021)\citenamefont
  {Fontana}, \citenamefont {Fitzpatrick}, \citenamefont {Ramo}, \citenamefont
  {Duncan},\ and\ \citenamefont {Rungger}}]{fontana2021evaluating}%
  \BibitemOpen
  \bibfield  {author} {\bibinfo {author} {\bibfnamefont {E.}~\bibnamefont
  {Fontana}}, \bibinfo {author} {\bibfnamefont {N.}~\bibnamefont
  {Fitzpatrick}}, \bibinfo {author} {\bibfnamefont {D.~M.~n.}\ \bibnamefont
  {Ramo}}, \bibinfo {author} {\bibfnamefont {R.}~\bibnamefont {Duncan}},\ and\
  \bibinfo {author} {\bibfnamefont {I.}~\bibnamefont {Rungger}},\ }\bibfield
  {title} {\bibinfo {title} {Evaluating the noise resilience of variational
  quantum algorithms},\ }\href {https://doi.org/10.1103/PhysRevA.104.022403}
  {\bibfield  {journal} {\bibinfo  {journal} {Physical Review A}\ }\textbf
  {\bibinfo {volume} {104}},\ \bibinfo {pages} {022403} (\bibinfo {year}
  {2021})}\BibitemShut {NoStop}%
\bibitem [{\citenamefont {Childs}\ \emph {et~al.}(2021)\citenamefont {Childs},
  \citenamefont {Su}, \citenamefont {Tran}, \citenamefont {Wiebe},\ and\
  \citenamefont {Zhu}}]{childs2021theory}%
  \BibitemOpen
  \bibfield  {author} {\bibinfo {author} {\bibfnamefont {A.~M.}\ \bibnamefont
  {Childs}}, \bibinfo {author} {\bibfnamefont {Y.}~\bibnamefont {Su}}, \bibinfo
  {author} {\bibfnamefont {M.~C.}\ \bibnamefont {Tran}}, \bibinfo {author}
  {\bibfnamefont {N.}~\bibnamefont {Wiebe}},\ and\ \bibinfo {author}
  {\bibfnamefont {S.}~\bibnamefont {Zhu}},\ }\bibfield  {title} {\bibinfo
  {title} {Theory of trotter error with commutator scaling},\ }\href
  {https://doi.org/10.1103/PhysRevX.11.011020} {\bibfield  {journal} {\bibinfo
  {journal} {Physical Review X}\ }\textbf {\bibinfo {volume} {11}},\ \bibinfo
  {pages} {011020} (\bibinfo {year} {2021})}\BibitemShut {NoStop}%
\bibitem [{\citenamefont {Ballance}\ \emph {et~al.}(2016)\citenamefont
  {Ballance}, \citenamefont {Harty}, \citenamefont {Linke}, \citenamefont
  {Sepiol},\ and\ \citenamefont {Lucas}}]{ballance2016high}%
  \BibitemOpen
  \bibfield  {author} {\bibinfo {author} {\bibfnamefont {C.~J.}\ \bibnamefont
  {Ballance}}, \bibinfo {author} {\bibfnamefont {T.~P.}\ \bibnamefont {Harty}},
  \bibinfo {author} {\bibfnamefont {N.~M.}\ \bibnamefont {Linke}}, \bibinfo
  {author} {\bibfnamefont {M.~A.}\ \bibnamefont {Sepiol}},\ and\ \bibinfo
  {author} {\bibfnamefont {D.~M.}\ \bibnamefont {Lucas}},\ }\bibfield  {title}
  {\bibinfo {title} {High-fidelity quantum logic gates using trapped-ion
  hyperfine qubits},\ }\href {https://doi.org/10.1103/PhysRevLett.117.060504}
  {\bibfield  {journal} {\bibinfo  {journal} {Physical Review Letters}\
  }\textbf {\bibinfo {volume} {117}},\ \bibinfo {pages} {060504} (\bibinfo
  {year} {2016})}\BibitemShut {NoStop}%
\bibitem [{\citenamefont {Kjaergaard}\ \emph {et~al.}(2020)\citenamefont
  {Kjaergaard}, \citenamefont {Schwartz}, \citenamefont {Braum\"{u}ller},
  \citenamefont {Krantz}, \citenamefont {Wang}, \citenamefont {Gustavsson},\
  and\ \citenamefont {Oliver}}]{kjaergaard2020superconducting}%
  \BibitemOpen
  \bibfield  {author} {\bibinfo {author} {\bibfnamefont {M.}~\bibnamefont
  {Kjaergaard}}, \bibinfo {author} {\bibfnamefont {M.~E.}\ \bibnamefont
  {Schwartz}}, \bibinfo {author} {\bibfnamefont {J.}~\bibnamefont
  {Braum\"{u}ller}}, \bibinfo {author} {\bibfnamefont {P.}~\bibnamefont
  {Krantz}}, \bibinfo {author} {\bibfnamefont {J.~I.-J.}\ \bibnamefont {Wang}},
  \bibinfo {author} {\bibfnamefont {S.}~\bibnamefont {Gustavsson}},\ and\
  \bibinfo {author} {\bibfnamefont {W.~D.}\ \bibnamefont {Oliver}},\ }\bibfield
   {title} {\bibinfo {title} {Superconducting qubits: Current state of play},\
  }\href {https://doi.org/10.1146/annurev-conmatphys-031119-050605} {\bibfield
  {journal} {\bibinfo  {journal} {Annual Review of Condensed Matter Physics}\
  }\textbf {\bibinfo {volume} {11}},\ \bibinfo {pages} {369--395} (\bibinfo
  {year} {2020})}\BibitemShut {NoStop}%
\bibitem [{Note3()}]{Note3}%
  \BibitemOpen
  \bibinfo {note} {The gate infidelities typically reported in the literature
  do not correspond directly to the error rates as defined in Eqs. (\ref
  {eq:depol}) and (\ref {eq:bitflip}), but for the purpose of the
  order-of-magnitude estimates here these discrepancies are
  insignificant.}\BibitemShut {Stop}%
\bibitem [{\citenamefont {Li}\ \emph {et~al.}(2018)\citenamefont {Li},
  \citenamefont {Petit}, \citenamefont {Franke}, \citenamefont {Dehollain},
  \citenamefont {Helsen}, \citenamefont {Steudtner}, \citenamefont {Thomas},
  \citenamefont {Yoscovits}, \citenamefont {Singh}, \citenamefont {Wehner}
  \emph {et~al.}}]{li2018crossbar}%
  \BibitemOpen
  \bibfield  {author} {\bibinfo {author} {\bibfnamefont {R.}~\bibnamefont
  {Li}}, \bibinfo {author} {\bibfnamefont {L.}~\bibnamefont {Petit}}, \bibinfo
  {author} {\bibfnamefont {D.~P.}\ \bibnamefont {Franke}}, \bibinfo {author}
  {\bibfnamefont {J.~P.}\ \bibnamefont {Dehollain}}, \bibinfo {author}
  {\bibfnamefont {J.}~\bibnamefont {Helsen}}, \bibinfo {author} {\bibfnamefont
  {M.}~\bibnamefont {Steudtner}}, \bibinfo {author} {\bibfnamefont {N.~K.}\
  \bibnamefont {Thomas}}, \bibinfo {author} {\bibfnamefont {Z.~R.}\
  \bibnamefont {Yoscovits}}, \bibinfo {author} {\bibfnamefont {K.~J.}\
  \bibnamefont {Singh}}, \bibinfo {author} {\bibfnamefont {S.}~\bibnamefont
  {Wehner}}, \emph {et~al.},\ }\bibfield  {title} {\bibinfo {title} {A crossbar
  network for silicon quantum dot qubits},\ }\href
  {https://doi.org/10.1126/sciadv.aar3960} {\bibfield  {journal} {\bibinfo
  {journal} {Science advances}\ }\textbf {\bibinfo {volume} {4}},\ \bibinfo
  {pages} {eaar3960} (\bibinfo {year} {2018})}\BibitemShut {NoStop}%
\bibitem [{\citenamefont {Harrigan}\ \emph {et~al.}(2021)\citenamefont
  {Harrigan}, \citenamefont {Sung}, \citenamefont {Neeley}, \citenamefont
  {Satzinger}, \citenamefont {Arute}, \citenamefont {Arya}, \citenamefont
  {Atalaya}, \citenamefont {Bardin}, \citenamefont {Barends}, \citenamefont
  {Boixo} \emph {et~al.}}]{harrigan2021quantum}%
  \BibitemOpen
  \bibfield  {author} {\bibinfo {author} {\bibfnamefont {M.~P.}\ \bibnamefont
  {Harrigan}}, \bibinfo {author} {\bibfnamefont {K.~J.}\ \bibnamefont {Sung}},
  \bibinfo {author} {\bibfnamefont {M.}~\bibnamefont {Neeley}}, \bibinfo
  {author} {\bibfnamefont {K.~J.}\ \bibnamefont {Satzinger}}, \bibinfo {author}
  {\bibfnamefont {F.}~\bibnamefont {Arute}}, \bibinfo {author} {\bibfnamefont
  {K.}~\bibnamefont {Arya}}, \bibinfo {author} {\bibfnamefont {J.}~\bibnamefont
  {Atalaya}}, \bibinfo {author} {\bibfnamefont {J.~C.}\ \bibnamefont {Bardin}},
  \bibinfo {author} {\bibfnamefont {R.}~\bibnamefont {Barends}}, \bibinfo
  {author} {\bibfnamefont {S.}~\bibnamefont {Boixo}}, \emph {et~al.},\
  }\bibfield  {title} {\bibinfo {title} {Quantum approximate optimization of
  non-planar graph problems on a planar superconducting processor},\ }\href
  {https://doi.org/10.1038/s41567-020-01105-y} {\bibfield  {journal} {\bibinfo
  {journal} {Nature Physics}\ }\textbf {\bibinfo {volume} {17}},\ \bibinfo
  {pages} {332--336} (\bibinfo {year} {2021})}\BibitemShut {NoStop}%
\bibitem [{\citenamefont {Depenbrock}\ \emph {et~al.}(2012)\citenamefont
  {Depenbrock}, \citenamefont {McCulloch},\ and\ \citenamefont
  {Schollw\"ock}}]{depenbrock2012nature}%
  \BibitemOpen
  \bibfield  {author} {\bibinfo {author} {\bibfnamefont {S.}~\bibnamefont
  {Depenbrock}}, \bibinfo {author} {\bibfnamefont {I.~P.}\ \bibnamefont
  {McCulloch}},\ and\ \bibinfo {author} {\bibfnamefont {U.}~\bibnamefont
  {Schollw\"ock}},\ }\bibfield  {title} {\bibinfo {title} {Nature of the
  spin-liquid ground state of the {$S=1/2$} {H}eisenberg model on the kagome
  lattice},\ }\href {https://doi.org/10.1103/PhysRevLett.109.067201} {\bibfield
   {journal} {\bibinfo  {journal} {Physical Review Letters}\ }\textbf {\bibinfo
  {volume} {109}},\ \bibinfo {pages} {067201} (\bibinfo {year}
  {2012})}\BibitemShut {NoStop}%
\bibitem [{\citenamefont {He}\ \emph {et~al.}(2017)\citenamefont {He},
  \citenamefont {Zaletel}, \citenamefont {Oshikawa},\ and\ \citenamefont
  {Pollmann}}]{he2017signatures}%
  \BibitemOpen
  \bibfield  {author} {\bibinfo {author} {\bibfnamefont {Y.-C.}\ \bibnamefont
  {He}}, \bibinfo {author} {\bibfnamefont {M.~P.}\ \bibnamefont {Zaletel}},
  \bibinfo {author} {\bibfnamefont {M.}~\bibnamefont {Oshikawa}},\ and\
  \bibinfo {author} {\bibfnamefont {F.}~\bibnamefont {Pollmann}},\ }\bibfield
  {title} {\bibinfo {title} {Signatures of {D}irac cones in a {DMRG} study of
  the kagome {H}eisenberg model},\ }\href
  {https://doi.org/10.1103/PhysRevX.7.031020} {\bibfield  {journal} {\bibinfo
  {journal} {Physical Review X}\ }\textbf {\bibinfo {volume} {7}},\ \bibinfo
  {pages} {031020} (\bibinfo {year} {2017})}\BibitemShut {NoStop}%
\bibitem [{\citenamefont {{Verdon}}\ \emph {et~al.}(2019)\citenamefont
  {{Verdon}}, \citenamefont {{Marks}}, \citenamefont {{Nanda}}, \citenamefont
  {{Leichenauer}},\ and\ \citenamefont {{Hidary}}}]{verdon2019quantum}%
  \BibitemOpen
  \bibfield  {author} {\bibinfo {author} {\bibfnamefont {G.}~\bibnamefont
  {{Verdon}}}, \bibinfo {author} {\bibfnamefont {J.}~\bibnamefont {{Marks}}},
  \bibinfo {author} {\bibfnamefont {S.}~\bibnamefont {{Nanda}}}, \bibinfo
  {author} {\bibfnamefont {S.}~\bibnamefont {{Leichenauer}}},\ and\ \bibinfo
  {author} {\bibfnamefont {J.}~\bibnamefont {{Hidary}}},\ }\href@noop {}
  {\bibinfo {title} {{Quantum Hamiltonian-Based Models and the Variational
  Quantum Thermalizer Algorithm}}} (\bibinfo {year} {2019}),\ \Eprint
  {https://arxiv.org/abs/1910.02071} {arXiv:1910.02071 [quant-ph]} \BibitemShut
  {NoStop}%
\bibitem [{\citenamefont {{Foldager}}\ \emph {et~al.}(2022)\citenamefont
  {{Foldager}}, \citenamefont {{Pesah}},\ and\ \citenamefont
  {{Hansen}}}]{foldager2022noise}%
  \BibitemOpen
  \bibfield  {author} {\bibinfo {author} {\bibfnamefont {J.}~\bibnamefont
  {{Foldager}}}, \bibinfo {author} {\bibfnamefont {A.}~\bibnamefont
  {{Pesah}}},\ and\ \bibinfo {author} {\bibfnamefont {L.~K.}\ \bibnamefont
  {{Hansen}}},\ }\bibfield  {title} {\bibinfo {title} {{Noise-assisted
  variational quantum thermalization}},\ }\href
  {https://doi.org/10.1038/s41598-022-07296-z} {\bibfield  {journal} {\bibinfo
  {journal} {Scientific Reports}\ }\textbf {\bibinfo {volume} {12}},\ \bibinfo
  {eid} {3862} (\bibinfo {year} {2022})}\BibitemShut {NoStop}%
\bibitem [{\citenamefont {McClean}\ \emph {et~al.}(2017)\citenamefont
  {McClean}, \citenamefont {Kimchi-Schwartz}, \citenamefont {Carter},\ and\
  \citenamefont {de~Jong}}]{mcclean2017hybrid}%
  \BibitemOpen
  \bibfield  {author} {\bibinfo {author} {\bibfnamefont {J.~R.}\ \bibnamefont
  {McClean}}, \bibinfo {author} {\bibfnamefont {M.~E.}\ \bibnamefont
  {Kimchi-Schwartz}}, \bibinfo {author} {\bibfnamefont {J.}~\bibnamefont
  {Carter}},\ and\ \bibinfo {author} {\bibfnamefont {W.~A.}\ \bibnamefont
  {de~Jong}},\ }\bibfield  {title} {\bibinfo {title} {Hybrid quantum-classical
  hierarchy for mitigation of decoherence and determination of excited
  states},\ }\href {https://doi.org/10.1103/PhysRevA.95.042308} {\bibfield
  {journal} {\bibinfo  {journal} {Physical Review A}\ }\textbf {\bibinfo
  {volume} {95}},\ \bibinfo {pages} {042308} (\bibinfo {year}
  {2017})}\BibitemShut {NoStop}%
\bibitem [{\citenamefont {Bravyi}\ \emph {et~al.}(2016)\citenamefont {Bravyi},
  \citenamefont {Smith},\ and\ \citenamefont {Smolin}}]{bravyi2016trading}%
  \BibitemOpen
  \bibfield  {author} {\bibinfo {author} {\bibfnamefont {S.}~\bibnamefont
  {Bravyi}}, \bibinfo {author} {\bibfnamefont {G.}~\bibnamefont {Smith}},\ and\
  \bibinfo {author} {\bibfnamefont {J.~A.}\ \bibnamefont {Smolin}},\ }\bibfield
   {title} {\bibinfo {title} {Trading classical and quantum computational
  resources},\ }\href {https://doi.org/10.1103/PhysRevX.6.021043} {\bibfield
  {journal} {\bibinfo  {journal} {Physical Review X}\ }\textbf {\bibinfo
  {volume} {6}},\ \bibinfo {pages} {021043} (\bibinfo {year}
  {2016})}\BibitemShut {NoStop}%
\bibitem [{\citenamefont {Peng}\ \emph {et~al.}(2020)\citenamefont {Peng},
  \citenamefont {Harrow}, \citenamefont {Ozols},\ and\ \citenamefont
  {Wu}}]{peng2020simulating}%
  \BibitemOpen
  \bibfield  {author} {\bibinfo {author} {\bibfnamefont {T.}~\bibnamefont
  {Peng}}, \bibinfo {author} {\bibfnamefont {A.~W.}\ \bibnamefont {Harrow}},
  \bibinfo {author} {\bibfnamefont {M.}~\bibnamefont {Ozols}},\ and\ \bibinfo
  {author} {\bibfnamefont {X.}~\bibnamefont {Wu}},\ }\bibfield  {title}
  {\bibinfo {title} {Simulating large quantum circuits on a small quantum
  computer},\ }\href {https://doi.org/10.1103/PhysRevLett.125.150504}
  {\bibfield  {journal} {\bibinfo  {journal} {Physical Review Letters}\
  }\textbf {\bibinfo {volume} {125}},\ \bibinfo {pages} {150504} (\bibinfo
  {year} {2020})}\BibitemShut {NoStop}%
\bibitem [{\citenamefont {Sun}\ \emph {et~al.}(2022)\citenamefont {Sun},
  \citenamefont {Endo}, \citenamefont {Lin}, \citenamefont {Hayden},
  \citenamefont {Vedral},\ and\ \citenamefont {Yuan}}]{sun2022perturbative}%
  \BibitemOpen
  \bibfield  {author} {\bibinfo {author} {\bibfnamefont {J.}~\bibnamefont
  {Sun}}, \bibinfo {author} {\bibfnamefont {S.}~\bibnamefont {Endo}}, \bibinfo
  {author} {\bibfnamefont {H.}~\bibnamefont {Lin}}, \bibinfo {author}
  {\bibfnamefont {P.}~\bibnamefont {Hayden}}, \bibinfo {author} {\bibfnamefont
  {V.}~\bibnamefont {Vedral}},\ and\ \bibinfo {author} {\bibfnamefont
  {X.}~\bibnamefont {Yuan}},\ }\bibfield  {title} {\bibinfo {title}
  {Perturbative quantum simulation},\ }\href
  {https://doi.org/10.1103/PhysRevLett.129.120505} {\bibfield  {journal}
  {\bibinfo  {journal} {Physical Review Letters}\ }\textbf {\bibinfo {volume}
  {129}},\ \bibinfo {pages} {120505} (\bibinfo {year} {2022})}\BibitemShut
  {NoStop}%
\bibitem [{\citenamefont {Sakurai}\ and\ \citenamefont
  {Napolitano}(2020)}]{sakurai1995modern}%
  \BibitemOpen
  \bibfield  {author} {\bibinfo {author} {\bibfnamefont {J.~J.}\ \bibnamefont
  {Sakurai}}\ and\ \bibinfo {author} {\bibfnamefont {J.}~\bibnamefont
  {Napolitano}},\ }\href {https://doi.org/10.1017/9781108587280} {\emph
  {\bibinfo {title} {Modern Quantum Mechanics}}}\ (\bibinfo  {publisher}
  {Cambridge University Press},\ \bibinfo {year} {2020})\BibitemShut {NoStop}%
\bibitem [{\citenamefont {Lieb}\ and\ \citenamefont
  {Mattis}(1962)}]{lieb1962ordering}%
  \BibitemOpen
  \bibfield  {author} {\bibinfo {author} {\bibfnamefont {E.}~\bibnamefont
  {Lieb}}\ and\ \bibinfo {author} {\bibfnamefont {D.}~\bibnamefont {Mattis}},\
  }\bibfield  {title} {\bibinfo {title} {Ordering energy levels of interacting
  spin systems},\ }\href {https://doi.org/10.1063/1.1724276} {\bibfield
  {journal} {\bibinfo  {journal} {Journal of Mathematical Physics}\ }\textbf
  {\bibinfo {volume} {3}},\ \bibinfo {pages} {749--751} (\bibinfo {year}
  {1962})}\BibitemShut {NoStop}%
\bibitem [{\citenamefont {Kempe}\ \emph {et~al.}(2001)\citenamefont {Kempe},
  \citenamefont {Bacon}, \citenamefont {Lidar},\ and\ \citenamefont
  {Whaley}}]{kempe2001theory}%
  \BibitemOpen
  \bibfield  {author} {\bibinfo {author} {\bibfnamefont {J.}~\bibnamefont
  {Kempe}}, \bibinfo {author} {\bibfnamefont {D.}~\bibnamefont {Bacon}},
  \bibinfo {author} {\bibfnamefont {D.~A.}\ \bibnamefont {Lidar}},\ and\
  \bibinfo {author} {\bibfnamefont {K.~B.}\ \bibnamefont {Whaley}},\ }\bibfield
   {title} {\bibinfo {title} {Theory of decoherence-free fault-tolerant
  universal quantum computation},\ }\href
  {https://doi.org/10.1103/PhysRevA.63.042307} {\bibfield  {journal} {\bibinfo
  {journal} {Physical Review A}\ }\textbf {\bibinfo {volume} {63}},\ \bibinfo
  {pages} {042307} (\bibinfo {year} {2001})}\BibitemShut {NoStop}%
\end{thebibliography}%

\appendix

\section{Symmetry, the spin gap, and decoherence-free subspaces}\label{sec:symmetry}
In this Appendix, we give some background on symmetry and the total spin operators, and add detail to the discussion in~\sec{sym}.

The total spin operator is defined by $\bS^2\equiv\bS^{(\mrm{tot})}\cdot\bS^{(\mrm{tot})}$, with $\bS^{(\mrm{tot})}=\sum_{i=1}^n\mathbf{S}^{(i)}$ [cf. \eq{HAFM}], and is related to the total spin quantum number $S$ by $\mathbf{S}^2\ket{\psi}=S(S+1)\ket{\psi}$ for eigenstates $\ket \psi$ of $\mathbf S^2$. The total magnetization operator in the $i$-direction is defined as $\bS_i^\mrm{(tot)}=\sum_{j=1}^n \left[\bS^{(j)}\right]_i$, and is related to the magnetization quantum number $S_z$ by $\bS_z^\mrm{(tot)}\ket\psi=S_z\ket{\psi}$ for eigenstates $\ket{\psi}$ of $\bS_z$. The operators $\bS^2$ and $\bS_z^\mrm{(tot)}$ commute, and are hence simultaneously diagonalizable. Acting with the ladder operator $\bS_{\pm}^\mrm{(tot)}=\bS_x^{\mrm{(tot)}}\pm \mrm i\, \bS_y^{\mrm{(tot)}}$ on a state with quantum number $S_z$ raises (lowers) the $S_z$ quantum number of that state with unity, given that the new value of $S_z$ lays between (or is equal to) $-S$ and $S$. It annihilates that state otherwise. The Hamiltonian $H$, the ansatz circuits $C$, and initial states $\psiinit\!\bra{\psi_\mrm{init}}$ in this paper commute with $\bS_x^{\mrm{(tot)}}$, $\bS_y^{\mrm{(tot)}}$, and $\bS_z^{\mrm{(tot)}}$ (and hence with $\bS^2$ and $\bS_{\pm}^{\mrm{(tot)}}$), and therefore $H$, $C$, and $\psiinit\!\bra{\psi_\mrm{init}}$ possess an $\mrm{SU}(2)$ symmetry. This means, among other things, that the ansatz circuits $C$ conserve the quantum numbers $S$ and $S_z$.  See, e.g., Ref. \cite{sakurai1995modern} for more background on total spin operators and symmetry.

For the HAFM on the chain, there is formal proof that the ground state lays within the $S=0$ sector, for example, via the Bethe ansatz \cite{franchini2017introduction} or through the more general result by Lieb and Mattis \cite{lieb1962ordering}. For the (tripartite) kagome lattice, such formal proof is unknown. There is, however, substantial evidence that $S=S_z=0$ for the ground state of the HAFM on (patches of) the kagome lattice \cite{lecheminant1997order,sindzingre2009low-energy,lauchli2011ground,yan2011spin,lauchli2019s,nakano2011numerical}.

We now prove the invariance of the variational energy $E(\theta)$ under the choice of two-qubit triplet state. The two-qubit triplet states are
\begin{align*}
  \ket{t_1}&=\ket{00},\\
  \ket{t_0}&=\bS_-^{(1,2)}\ket{t_1}/\sqrt{2}=(\ket{01}+\ket{10})/\sqrt{2},\\
  \ket{t_{-1}}&=\bS_-^{(1,2)}\bS_-^{(\mrm{1,2})}\ket{t_1}/2=\ket{11},
\end{align*}
with $\bS_\mp^{(1,2)}$ the lowering (raising) operator on two qubits.
  Without loss of generality, change the first singlet of the dimer covering $\ket{\psi_\mrm{init}}$ into any of the triplet states, and denote the remainder of the initial state, which is a $(S=0,S_z=0)$ dimer covering on $n-2$ qubits, by $\ket{\mrm{dim}}$. Thus, we obtain the initial state $2^{-m/2}\left[\bS_-^{(1,2)}\right]^m\ket{t_1}\ket{\mrm{dim}}$, with $m\in\{0,1,2\}$ specifying the specific triplet state. Note, $\bS^{(1,2)\dagger}_\mp=\bS^{(1,2)}_\pm$ and $\bS^\mrm{(tot)\dagger}_\mp=\bS^\mrm{(tot)}_\pm$. Then,
	\begin{align*}
		E(\theta)&=2^{-m}\bra{t_1}\left[\bS_+^{(1,2)}\right]^m\bra{\mrm{dim} }C^\dagger H C \left[\bS_-^{(1,2)}\right]^m\ket{t_1}\ket{\mrm{dim}}\\
		&=2^{-m}\bra{t_1}\bra{\mrm{dim} } \left[\bS_+^{(\mrm{tot})}\right]^m C^\dagger H C \left[\bS_-^{(\mrm{tot})}\right]^m \ket{t_1}\ket{\mrm{dim}}\\
		&=\bra{t_1}\bra{\mrm{dim} }C^\dagger H C \ket{t_1}\ket{\mrm{dim}},
	\end{align*}
	which does not depend on $m\in\{0,1,2\}$.

Under the influence of noise, the ansatz circuit $C(\theta)$ will generally not produce a pure state $\ket\theta$, but rather a mixed state $\rho_\theta$. If the noise breaks the $\mrm{SU}(2)$ symmetry of $C(\theta)$, $\KVQE_{G/K}$ and CVQE (running with $S=1$ initial states) may abuse this noise to put amplitude on $S=0$ states, thus obtaining unjustly low variational energies. This may be dealt with by symmetry verification \cite{bonet2018low-cost}. Alternatively, a penalty term $A_{P_1}(\bS^2-2\id)^2$, with $A_{P_1}> 0$, may be added to $H$ \cite{mcclean2016theory,ryabinkin2018constrained,kuroiwa2021penalty}. In the Pauli basis, this penalty term has $O(n^4)$ terms, and may hence be costly to compute in every iteration of $\KVQE_{G/K}$ and CVQE.

We now give a concise introduction to Decoherence-Free Subspaces (DFSs) \cite{palma1996quantum,zanardi1997error,duan1998reducing,lidar1998decoherence} and a previous paper on DFSs in the context of VQEs. To introduce DFSs \cite{kempe2001theory}, consider a general system-bath interaction $\tilde H=H_S+H_B+H_{SB}$, where $H_{S(B)}$ acts nontrivially on the system (bath) only. Without loss of generally, $H_{SB}=\sum_\alpha \mc S_\alpha \mc B_\alpha$, where $\mc S_\alpha (\mc B_\alpha)$ acts nontrivially on the system (bath) only. A DFS is a system subspace containing those states $\ket\psi_S$ for which \begin{equation}\label{eq:dfs}
	    \mc S_\alpha \ket\psi_S=\la_\al \ket\psi_S,
	\end{equation}
    for all $\al$, and with $\la_\al$ independent of $\ket\psi_S$. States in a DFS are not affected by the system-bath coupling at all; for an initial product state $\ket\Psi=\ket{\psi}_S\ket{\varphi}_B$, it follows  $\tr_B(\ee^{-\ii t\tilde H}\ketbra{\Psi}{\Psi}\ee^{\ii t\tilde H})=\ee^{-\ii t H_S}\ket{\psi}_S\bra{\psi}_S\ee^{\ii t H_S}$. If, furthermore, $H_S$ does not take $\ket\psi_S$ outside the DFS, states in a DFS remain in the DFS and evolve unitarily despite the nontrivial system-bath coupling. This is generalized straightforwardly to cases where $\tilde H$ is time dependent~(with known time dependence of $H_S$).

	In Ref. \cite{kokail2019self}, a programmable analog quantum simulator is used to simulate the lattice Schwinger model, where the states occurring during their ansatz state preparation remain an approximate eigenstate of $\bS_z^{\mrm{(tot)}}$ and a charge conjugation and spatial reflection (CP) operation after modifications to their variational ansatz.

\vspace{1em}
\section{The fSim gate}\label{sec:fSim}
The native, parametrized two-body gate of Google AI Quantum, called the `fermionic simulation' gate, reads
	\begin{equation*}
		\fSim(\theta,\phi)=
		\left(\begin{array}{cccc}
			1	&	0	&	0	&	0	\\
			0	&	\cos(\theta)	&	-\ii \sin(\theta)	&	0	\\
			0	&	-\ii \sin(\theta)	& \cos(\theta)	&	0	\\
			0	&	0	&	0	&	\ee^{-\ii \phi}
		\end{array}\right),
	\end{equation*}
	and has been demonstrated experimentally for all $\theta\in[0,\pi/2]$ and $\phi\in[-\pi,\pi]$ \cite{foxen2020demonstrating}. Equation (\ref{eq:HEIS_as_fSim}) shows how the $\HEIS$ gate is implemented using the fSim gate. This relation shows that during the unconstrained optimization of the $\KVQE_{G/K}$ or CVQE variational parameters, fSim gates may occur with a parameter $\theta$ that falls outside the range for which implementation has been demonstrated. Nevertheless, $\mrm{fSim}(\theta,\phi)$ gates for general $\theta\in[-\pi,\pi)$ can  be implemented by a single fSim gate after remapping $\theta$ and inserting appropriate single-qubit $Z$ gates, as is shown by the identity
	\begin{align*}
		\fSim(\theta,\phi)=\left\{
		\begin{array}{ll}
			Z_0Z_1\,\fSim(\theta-\pi,\phi)	&	:-\pi \leq \theta < - \pi/2 \\
			Z_0\, \fSim(-\theta,\phi)\,Z_0	&	:-\pi/2 \leq\theta<0 \\
			\fSim(\theta,\phi)	&	:0 \leq \theta < \pi/2 \\
			Z_0\, \fSim(-\theta+\pi,\phi)\,Z_1	&	:\pi/2 \leq \theta < \pi
		\end{array}
		\right.,
	\end{align*}
with $Z_0=Z\otimes \id$ and $Z_1=\id \otimes Z$.

\end{document}